\documentclass[preprint,12pt]{elsarticle}

\usepackage[english]{babel}
\usepackage[strings]{underscore}
\usepackage{microtype}
\usepackage{graphicx}
\usepackage{subcaption}
 \graphicspath{{./img/}}
 \DeclareGraphicsExtensions{.pdf}
\usepackage{booktabs}
\usepackage{multirow}
\usepackage{amsmath}
\usepackage[export]{adjustbox}
\usepackage{algcompatible}
\usepackage{lscape}
\usepackage[noend]{algpseudocode}
\usepackage{algorithm}
\usepackage[table]{xcolor}
\usepackage{xr}
\makeatletter
\newcommand*{\addFileDependency}[1]{
  \typeout{(#1)}
  \@addtofilelist{#1}
  \IfFileExists{#1}{}{\typeout{No file #1.}}
}
\makeatother

\newcommand*{\myexternaldocument}[1]{
    \externaldocument{#1}
    \addFileDependency{#1.tex}
    \addFileDependency{#1.aux}
}

\myexternaldocument{./additional}

\usepackage{xcolor}

\usepackage{amssymb}
\newcommand*{\smallimg}[1]{%
    \raisebox{-.5\height}{%
        \includegraphics[
        width=0.04\textwidth,
        valign=t,
        keepaspectratio,
        ]{#1}%
    }%
}

\journal{Journal of Parallel and Distributed Computing}

\begin{document}

\begin{frontmatter}



\title{Towards an Efficient Combination of Adaptive Routing and Queuing Schemes in Fat-Tree Topologies}

\author[uclm]{Jose Rocher-Gonzalez}
\ead{jose.rocher@uclm.es}
\author[uclm]{Jesus Escudero-Sahuquillo}
\ead{jesus.escudero@uclm.es}
\author[uclm]{Pedro J. Garcia}
\ead{pedrojavier.garcia@uclm.es}
\author[uclm]{Francisco J. Quiles}
\ead{francisco.quiles@uclm.es}
\author[intel]{Gaspar Mora}
\ead{gaspar.mora.porta@intel.com}

\address[uclm]{Departamento de Sistemas Inform\'aticos, Universidad de Castilla-La Mancha, Spain.}
\address[intel]{Intel Corporation, Santa Clara, USA.}

\begin{abstract}
The interconnection network is a key element in High-Performance Computing (HPC) and Datacenter (DC) systems whose performance depends on several design parameters, such as the topology, the switch architecture, and the routing algorithm.
Among the most common topologies in HPC systems, the Fat-Tree offers several shortest-path routes between any pair of end-nodes, which allows multi-path routing schemes to balance traffic flows among the available links, thus reducing congestion probability.
However, traffic balance cannot solve by itself some congestion situations that may still degrade network performance.
Another approach to reduce congestion is queue-based flow separation, but our previous work shows that multi-path routing may spread congested flows across several queues, thus being counterproductive.
In this paper, we propose a set of restrictions to improve alternative routes selection for multi-path routing algorithms in Fat-Tree networks, so that they can be positively combined with queuing schemes.
\end{abstract}


\end{frontmatter}


\section{Motivation}
 \label{sec:motivation}

Nowadays, interconnection networks have become key elements in
High-Performance Computing (HPC) and Datacenters (DC) systems, 
where they connect thousands of computing and storage nodes. Therefore, the global performance of any HPC or DC system is directly coupled 
with the performance of its interconnection network. 
This makes critical to carefully design all the aspects of the interconnection network 
such as the topology, routing, power efficiency, etc. 
so that the network is not the bottleneck of the whole system.

One key design aspect of interconnection networks is the topology.
Efficient topologies should offer full connectivity, 
cost-effectiveness, high-bisection bandwidth, path diversity, etc
\cite{Dally04book}.
The properties of a given topology should be leveraged
by efficient routing algorithms, mainly  balancing traffic among the available paths, in order to reduce the probability
of hot-spot appearance.

Fat-Trees 
are one of the two most common topologies in the supercomputers
of the TOP500 list \cite{Top500}. 
In particular, Fat-Trees offer full bisection bandwidth, inherent fault tolerance capabilities, 
and path diversity, among other advantages. 
Moreover, their structure is naturally deadlock-free between any pair of end-nodes. 
These features are leveraged by several efficient routing algorithms, 
either deterministic \cite{Zahavi2010}
oblivious \cite{ObliviousFT} or adaptive
\cite{DBLP:journals/jsac/ZahaviKK14,DBLP:conf/hoti/GeoffrayH08,DBLP:conf/sc/KimDA06}.

However, even in networks with efficient topologies and routing algorithms,
congestion may appear due to packet flows intensely and persistently contending to use the network resources, then severely degrading network performance.
In the lossless networks used in HPC and DC systems, congestion spreads 
due to the backpressure of flow control, 
forming what is known as a ``congestion tree'' \cite{DBLP:conf/hipeac/GarciaFDJQN05}, that grows from the point where congestion originates (the root)
towards the end-nodes (leaves).
Note that congestion is favored by ``many to one" 
traffic patterns, i.e., by oversubscribed end-nodes (``incast'' congestion), but also by flows contending at the inner network (``in-network'' congestion).

Regardless its origin, congestion may lead to non-congesting flows advancing
at the same pace as congesting ones, due to the 
\emph{Head-of-Line (HoL) blocking} effect \cite{DBLP:journals/tcom/KarolHM87}.
In general, HoL blocking happens when a packet stuck at the head of a buffer or 
queue blocks the packets stored behind in that queue,
even if the latter request free resources.
This effect is especially severe in lossless networks, 
where congesting packets cannot be discarded, hence they cause HoL blocking to non-congesting ones if they share queues.
Moreover, in lossless networks HoL blocking may appear either at the switch where congestion originates 
(\emph{low-order HoL blocking}\cite{DBLP:journals/tcom/KarolHM87})
or at other switches, due to the backpressure of flow control from that switch 
(\emph{high-order HoL blocking}\cite{Jurczyk96phenomenonof}).
Another negative effect of congestion is \emph{buffer hogging} \cite{bufferhogging}
which appears when congesting packets 
hog all the the space of a given buffer.

Among the solutions proposed to deal with congestion and its effects, several techniques implemented in real systems are based
on throttling packet injection at the sources upon congestion notification \cite{InfiniBand2015,rfc3168}. 
However, they do not scale with network size because the notification delay depends on the network diameter, so the sources may react too late, or receive obsolete information \cite{Escudero11icpp}. 
Other methods 
proposed recently \cite{DBLP:conf/sc/JiangDD15, DBLP:conf/hpca/JiangBMD12} use speculative messages that can be dropped under a congestion scenario,
combined with a reservation handshake protocol between source and destination end-nodes.

Another efficient and simple approach to reduce the effects of congestion 
is separating packet flows into different queues 
(or Virtual Channels \cite{Dally98}) at each switch port.
In this way, the sharing of buffer space among 
congesting and non-congesting flows may be reduced, and so the HoL blocking.
This queue-based flow-separation approach is followed by many proposals, usually known as
\emph{queuing schemes}, that differ on the specific policy followed to map flows to queues.
The most prominent of them are analyzed in Section \ref{sec:background:congestion_management}.

Besides, the use of multi-path (oblivious or adaptive) routing has been
considered to deal with congestion. Actually, in some cases these routings may hinder the formation of congested points, but they may also be useless, like in
incast congestion situations. Moreover, these routings may lead to dynamic changes
in the shape of congestion trees, making them more difficult to solve by other congestion
management mechanisms that may be active simultaneously. In that regard, we performed a previous study on the combination of multi-path routing algorithms with some popular queuing schemes in Fat-Trees \cite{rocherHOTI2017}. 
In that study, the experiment results show how multi-path 
routings spread congestion trees throughout the network under heavy congestion scenarios,
such that a higher number of queues are affected. 
Indeed, some combinations of queuing schemes with multi-path routings are 
counterproductive, as they are outperformed by deterministic routing.
Based on that preliminary study, 
this paper further explores several ways to restrict 
the set of routes that can be selected by adaptive routing, analyzing how this impacts on the spreading of 
congestion trees, on the efficiency of queuing schemes, and ultimately, on network performance.
In summary, the main contributions of this paper are:

\begin{itemize}

\item We propose restrictions to the path selection of adaptive routing, such as 
limiting adaptivity to a given network stage,
reducing the set of output ports that can be selected, or modifying the threshold
of queue occupancy that triggers adaptivity.

\item We study how the combined use of several of these restrictions makes the network to behave better.

\item We analyze the impact of applying the proposed restrictions to multi-path routing on different queuing schemes used to reduce HoL blocking.

\item We evaluate different combination of queuing schemes with restricted
adaptive routings under incast and in-network congestion scenarios.

\end{itemize}

The rest of the paper is organized as follows.
Section \ref{sec:background} provides some background on Fat-trees, routing algorithms and queuing schemes. Based on our previous work, Section \ref{sec:problem} analyzes the problems that arise upon combining adaptive routing with queuing schemes. 
Section \ref{sec:improving} describes the proposed restrictions to adaptive routing. 
In Section \ref{sec:evaluation}, 
the proposals are evaluated through simulation results. 
Finally, we conclude the study in Section \ref{sec:conclusion}.

\section{Background}
 \label{sec:background}

\subsection{Fat-Tree Topologies Properties}
 \label{sec:background:FatTrees}

Fat-Trees 
are a popular family of multistage
network topologies widely used in HPC systems.
Indeed, Fat-Trees define a variety of connection patterns,
but not all of them are suited to be implemented in real systems.
For economical reasons, many systems use Real-Life-Fat-Trees (RLFTs),
a sub-class of Parallel Ports Generalized Fat-Trees (PGFTs),  \cite{DBLP:conf/ipps/Zahavi11}, which offer constant bisection bandwidth (CBB) between stages at a reduced cost per switch port.
Switches in a RLFT have the same port count $P$, each switch using $K$ ports ($K=P/2$) to connect with other switches in the next stage ($K$ being known as the switch arity).
The number of end-nodes $N$ in a $T$-stage RLFT is $N=2(K^T)$,
and the number of switches $S=(N\cdot(2T-1))/2K$.

Fat-Tree topologies in general, and RLFTs in particular bear an interesting property:
if shortest-path routing is used, the topology is naturally deadlock-free \cite{Dally04book}. 
Another important property is the shortest-path redundancy or path diversity: there are several shortest paths to choose between any pair of end-nodes. Note that a minimal route between any two end-nodes in a $T$-stage Fat-Tree can visit at most $2T-1$ switches.

\begin{figure*}[!htb]
\centering
\includegraphics[width=.95\columnwidth]{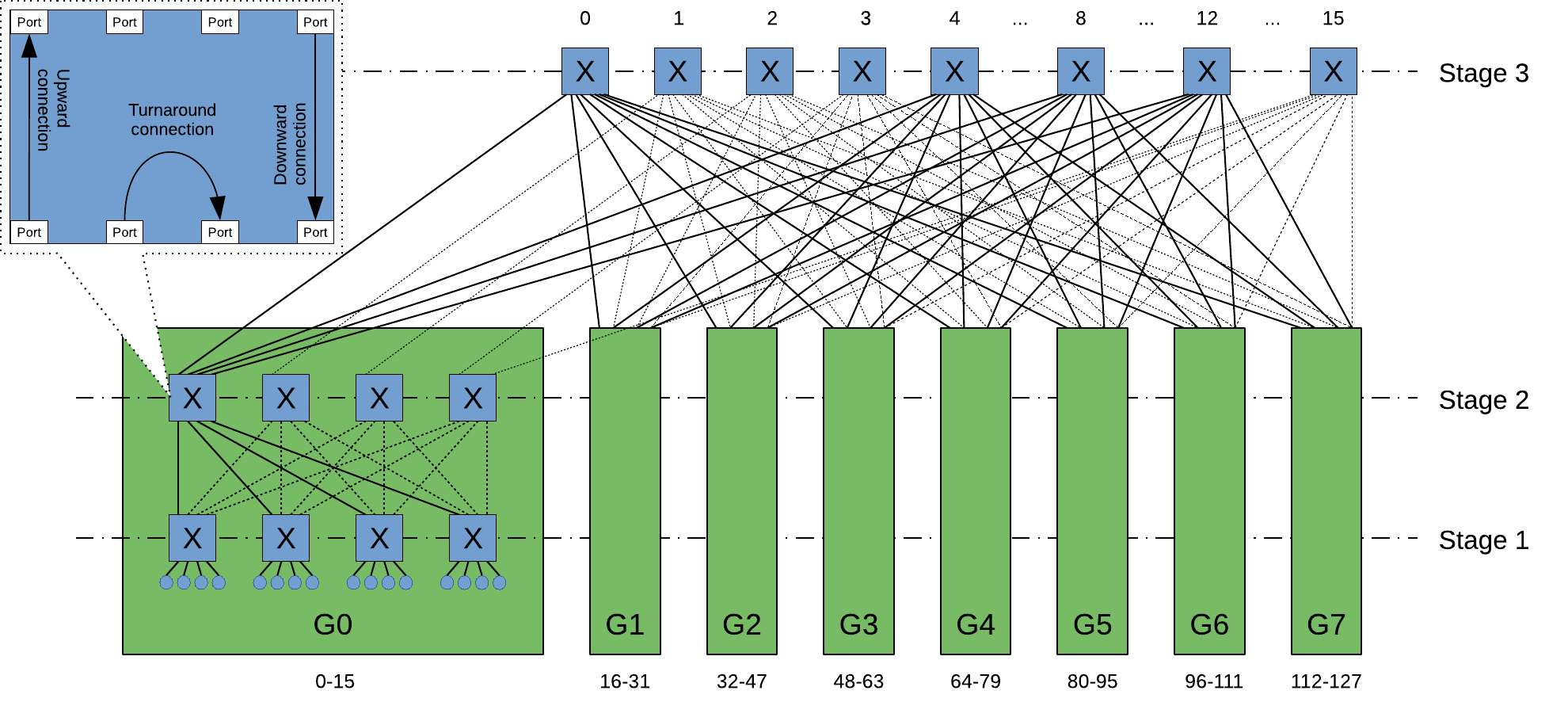}
\caption{A 3-stage RLFT built from 8-port switches that interconnects 128 end-nodes. Each group $G_i$ interconnects 16 end-nodes through 1st- and 2nd-stage switches, and is connected to other groups through 3rd-stage switches. }
\label{fig_RLFT_paths}
\end{figure*}

Shortest-path routing in RLFTs can be divided into an upward phase, a turn-around step, and a downward phase.
First, in the upward phase, any of the $K$ ports of a switch leading to the next stage can be selected to route a packet, and so until the packet reaches the turn-around step of the path. Then, there is only one downward path possible for the packet to reach its destination. Figure \ref{fig_RLFT_paths} shows a 3-stage RLFT, with a zoom to a 2nd-stage switch depicting examples of upward, downward, and turn-around steps of paths performed through internal connections between ports of that switch.


For the sake of clarity, Figure \ref{fig_RLFT_paths} does not show all the links between stages, but mainly (as thicker lines) most of the links supporting paths that cross the leftmost 2nd-stage switch. Nevertheless, it can be seen that there exist multiple shortest-paths between any pair of end-nodes.
These paths diverge at some switch in the upward phase, and later converge before reaching their final destination. Note that the number of shortest paths depends on the switch arity $K$, and on the number of stages $T$. 
Specifically, the maximum path diversity between any pair of end-nodes in a RLFT is $K^{T-1}$ paths.
Note also that some shortest paths with different destination end-nodes may overlap at switch ports and links.
This is relevant if one of the overlapping paths becomes congested, as this could delay the packets following the other paths.
As explained in the next section, the routing algorithm selects the shortest paths finally allowed between any pair of nodes, which furthermore determines the probability of different shortest paths overlapping at ports/links.

\subsection{Routing Algorithms for Fat-Trees}
 \label{sec:background:routing}  
The Fat-Tree shortest-path properties explained above are leveraged by efficient routing algorithms, that can be classified based on the criteria used to select a
specific path among the possible ones between two given end-nodes.

On the one hand, deterministic routing algorithms always select the same path between any given pair of end-nodes, 
and they do not need any information about the network state or traffic conditions.
Nevertheless, these algorithms can take advantage of the path diversity to distribute evenly the routes, so that the number of paths overlapping at each port/link is balanced, at least among port/links at the same network stage. 
This approach is followed by routing algorithms such as
$D$-mod-$K$ \cite{Zahavi2010}, it is very popular as it is easy to implement, requiring a small number of resources and simple operations to determine the path. 
Moreover, it is very efficient for uniform-traffic scenarios and guarantee in-order packet delivery.
However, its performance degrades under congestion situations (leading to HoL blocking), or bursty traffic, or in the case of failures.

On the other hand, multi-path (oblivious or adaptive) routing algorithms may select different paths to route packets from a given source to a given destination. For each packet, a path is selected from a set of available ones between the corresponding pair of end-nodes. 
Oblivious routing algorithms select the paths without regard for network state \cite{Dally04book}. 
Note that deterministic routing can be viewed as an oblivious one, 
whose set of available paths between any pair of end-nodes consists of one
path\footnote{For the sake of clarity, hereafter we will refer to oblivious routing only 
to indicate multi-path routing, i.e. when the number of available paths in the set is greater than one.}.
In general, oblivious routings select paths based on random or round-robin policies, 
trying to balance traffic among the available paths. 
However, as this is done ``blindly'', they cannot react against failures or congestion, once it appears.

On the contrary, adaptive routing algorithms \cite{DBLP:journals/jsac/ZahaviKK14,DBLP:conf/hoti/GeoffrayH08,DBLP:conf/sc/KimDA06} base path selection on some knowledge of network state.
This information in general reflects traffic conditions and allows selecting routes that avoid link failures or congested points.
Indeed, some adaptive routing implementations behave as deterministic when no contention or failures are detected. 
Adaptive algorithms may be complex to implement, depending on the policy to select the paths, 
which can be based on local (link/switch scope) and/or global (network scope) information.
There are several commercial interconnect solutions that implement adaptive routing, such as InfiniBand \cite{haramaty2017adaptive} or Cray's Slingshot\cite{CrayWhitePaper}.

However, adaptive routing algorithms may be useless or even counterproductive under congestion situations. Indeed, in the case of incast congestion, the congesting flows will keep meeting at the congested end-node regardless the followed routes. Under in-network congestion, changing the routes of the congesting flows may move the congested point to other place in the network. In general, if the routes of  congesting flows are adapted, the congestion trees may reach more network ports, making the situation more difficult to solve. The impact of this problem is explained in more detail in Section \ref{sec:problem}

\subsection{Hol-Blocking Reduction through Queuing Schemes}
 \label{sec:background:congestion_management}

Queuing schemes 
divide the buffer space at each switch port into different queues (usually implemented through Virtual Channels (VCs) \cite{Dally87tc}  
or \emph{Virtual Lanes} (VLs) \cite{InfiniBand2015}) where packet flows can be stored separately.

Some proposed queuing schemes \cite{Katevenis98hpca}, 
\cite{GasparRECNIQ}, 
explicitly identify the congesting flows and dynamically isolate them in special queues, so that HoL bloking caused by congesting flows to non-congested ones is prevented. However, 
the implementation of these schemes is not feasible as  
they require resources not supported (as far as we know) by today's commercial interconnects.
On the other hand, other schemes map different flows to different queues 
according to a static policy, i.e. defined before packet injection,
regardless of network state and traffic conditions. 
Most of these ``static'' queuing schemes reduce HoL blocking only partially, but are viable in current components. Some static schemes, such as VOQnet \cite{Dally98}, VOQsw \cite{Tamir92tc}, 
DAMQs \cite{Tamir92tc},
DBBM \cite{Nachiondo10pds}, 
or DSBM \cite{Olesinski09icc} 
are topology and routing agnostic. 
By contrast, other static schemes like 
vFtree \cite{Guay11ipdps}, and
Flow2SL\cite{Escudero14jpdc} are tailored to specific routing algorithms and network topologies, such that HoL blocking is reduced 
more efficiently than agnostic schemes while requiring similar resources .
Indeed, the latter 
schemes take advantage of topology and routing knowledge to separate flows as much as possible with the available queues.

In this paper, we focus on three popular static queuing schemes suitable for Real-Life-Fat-Trees:
DBBM, Flow2SL, and vFtree.
More precisely, 
for a network using switches with \emph{Q} queues per port, DBBM (Destination-Based Buffer Management) maps a packet
addressed to destination \emph{D} to the queue given by the modulo expression $D~mod~Q$. Note that the set of destinations (so, of packet flows) mapped to a given queue does not vary.

On the other hand, both vFtree and Flow2SL 
assume a Fat-Tree topology with $D$-mod-$K$ \cite{Zahavi2010} routing algorithm. 
For a network using switches with \emph{Q} queues per port, vFtree 
maps to the same queue
the packet flows that have identical source switch and are addressed to the same destination switch,
while flows that have the same source switch but are addressed to
\emph{Q} different/consecutive destination switches are mapped to different queues.
As Q is a reduced number, 
several source-destination pairs are mapped to the same queue.
Unfortunately, vFtree is effective only 
in Fat-Trees with fewer than three stages, so Flow2SL was proposed to overcome the vFtree flaws \cite{Escudero14jpdc}.
Flow2SL defines as many groups of consecutive destination end-nodes
as the number of queues available at each port (\emph{Q}). 
Then, 
the packet flows that have the same source group and the same destination group are mapped to the same queue,
while flows that have the same source group but
different destination groups are mapped to different queues.
Note that 
the number of flows
sharing queues decreases (and so HoL blocking) along any path due to the destination balance per link defined by $D$-mod-$K$.

Furthermore, some switch architectures
are less prone to suffer from HoL-blocking. 
For instance, \emph{Output-queued (OQ) switches} implement buffers at output ports, 
so preventing the low-order HoL blocking totally, but requiring a switch speedup equal to the port count, 
which makes them expensive or even unfeasible. 
Besides, \emph{Input-queued (IQ) switches} may implement \emph{Virtual Output Queues}
(VOQs) \cite{Tamir92tc} through demultiplexed entries from the input-port buffers to the internal switch crossbar, 
also preventing low-order HoL blocking.
However, these architectures still can suffer from high-order HoL blocking, hence they may also implement some additional queuing scheme.
In our study we assume the use of IQ switches, either with or without VOQs, as these are usual architectures in commercial interconnects.



\section{Problem Statement}
 \label{sec:problem}

In a previous work \cite{rocherHOTI2017}, we studied how the combination of adaptive or oblivious routing (see Section \ref{sec:background:routing}) with some queuing schemes (see Section \ref{sec:background:congestion_management}) may be counterproductive. Indeed, multi-path routing algorithms may spread congested flows through a higher number of paths than deterministic routing, such that these flows reach more ports and queues, and so produce more HoL blocking.
This can happen in Fat-Tree-like topologies due to their shortest-path diversity (see Section \ref{sec:background:FatTrees}), that allows adaptive and oblivious routings varying the route between two end-nodes. 
Deterministic routing algorithms like $D$-mod-$K$ minimize this problem since the routes do not vary, hence congested flows reach a more reduced number of ports and queues.

\begin{figure*}[!htb]
\centering
\includegraphics[width=.9\columnwidth]{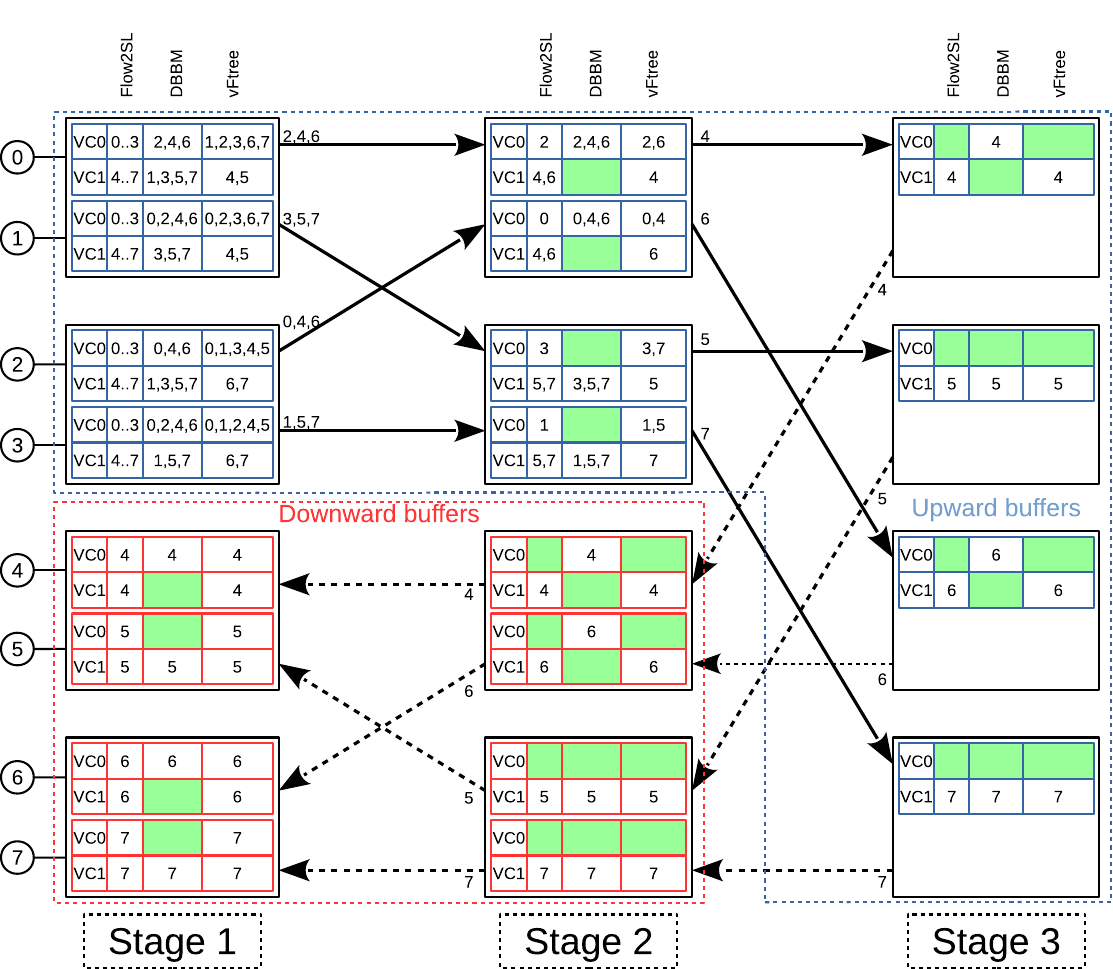}
\caption{An $8$-node Fat-Tree using $D$-mod-$K$ routing and different queuing schemes (Flow2SL, DBBM and vFtree) with $2$ VCs per input port buffer (blue layouts for upward input buffers; red layouts for downward ones). Links crossed in the upward sense are shown as bold arrows while those crossed in the downward sense as dashed arrows.}
\label{fig_RLFT_3_2_destro}
\end{figure*}

\figurename~\ref{fig_RLFT_3_2_destro} shows the mapping of packet destinations to queues (Virtual Channels, VCs) performed by the static queuing schemes Flow2SL, DBBM and vFtree in a $8$-node $3$-stage Fat-Tree using $D$-mod-$K$. In this figure, the output ports are labeled (alongside the connected link) with the destinations of packet routes allowed to pass through them according to $D$-mod-$K$.
Buffers at switch input ports used in the upward phase are called ``upward buffers'' and those used in the downward phase are called ``downward'' buffers.
We assume two VCs per buffer, and that packets are sent from end-nodes $0$-$3$ to nodes $0$-$7$.
Note that $D$-mod-$K$ routing balances all the routes among the links in the upward phase, such that there is a maximum of three destinations of routes crossing an output port in the first stage, and only one destination per output port for the second and third stages.
Also, in the downward phase, there is only one destination of routes crossing the switch output ports.
Note that the different queuing schemes lead to different destination-to-VC mappings, that do not exhibit the same effectiveness in storing separately packet flows with different destinations. For instance, at the 2nd-stage upward buffers, DBBM maps all the packet destinations to the same VC, while wasting the other one. By contrast, Flow2SL and vFtree are able to separate packet flows as much as possible, minimizing the number of destinations mapped to the same VC, as they take advantage of the $D$-mod-$K$ destination balance. 



\begin{figure*}[!htb]
\centering
\includegraphics[width=.9\columnwidth]{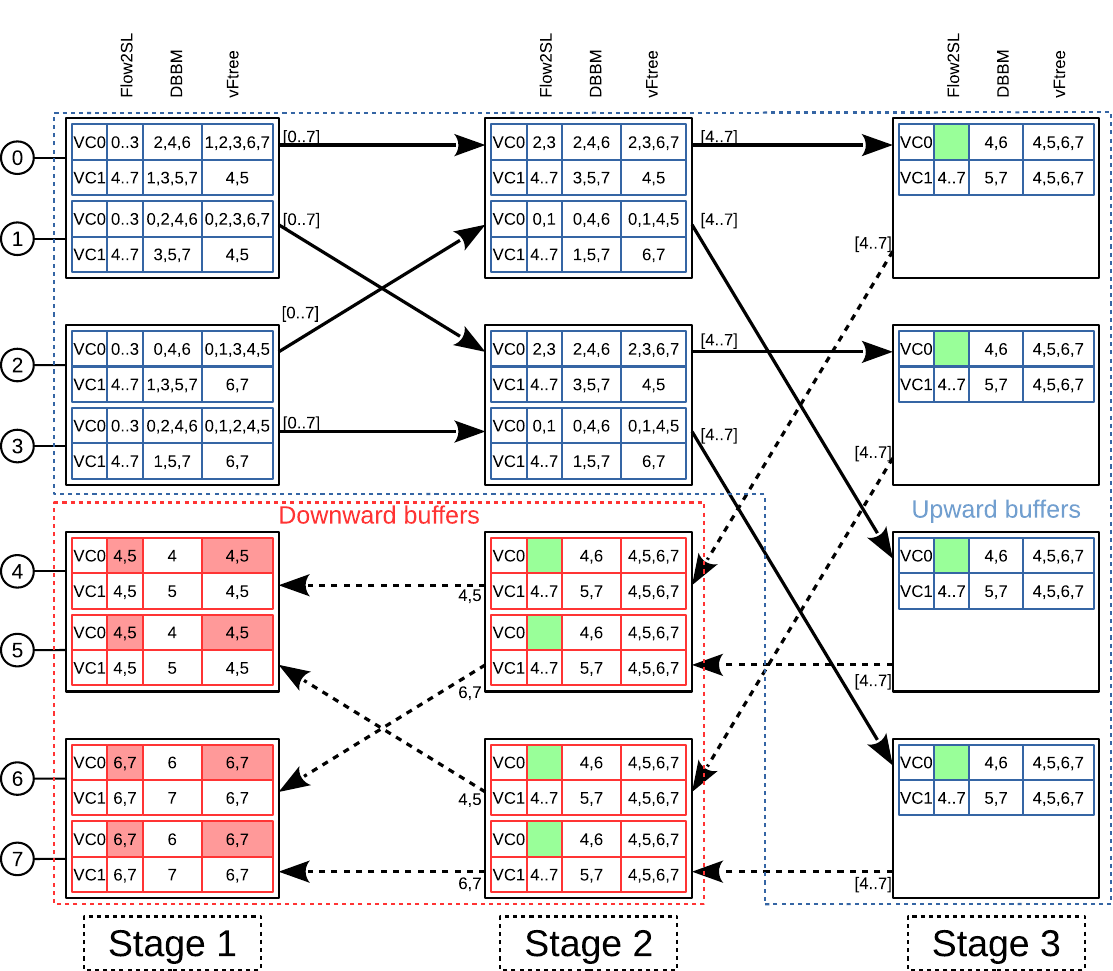}
\caption{An $8$-node Fat-Tree using adaptive/oblivious routing and different queuing schemes (Flow2SL, DBBM and vFtree) with $2$ VCs per input port buffer (blue layouts for upward input buffers; red layouts for downward ones). Links crossed in the upward sense are shown as bold arrows while those crossed in the downward sense as dashed arrows.}
\label{fig_RLFT_3_2_adapt}
\end{figure*}

By contrast, when adaptive or oblivious routing is used, 
switch input ports receive packets following a larger number of shortest paths, compared to $D$-mod-$K$.
Consequently, at a given switch port, the available VCs may have to store packets addressed to a larger number of destinations, such that the queuing schemes will not be able to reduce too much the number of destinations sharing VCs.
\figurename~\ref{fig_RLFT_3_2_adapt} shows the same Fat-Tree network as Figure \ref{fig_RLFT_3_2_destro}, but now using adaptive or oblivious routing, instead of $D$-mod-$K$.
Notice that now the destinations labeling any output port (and link) are more than for $D$-mod-$K$, as any shortest path connecting any source/destination pair is now available.
Therefore, the input buffers have to store packets addressed to more destinations,
so spoiling the effectiveness of the queuing schemes.
Indeed, although the specific impact of this problem vary for the different schemes, in general the number of destinations sharing VCs increases, thus a congested destination would affect more queues and more flows.  
Moreover, in the third stage, vFtree maps all the reachable destinations to all the available VCs, allowing that packets addressed to a single congested destination hog the buffer space.
In the first- and 2nd-stage upward buffers, Flow2SL performs a reasonable mapping, but in the 3rd-stage upward buffers and in the 2nd-stage downward buffers, it maps all the destinations to a single VC, and all the destinations to all the VCs in the first-stage downward buffers.

In summary, although oblivious routing may be used with the aim of delaying the appearance of congestion, and adaptive routing with the aim of bypassing congested areas, using such routings actually may end up dispersing congested flows throughout the network, so creating new congestion situations in other areas, and spoiling the effectiveness of queuing schemes in separating congesting flows.
In order to solve this problem, it would be necessary to configure the multi-path routing so that congestion trees do not spread such that they jeopardize the queuing schemes.
In the following section, we explain our approach to achieve this.

\section{Efficient combination of adaptive routing and queuing schemes}
\label{sec:improving}

In this section, 
we propose a new approach that uses several criteria to restrict the set of alternative shortest-path routes that can be selected by adaptive routing algorithms, and leverages the queuing schemes to reduce HoL blocking.
In the following sections, we provide a general overview of this approach, then we describe the proposed restrictions, the impact of using them together, and how to include this approach in commercial switch fabrics.

\subsection{General Overview}
\label{sec:improving:overview}

First of all, we assume Real-Life-Fat-Tree (RLFT) networks (see Section \ref{sec:background:FatTrees}), using input-queued (IQ) switches and credit-based flow control, as they are widely used in lossless networks for HPC  systems\footnote{Note that the switch architecture and flow control policy are not restricted. We could assume other flow-control policies, such as store-and-forward or priority flow control (PFC), the latter used in lossless Ethernet networks \cite{WhitePaper}.}.
Moreover, our approach assumes that switches always use deterministic routing (e.g. $D$-mod-$K$) by default, but when congestion appears it is applied a modified version of an adaptive routing algorithm\footnote{Note that deterministic routing provides similar results as adaptive routing when congestion levels are moderate \cite{Gomez07ipdps}.}.
Basically, our approach is based on the observation that current path selection policies base their decisions on the information from the free buffer space at neighboring switches, and these decisions may lead to congestion spreading in the network (see Section \ref{sec:problem}).
To deal with this problem, we need to answer an important question: \emph{what are the most significant factors impacting on the congestion trees spreading?}

At a given switch input port, the path selection policy looks at the available credits information for all the neighboring switches, which can be potentially selected to route a packet.
Then, it selects the output port whose neighboring switch input port has the highest number of available credits.
Note that the available credits in the neighboring switches show the \emph{queue occupancy} and so the level of congestion in the alternative routes.
Based on this information, maybe we do not want to select alternative routes for those packets, because it is highly probably that they are contributing to generate a congestion situation, or maybe we want to delay as much as possible the selection of these alternative routes.
Moreover, inside a switch the number of output ports connecting to neighboring switches, which can be potentially selected to route a packet, determines the congestion spreading probability.
The lower this number, the lower the probability to spread congestion trees.
This number depends on both the \emph{switch port count} and the \emph{number of destinations being routed through a switch port}.
Specifically, we can distinguish the following factors impacting on the congestion trees spreading:

\begin{enumerate}
    \item The \textbf{occupancy level of switch queues (or VCs) of neighboring switches} provides information to identify congesting flows clogging a given switch buffer in the network.
    Current path selection policies measure this level and select the route whose associated queue has the lowest occupancy.
    We need to modify this behavior, so that we delay as much as possible the routing of packets through alternative routes, and, meanwhile, leverage the available queuing scheme to reduce HoL blocking.
    
    \item The \textbf{number of destinations being routed through a switch port} is larger when adaptive routing is used (see Section \ref{sec:problem}), compared to deterministic routing.
    This increases the probability of spreading congestion trees.
    We need to lower this number in order to reduce the number of destinations mapped to queues by the queuing schemes, which implies to reduce the number of alternative routes available in switches at some stages. 
    One solution to this consists on applying adaptivity only in one stage in the RLFT topology.
    
    \item The \textbf{number of switch port counts used to route packets} is an indicator of the number of alternative routes that can be selected at a given switch, where the path selection function may select as many alternative routes as the switch port counts.
    In RLFT topologies, this number depends on the switch arity $K = P/2$ in the upward routing phase, while in the downward stage there is only one possible output port to be selected.
    As this number of port counts determines the potential congestion spreading degree at a given switch, we need to bias these routes by reducing the number of ports that the path selection function may choose.
\end{enumerate}

Based on these factors, we have proposed a set of criteria to restrict the number of routes that may be chosen by the path selection function of the adaptive routing algorithm, so that we reduce the adaptivity degree.
In addition, these criteria will help the queuing schemes to decrease the number of different destinations mapped to queues, thus reducing HoL blocking probability.
However, note that reducing adaptivity too much may hinder the path selection function from selecting alternative routes, so that we need to configure these criteria to guarantee certain level of adaptivity.

On the other hand, apart from the factors impacting the congestion spreading, described before, there is another important question that we need to address when using the proposed criteria: \emph{To what extent can the network performance improve if several of these criteria are used at the same time?}
It is not easy to answer this question, since we are looking to combine a set of different criteria with popular queing schemes in an efficient manner.
In some cases, the congestion spreading is reduced even more when we combine these criteria, but in other situations, the combination of these criteria may be counterproductive, as we analyze in Section \ref{sec:evaluation}.

\subsection{Restrictions to the path selection of adaptive routing.}
\label{sec:improving:restrictions}

Based on the three factors described before, we have proposed a set of criteria that restrict the number of alternative routes which can be chosen by the path selection policy.
Specifically, the first criterion is to use a \emph{triggering threshold} that measures the queues (or VCs) occupancy of neighboring switches, so that it ``triggers'' alternative path selection only if the occupancy of a specific queue exceeds this threshold.
We also propose a second criterion to \emph{limit the adaptivity to a single topology stage}, which will enforce the routing algorithm to only select among alternative routes in one stage in the topology.
Finally, we propose a third criterion to \emph{restrict adaptivity per switch port counts} so that we allow the routing to select among a subset of the output ports to route packets.
In the following subsections, we describe these criteria and the options to configure them.

\subsubsection{Triggering Threshold}
\label{sec:improving:restrictions:triggering}

According to factor \#1 in Section \ref{sec:improving:overview}, we need to measure the queue occupancy of neighboring switches.
We assume that, instead of always selecting alternative routes, the routing path-selection function uses deterministic routing by default (e.g., $D$-mod-$K$), until the occupancy of the ``deterministic'' queue reaches the ``triggering'' threshold, then selecting an alternative route.
If we tune this threshold properly, we can reduce the congestion trees spreading, since congestion situations do not remain forever in the network, and we can mitigate the impact of congesting packets being stored in a given queue, thanks to the use of a queuing scheme to reduce HoL blocking.

In our model, we measure the number of available credits\footnote{Note that this flow control policy is used by some of the network technologies available in the market, such as InfiniBand.} per queue in the neighboring switches, so that we use the triggering threshold based on this information.
In other words, when the number of available credits in some queue in the neighboring switch decreases until certain level, we assume that the queue occupancy has reached the triggering threshold, and then adaptive routing can be applied.
Specifically, we have defined the following options to configure the \emph{triggering threshold}:

\begin{itemize}
\item\textbf{No threshold (NoTH).} The path selection function selects the output port with the highest number of free credits. 
This is the default behavior of fully-adaptive routing algorithms, which ends up routing congested packets through non-congested routes.
We have defined this option since the following criteria may not combine too well with the triggering threshold when a particular queuing scheme is used.

\item\textbf{One threshold (TH).} When this option is used, the path selection function uses $D$-mod-$K$ routing until the selected output port decreases the number of available credits in the next switch queue until a certain threshold, known as the \emph{low triggering threshold} (LTTh).
In this moment, the path selection function chooses an alternative route among the available ones, until the original port selected by $D$-mod-$K$ recovers a number of available credits that is higher than the LTTh value.
Note that this value determines the sensitiveness to select alternative routes when congestion appears.
The tuning of the LTTh value is discussed in Section \ref{sec:evaluation:parameters}.

\item\textbf{Two thresholds (2TH).} In this case, the path selection function is also triggered when the output port selected by $D$-mod-$K$ decreases its available credits until the LTTh value.
By contrast, alternative routes are selected until the port initially chosen by $D$-mod-$K$ releases the available credits to a number higher than the LTTh value, which is given by the \emph{high triggering threshold} (HTTh).
When this happens, the path selection function uses $D$-mod-$K$ again.
Hence, this option maintains the alternative paths selection during more time, compared to the TH option, which means that the 2TH option relaxes the adaptivity restriction, compared to the TH option.
\end{itemize}

Consequently, by applying the TH and 2TH options, we achieve that packets belonging to congesting flows remain in the same queue for longer, instead of being routed immediately through an alternative route.
This prevents congestion trees from spreading quickly throughout the network, as it happens when full adaptivity is used.
Moreover, whereas a packet is stored at a given queue, the queuing scheme separates congesting from non-congesting packets, so that HoL blocking is reduced.
The \emph{triggering criterion} can be used at every switch in the Fat-Tree, regardless its position in the topology.

\subsubsection{Limit the adaptivity to a single topology stage}
\label{sec:improving:restrictions:stages}

\begin{figure*}[!h]
\begin{subfigure}{0.49\textwidth}
\includegraphics[width=0.9\columnwidth,left]{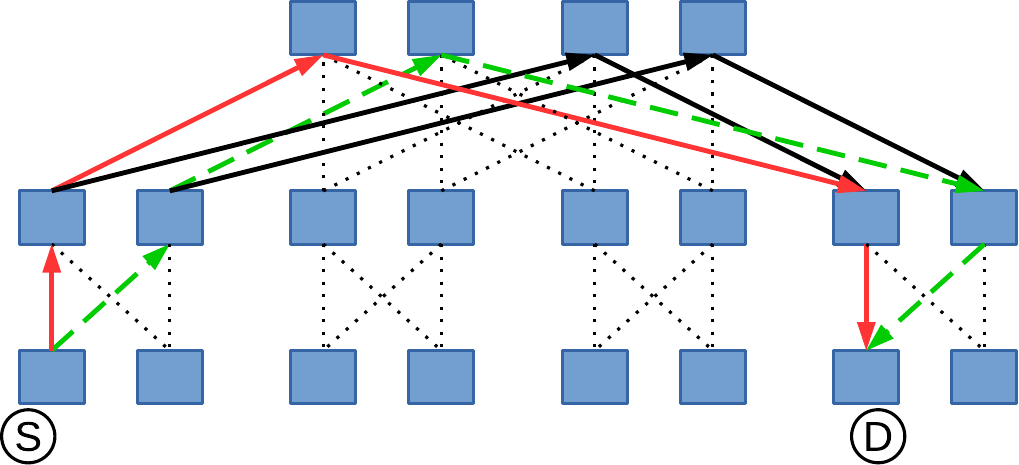}
\caption{First stage (1S).}
\label{fig_adaptive_path_fs}
\end{subfigure}
\begin{subfigure}{0.49\textwidth}
\includegraphics[width=0.9\columnwidth,right]{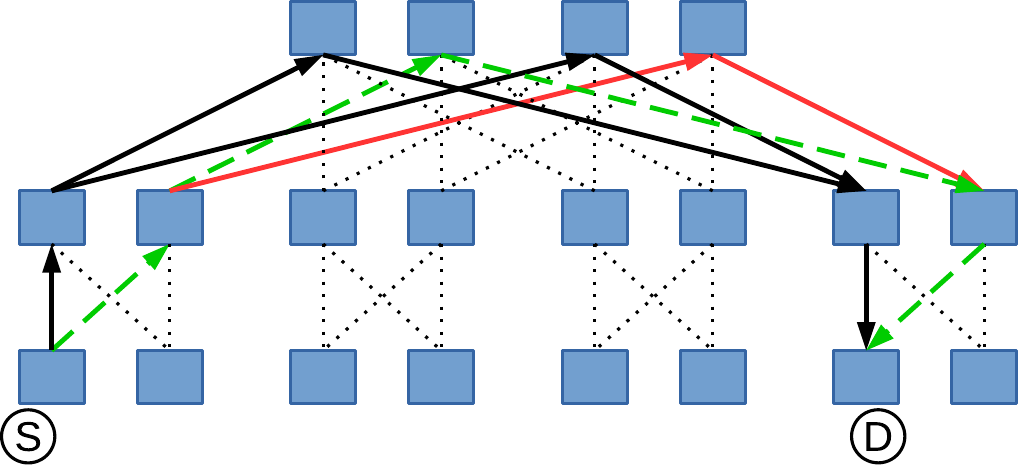}
\caption{Second stage (2S).}
\label{fig_adaptive_path_ss}
\end{subfigure}
\caption{Adaptive routing limited to a single stage in a $3$-stage RLFT ($K=2$). Green dashed lines represent the deterministic routing paths. Red bold lines represent the alternative path to that selected by deterministic routing. Black bold lines represent those shortest-paths not selected by the routing algorithm.}
\label{fig_traffic_paths}
\end{figure*}

According to factor \#2 described in Section \ref{sec:improving:overview}, we want to reduce the number of destinations being routed through a switch port, so that the number of destinations mapped to queues by the queuing schemes is lower.
In other words, we want to enforce the routing algorithm to select among alternative routes only in some stages in the RLFT topology.
For instance, the routing adaptivity could be applied in the first stage (1S), the second stage (2S), etc.
Note that fully adaptive routing algorithms (see Section \ref{sec:problem}) apply adaptivity in all the stages (*S), which provokes that congestion may be spread and queuing schemes being useless.
\figurename~\ref{fig_traffic_paths} shows an example of the 1S and 2S options when they are used in a 3-stage RLFT topology.

In Figure \ref{fig_adaptive_path_fs} the alternative paths selection is limited to the first stage, i.e., the 1S option.
Note that, in the first stage, there are two possible routes from source $S$, given by $K=2$, to reach destination end-node $D$, so that congesting flows could be spread through the two links in the upward direction from the switch connected to source $S$.
In the second stage, it is applied deterministic routing (e.g., $D$-mod-$K$), so that there is only one colored route from the two leftmost switches in the second stage to switches in the third stage.
Note that the bold arrows represent those possible paths between second and third stage which are not selected, due to the routing restriction.
In Figure \ref{fig_adaptive_path_ss}, adaptivity is only used in the second stage, while deterministic routing is applied between first and second stages.
Again, there are two different routes per switch (given by $K=2$) between 2nd and 3rd stages for source $S$ to communicate with destination $D$.
However, the number of possible routes reaching the second stage is lower than that of Figure \ref{fig_adaptive_path_fs} as we move upwards in the topology.
Note that the number of shortest paths crossing through a given port in the second stage is more reduced for the 2S option, compared to when we use the 1S option.
Hence, when we apply the 1S or 2S options, we reduce the number of routes that can be selected to cross through a given switch port, and so the number of destinations mapped to the different queues performed by the queuing scheme.
Therefore, the use of this restriction reduces one the problems of combining adaptive routing with queuing schemes, described in Section \ref{sec:problem}, since, as we reduce the number of destination mapped per queues, then we lower the HoL blocking probability.

\subsubsection{Restrict adaptivity per switch port counts}
\label{sec:improving:restrictions:ports-count}

Considering factor \#3 described in Section \ref{sec:improving:overview}, we want to bias the number of ports used to choose alternative routes.
This means that we will consider a subset of the $K$ ports available if we are routing in the upward direction, while we will select only one port in the turn-around step or downward phase.
Depending on the routing phase, the path selection function needs to select among $K$ ports (upward phase) or one port (downward phase).
Moreover, in the last stage, the routing performs the turn-around step from one input port to one output port.
Note that if fully-adaptive or oblivious routing is used, all the $K$ ports are candidates for the path selection function in the upward phase.
On the contrary, if deterministic routing is used (e.g., $D$-mod-$K$), only one output port will be selected both in the upward and downward phase.

We propose to reduce the number of $K$ ports that can be selected to route packets by means of a new restriction.
Specifically, the path selection function will choose among
$K/\Delta$ output ports in the upward direction ($2 \leq \Delta \leq K$).
If $\Delta=1$ this restriction is not applied.
In other words, note that $\Delta$ is the number of not selected ports in between two candidate ports that can be chosen to route the same packet by the adaptive routing algorithm.
In this way, the path selection function will route the packets through a more reduced number of ports, so that the adaptivity degree is reduced.
Note that the larger $\Delta$, the more restricted the adaptivity.
Circles colored in grey (i.e., NoTH, *S and K) show the regular options to configure fully-adaptive routing algorithms, i.e., without using any restriction to the path selection policy.

\subsection{Using proposed restrictions at the same time}

\begin{figure*}[htb!]
\centering
\includegraphics[width=0.5\columnwidth]{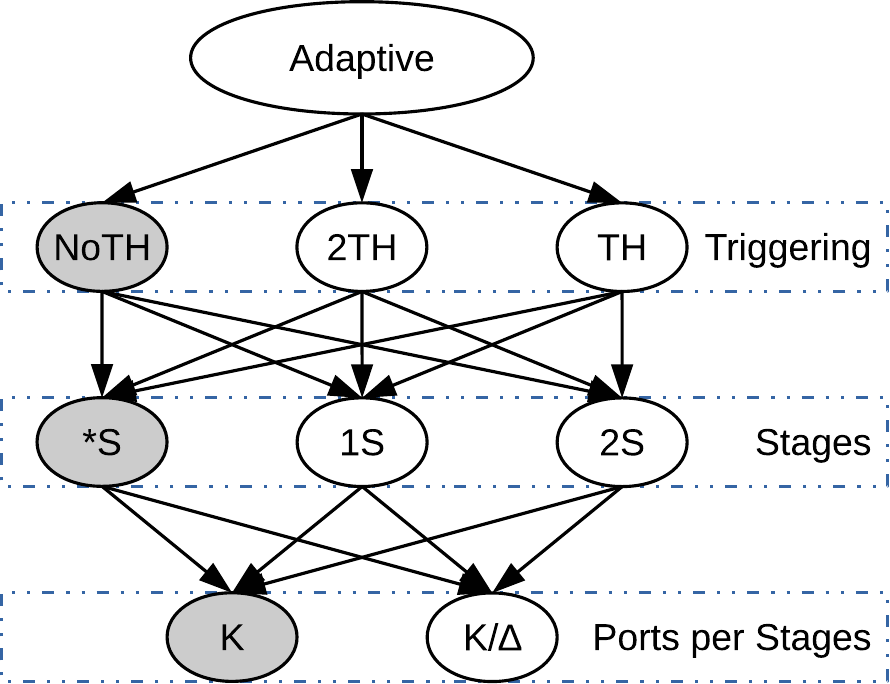}
\caption{Possible combinations of the proposed restrictions. The options for each restriction are more aggressive as we move towards the right in the squares.}
\label{fig:combinations}
\end{figure*}

As we have described before, it is possible to combine the different criteria to further modify the adaptive path selection policies, while leveraging the use of queuing schemes.
\figurename~\ref{fig:combinations} shows a diagram of the proposed restrictions, organized in horizontal dashed squares.
Inside these squares, we can see the proposed options that can be used to configure these restrictions.

Note that $\Delta = 1$ for the $K$ option in a grey circle.
This figure also shows all the possible combinations among the proposed restrictions.
Arrows create a descending path linking one option per criteria, so that only one option per criteria can be used.
For instance, we could configure our adaptive routing algorithm to use the TH, 1S and $K/\Delta$ options.
As another example, we can combine any of the \emph{Stages} options with the $K/\Delta$ option of the \emph{Ports per Stages} criterion, in order to reduce the number destinations being mapped to queues by the queuing schemes at a given stage. 
In this way, the paths through where congestion can be spread are reduced.
Note that we could restrict even more adaptivity, if we combine the \emph{Ports per Stages} and the \emph{Stages} criteria (e.g., 1S option), with the \emph{triggering} criterion.
In this way we can select among $K/\Delta$ output ports per switch in the upward phase, only in the first stage, then applying $D$-mod-$K$ in the other stages.

Algorithm \ref{algorithmAdaptive} shows the pseudo-code used to apply the proposed restrictions combined in the same routing logic.

\algrenewcommand\algorithmicindent{2em}%
\algnewcommand\algorithmicforeach{\textbf{for each}}
\algdef{S}[FOR]{ForEach}[1]{\algorithmicforeach\ #1\ \algorithmicdo}
\algnewcommand\algorithmicswitch{\textbf{switch}}
\algnewcommand\algorithmiccase{\textbf{case}}
\algdef{SE}[SWITCH]{Switch}{EndSwitch}[1]{\algorithmicswitch\ #1\ \algorithmicdo}{\algorithmicend\ \algorithmicswitch}%
\algdef{SE}[CASE]{Case}{EndCase}[1]{\algorithmiccase\ #1}{\algorithmicend\ \algorithmiccase}%
\algtext*{EndSwitch}%
\algtext*{EndCase}%
\newcommand{\vars}{\texttt} 
\newcommand{\func}{\textsf}
\let\oldReturn\Return
\renewcommand{\Return}{\State\oldReturn}


  \begin{algorithm}[htb]
   \scriptsize
   \caption{Restricted Path Selection Function}
  \label{algorithmAdaptive}
  \hspace*{\algorithmicindent}\textbf{Input:} \\
  \hspace*{\algorithmicindent}\hspace*{\algorithmicindent}\vars{Destination}: Packet destination.\\
  \hspace*{\algorithmicindent}\hspace*{\algorithmicindent}\vars{Queue}: Queue assigned to the packet by the static queuing scheme.\\
  \hspace*{\algorithmicindent}\hspace*{\algorithmicindent}\vars{Triggering}: Options to configure the triggering restriction (NoTH, TH and 2TH).\\  
  \hspace*{\algorithmicindent}\hspace*{\algorithmicindent}\vars{Stages}: Enables routing adaptivity (*S, 1S, 2S, etc.).\\
  \hspace*{\algorithmicindent}\hspace*{\algorithmicindent}\vars{Delta}: Apply the ports per Stages restriction.\\
  \hspace*{\algorithmicindent}\textbf{Output:}\\
  \hspace*{\algorithmicindent}\hspace*{\algorithmicindent}\vars{Port}: Output port selected.\\
  
  \begin{algorithmic}[1]
     \scriptsize
  
  
  \Function{RestrictedPathSelection}{Destination, Queue, Stages, Delta, Triggering}
  \State{\vars{Port} $\leftarrow$ \func{dmodkUpstage}(\vars{Destination})}  
  \If{\vars{Stages}} \bf \# If true, it builds the 'OutputPortList' with the possible output ports
  \State{\vars{ModuloResult} $\leftarrow$ \vars{Destination} $\%$ \vars{Delta}}
  \State{\vars{OutPortList.RemoveIf(!ModuloAndCompare(Delta,ModuloResult))}}
  \State{\vars{OutPortList.SortBy(Port)}}

 \Switch{\vars{Triggering}} \bf \# This part selects the triggering options
  \Case{\vars{NoTH}}
  \State{\vars{MaxCredits} $\leftarrow$ $0$}
  \ForEach{\vars{i} in \vars{OutPortList}}
   \If{\vars{CreditsIn[i][Queue]} $>$ \vars{MaxCredits}}
   \State{\vars{MaxCredits} $\leftarrow$ \vars{CreditsIn[i][Queue]}}
   \State{\vars{Port} $\leftarrow$ \vars{i}}
    \If{\vars{MaxCredits} = \vars{AllCreditsInQueue}}
    \State{break}
    \EndIf
   \EndIf
  \EndFor
\State{break}
\EndCase

\Case{\vars{TH}}
  \If{\vars{CreditsIn[Port][Queue]} $<$ \vars{LTTh}}
   \State{\vars{MaxCredits} $\leftarrow$ \vars{LTTh}}
  \ForEach{\vars{i} in \vars{OutPortList}}
   \If{\vars{CreditsIn[i][Queue]} $>$ \vars{MaxCredits}}
   \State{\vars{MaxCredits} $\leftarrow$ \vars{CreditsIn[i][Queue]}}
   \State{\vars{Port} $\leftarrow$ \vars{i}}
    \If{\vars{MaxCredits} = \vars{AllCreditsInQueue}}
    \State{break}
    \EndIf
   \EndIf
  \EndFor
  \EndIf
  \State{break}
\EndCase


 \Case{\vars{2TH}}
  \If{\vars{CreditsIn[Port][Queue]}$<$\vars{LTTh} $OR$ \par
  \hskip\algorithmicindent \hskip\algorithmicindent \hskip\algorithmicindent
  \hskip\algorithmicindent
  \vars{PortCongested[Port][Queue]}} 
   \State{\vars{MaxCredits} $\leftarrow$ \vars{LTTh}}
   \If{\vars{CreditsIn[Port][Queue]}$<$\vars{HTTh}}
    \State{\vars{PortCongested[Port][Queue]} $\leftarrow$ True}
  \ForEach{\vars{i} in \vars{OutPortList}}
   \If{\vars{CreditsIn[i][Queue]} $>$ \vars{MaxCredits}}
   \State{\vars{MaxCredits} $\leftarrow$ \vars{CreditsIn[i][Queue]}}
   \State{\vars{Port} $\leftarrow$ \vars{i}}
    \If{\vars{MaxCredits} = \vars{AllCreditsInQueue}}
    \State{break}
    \EndIf
    \EndIf
   \EndFor
   \Else{ \vars{PortCongested[Port][Queue]} $\leftarrow$ False}
   \EndIf
   \EndIf
\EndCase
  \EndSwitch
\EndIf
 \Return{Port}
   \EndFunction
  \end{algorithmic}
\end{algorithm}

After receiving the input parameters (i.e., the specific packet \emph{Destination}, \emph{Queue} where the static queuing scheme has stored that packet, and the \emph{Triggering}, \emph{Stages} and \emph{Delta} options), the algorithm builds the 'OutputPortList' including all the possible output ports that may be selected to route the packet addressed to \emph{Destination}.
Note that if the restriction parameters are empty the routing logic would work in deterministic mode, and if only the Stages option is set to '*S' then the logic enables the fully adaptive mode.

\subsection{Discussion on the effects of using the proposed approach}
\label{sec:improving:effects}

In this section, we analyze how the proposed restrictions for the path selection function leverage the queuing schemes, so that they preserve their efficiency when reducing HoL blocking.
We focus on the number of destinations being routed through a given switch output port, and compare this number with that of deterministic, fully-adaptive and oblivious routing algorithms.
Table \ref{tab:destperlink} shows this analysis for a 3-stage RLFT topology, with different configurations for routing algorithms using the proposed restrictions.

\begin{table}[!hbt]
 \centering
 \caption{Number of destinations routed through an output port in a $3$-stage RLFT, using different routing algorithms. We assume that only shortest paths are considered. $E$ is for end-nodes, and $SX$ is for switches at stage $X$. $U$ denotes upward phase and $D$ downward phase. $N$ is the total number of end-nodes.}
 \label{tab:destperlink}
\resizebox{.99\textwidth}{!}{%
\begin{tabular}{@{}l|cccccc@{}}
\toprule
Routing and Restrictions     & EU     & S1U                & S2U                      & S3D                           & S2D                     & S1D \\ \midrule
Deterministic  ($D$-mod-$K$) & $N-1$    & $\dfrac{N-K}{K}$   & $\dfrac{N-K^2}{K^2}$     & $\dfrac{N/K^2}{2K} = 1$       & $1$                        & $1$ \\ \midrule
Fully adaptive and oblivious\\ (no restrictions) 	& $N-1$    & $N-K$              & $N-K^2$                  & $\dfrac{N}{2K} = K^2$         & $\dfrac{N}{2K^2} = K$      & $1$ \\ \\ \midrule 
Adaptive Stage 1 (1S)  	    & $N-1$    & $N-K$              & $\dfrac{N-K^2}{K}$       & $\dfrac{N}{2K^2} = K$         & $\dfrac{N}{2K^2} = K$      & $1$ \\ \\
Adaptive Stage 2 (2S)  		& $N-1$    & $\dfrac{N-K}{K}$   & $\dfrac{N-K^2}{K}$       & $\dfrac{N}{2K^2} = K$         & $\dfrac{N}{2K^3} = 1$      & $1$ \\ \\\midrule
All stages (*S) and $K/\Delta$	& $N-1$    & $\dfrac{N-K}{K/\Delta}$ & $\dfrac{N-K^2}{(K/\Delta)^2}$ & $\dfrac{N}{2K/\Delta^2} = K^2/\Delta^2$ & $\dfrac{N}{2K^2/\Delta} = K/\Delta$  & $1$ \\ \\
Adaptive Stage 1 (1S) and $K/\Delta$	& $N-1$    & $\dfrac{N-K}{K/\Delta}$ & $\dfrac{N-K^2}{K^2/\Delta}$   & $\dfrac{N}{2K^2/\Delta} = K/\Delta$     & $\dfrac{N}{2K^2/\Delta} = K/\Delta$  & $1$ \\ \\

Adaptive Stage 2 (2S) and $K/\Delta$ 	& $N-1$    & $\dfrac{N-K}{K}$   & $\dfrac{N-K^2}{K^2/\Delta}$   & $\dfrac{N}{2K^2/\Delta} = K/\Delta$   & $\dfrac{N}{2K^3} = 1$      & $1$ \\

\bottomrule
\end{tabular}}
\end{table}

In this table, columns show the \emph{Routing} algorithm and the different network elements: end-node (E) or switch (S), the stage (1, 2 or 3) and the routing phase ('U' means upward and 'D' means downwards).
Note that deterministic routing (i.e., $D$-mod-$K$) does not apply any adaptivity degree, while fully adaptive or oblivious routing leverage all the possible shortest paths between two end-nodes in the network.
For instance, the column of network element 'S2U' shows the destinations crossing through an output port in a switch placed in the second stage, in the upward phase.
Note that if we apply all the proposed restrictions in an aggressive way, the routing algorithm will tend to avoid the selection of alternative routes, so that it will behave as deterministic routing.

As we have mentioned before, we want to preserve a certain degree of adaptivity in the routing configuration.
As the same time, we need to leverage the queuing schemes to reduce HoL blocking, but without spreading too much the congestion trees.
Consequently, we need to set a trade-off between the adaptivity degree and the use of proposed criteria to reduce this degree.
In addition to this, depending on the selected queuing scheme, we would need to configure the routing algorithm with different restrictions criteria.

For instance,  queuing schemes such as DBBM and vFtree are more suitable to use adaptive routing in the first stage of the topology (see \figurename~\ref{fig_RLFT_3_2_adapt} in Section \ref{sec:problem}).
DBBM is able to map all the possible shortest path destinations among all the VCs in a fair manner. 
Also, vFtree destinations are mapped with an optimal distribution among VCs, since it was designed specially for two-stage RLFTs.
By contrast, using adaptive routing in the second stage is counterproductive to both DBBM and vFtree.
The reason is that they do not share destinations among VCs since the $D$-mod-$K$ routing algorithm is used in the first stage. 
But, in the second stage there are several destinations mapped to the same VC, since adaptive routing is used, thus reducing the queuing scheme efficiency.
Hence, when adaptive routing is applied in the second stage congesting packets may collapse all the VCs.
On the other hand, Flow2SL maps destinations in a properly way among VCs in the forward direction, i.e., VCs do not share packets with the same destination.
However, it was designed taking into account the properties of $D$-mod-$K$ routing algorithm, and it will have issues when adaptive routing is used and congestion appears in the downward phase.
We have observed that, with adaptive routing, Flow2SL maps all the destinations to all the VCs in the downward phase (see Section \ref{sec:problem}), so that Flow2SL could improve its performance if the number of paths being mapped to queues is reduced.

\section{Performance Evaluation}
 \label{sec:evaluation}

In this section, we evaluate the proposed criteria used to restrict multi-path routing path selection when combined with static queuing schemes, compared to deterministic routing and ``clasic'' adaptive routing algorithms.
First, we describe the simulation model and the assumed network configurations for the experiments performed.
Next, we analyze the results obtained from these experiments, performed under synthetic traffic scenarios generating different realistic congestion situations in the network.

\subsection{Simulation Model}
 \label{sec:evaluation:omnet}
 
The simulation model has been developed using the event-driven OMNeT++ framework \cite{OMNeTweb}, which is written in the C++ language.
On top of this framework engine, we have used a custom-made network model already available in our lab, which as been widely used and tested. 
Further details of this model have been disseminated before through several publications \cite{Yebenes13PDP}. 
Basically, we have extended this network model to support adaptive routing algorithms and the restrictions described in Section \ref{sec:improving}.
We have also modeled synthetic traffic patterns generating different congestion scenarios in the network.
Note that our simulation model has been widely tested in the past, and even validated against real networks (e.g., InfiniBand).

\begin{figure*}[htb!]
\centering
\includegraphics[width=0.4\columnwidth]{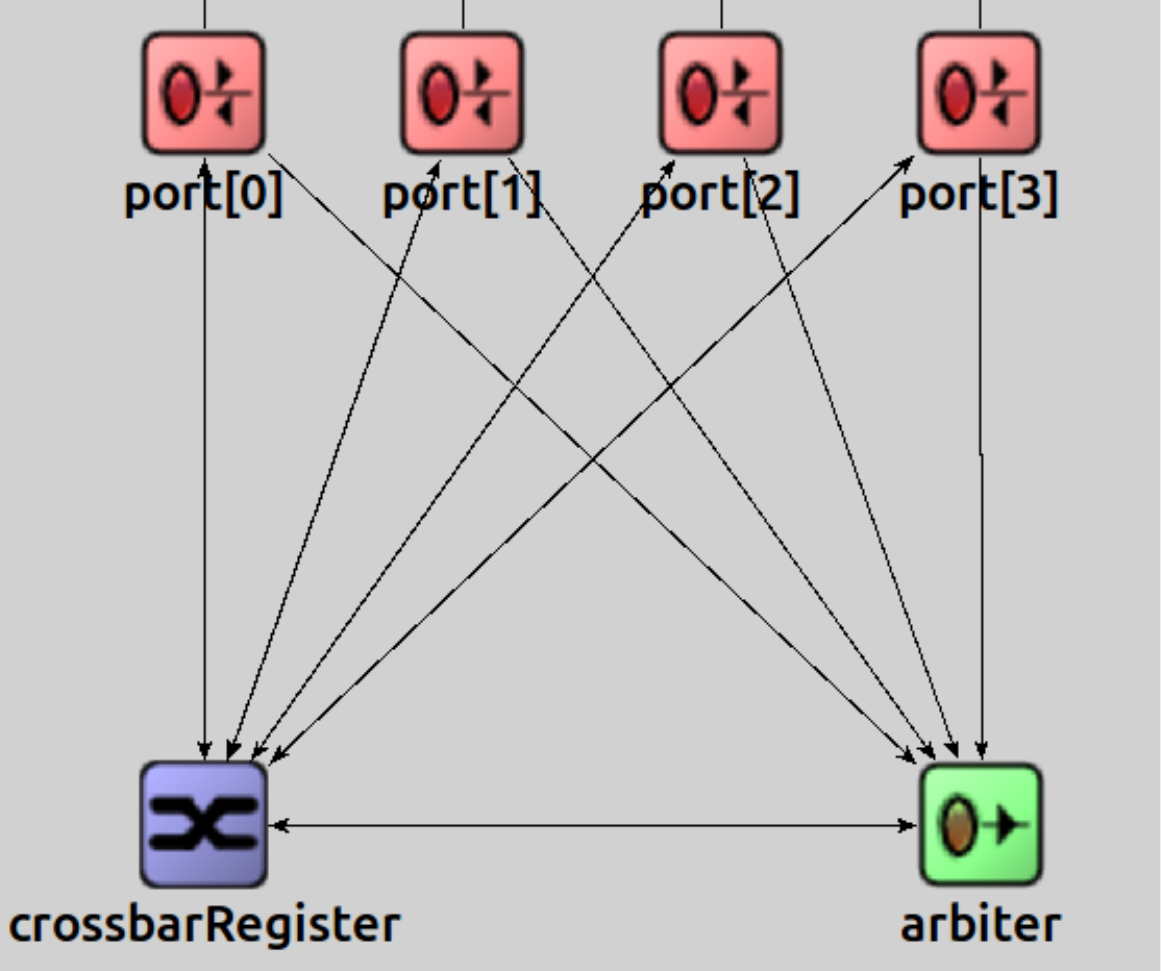}
\caption{NED diagram of the switch model used in the OMNeT++-based simulator.}
\label{fig:omnet}
\end{figure*}

For this work, we assume the input-queued (IQ) switch module, available in the model, which is a ``compound module'' made of several ``simple modules'', such as an array of \emph{port} modules, the \emph{crossbarRegisters} module, and the \emph{arbiter} module.
\figurename~\ref{fig:omnet} shows a NEtwork Description (NED) diagram of the assumed switch model\footnote{Note that the OMNeT++ framework uses the NED language to define the simple and compound modules, and the connections among them using ``gates''. Each of these modules is implemented in C++.}.

Specifically, the \emph{port} module models the credit-based flow control and the Virtual Channels (VCs), and it keeps track of the incoming and outgoing packets to the switch.
At each \emph{port} module, we have modeled buffers by using
the class \texttt{cPacketQueue} provided by the OMNeT++ framework.
The \emph{port} module also models Virtual Output Queues (VOQs), which can be enabled or not in the simulation.
This module also includes the routing functionality applied to the incoming packets, which calculates their output port, based on the routing algorithm, and the VOQ (if VOQs are enabled).
We have extended the model to support adaptive routing algorithms, their different path selection policies, and the restrictions proposed in Section \ref{sec:improving}.
Note that these policies are aware of the credits information available at each \emph{Port} module, and they will be included in the \emph{Arbiter} module, which performs the path selection policy.
Moreover, the \emph{CrossbarRegisters} module models packets ready to get the crossbar, once they have been routed and it is has been checked that there is space in the neighboring switches to forward them.
Note that packets that do not fulfill these conditions will not be set as ready, and will not be forwarded.
Finally, the \emph{Arbiter} module models the arbitration algorithm.
We have assumed the 3-stage \emph{iSlip} algorithm \cite{iSLIP}, widely used in commercial switch implementations.
As we have described before, the arbiter has the global view of the credits information at all the ports of the switch, so that it also models the path selection policies, and the restrictions that we propose in this paper.

Regarding the network topologies, the simulator offers many of them, such as direct, indirect, hybrid, etc.
For this study, we have modeled a $3$-stage Fat-Tree network topology interconnecting $11664$ end-nodes, using $36$-port switches (i.e., $P=36$).
Note that $K$ is the arity and it is half the switch port counts, i.e. $K=P/2$.
We have modeled the routing algorithms described before, i.e. $D$-mod-$K$, oblivious and restricted and non-restricted
(i.e., fully) adaptive routing.
Regarding the restricted adaptive routing, we have modeled the triggering thresholds (TH and 2TH), stages restrictions (1S and 2S) and port per stages restrictions (K/$\Delta$).
We have also modeled the unrestricted paths (AS and K), required by the fully adaptive routing algorithm. 

Regarding the static queuing schemes, we have modeled Flow2SL, DBBM and vFtree, using the following switch buffer organizations:

\begin{itemize}
\item \emph{1-VC}. This is the simplest buffer organization configured with a single VC.
We have modeled this scheme to show the switch architecture and routing algorithms performance, without any HoL blocking prevention.
\item \emph{DBBM-3VC, Flow2SL-3VC, and vFtree-3VC}. Static queuing schemes using $3$ VCs to separate traffic flows.
\item \emph{1VC-VOQ}. This buffer organization implements VOQs for low-order HoL blocking prevention, but it does not implement any queuing scheme.
\item \emph{DBBM-3VC-VOQ, Flow2SL-3VC-VOQ, and vFtree-3VC-VOQ}. We also model stating queuing schemes using $3$ VCs combined with VOQs.
\end{itemize}

The size of the input buffer at each switch and end-node port is $192$KB and packet MTU is $4$ KB (i.e. $48$ packets are stored per buffer).
Regarding the links configuration, we assume serial full-duplex links with $100$ Gbps of link bandwidth and $6$ nanoseconds of link propagation delay.

\begin{figure*}[!h]
\begin{subfigure}{0.49\textwidth}
\includegraphics[width=0.9\columnwidth,left]{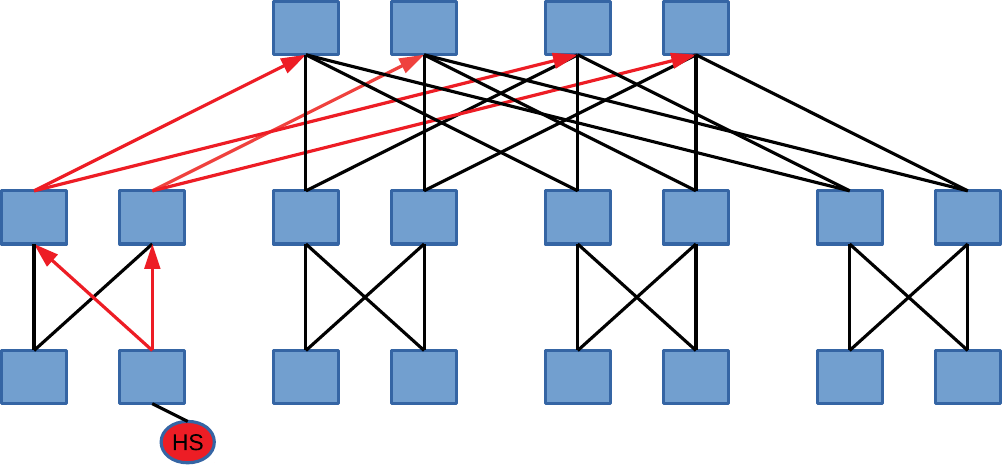}
\caption{1 Hot-Spot generated by downward traffic (incast congestion).}
\label{fig_traffic_HS1}
\end{subfigure}
\begin{subfigure}{0.49\textwidth}
\includegraphics[width=0.9\columnwidth,right]{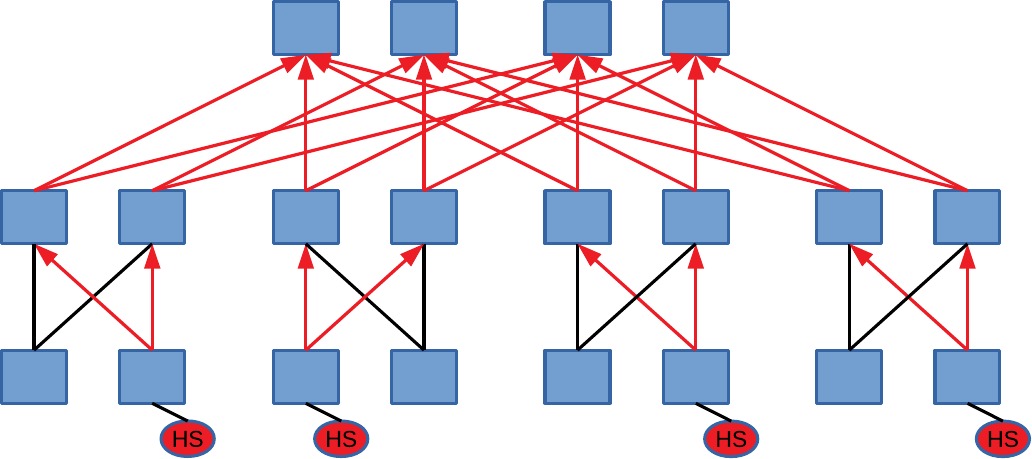}
\caption{4 Hot-Spots generated by downward traffic (incast congestion).}
\label{fig_traffic_HS4}
\end{subfigure}

\centering
\begin{subfigure}[!h]{0.49\textwidth}
\includegraphics[width=0.9\columnwidth,center]{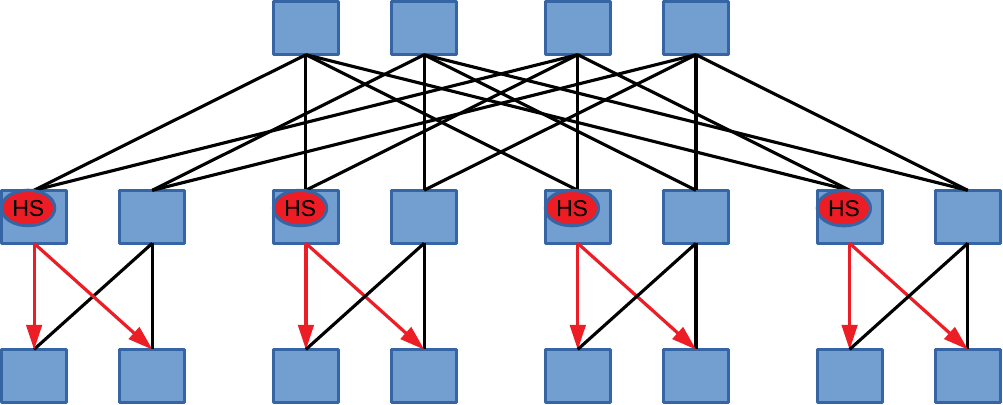}
\caption{Intermediate Hot-Spot generated by upward traffic (in-network congestion).}
\label{fig_traffic_HSW}
\end{subfigure}

\caption{Examples of the modeled hot-spot traffic scenarios in a $3$-stage Fat-Tree ($k=2$) using adaptive routing. Red arrows show the backpressure generated by congestion, and the directions towards where congestion trees grow.}
\label{fig_traffic_scenarios}
\end{figure*}

Regarding the traffic workloads, we have extended the simulation tool to model several types of communication patterns that lead to congestion scenarios, which are shown in \figurename~\ref{fig_traffic_scenarios}.
The congestion roots (hot-spots) are depicted by means of red circles labeled with ``HS''.
We assume that these hot-spots can be generated close to the end nodes (see \figurename~\ref{fig_traffic_HS1} and \figurename~\ref{fig_traffic_HS4}) or at the inner part of the network (see \figurename~\ref{fig_traffic_HSW}). In this way we consider the two main generic types of congestion scenarios, i.e. incast and in-network congestion. 
In these figures, we also show the congestion trees, represented by means of red arrows, growing upstream from the roots.
More precisely, \figurename~\ref{fig_traffic_HS1} shows a hot-spot scenario 
where a single hot-spot is generated near to an end-node due to a many-to-one traffic pattern.
Note that, the congestion tree can be initially formed in the upward-path, 
but it spreads to all directions due to the adaptive routing algorithm.
In the second scenario (\figurename~\ref{fig_traffic_HS4}), we model four congestion trees 
generated by several many-to-one traffic patterns.
In this scenario, the congestion is also spread to the upward paths. 
The last scenario, represented in \figurename~\ref{fig_traffic_HSW}, generates hot-spots 
in switch output ports in the second stage of the Fat-Tree when $D$-mod-$K$ routing is used.
The goal of this traffic pattern is to explore how adaptive routing behaves when congestion situations are originated in the inner part of the network.
To implement this scenario, the hot-spot is generated computing the set of destinations 
that cross an output port using $D$-mod-$K$ routing. 

In order to model these traffic patterns in the simulator, we have implemented three scenarios.
The first one, generates a single hot-spot (see \figurename~\ref{fig_traffic_HS1}), which we have configured using two options.
In the first option, called ``HS10-1'' the hot-spot end-node receives traffic generated by 10\% of the network end-nodes, while 90\% of the end-nodes generate packets following the uniform traffic pattern. 
In the second option, the hot-spot end-node receives traffic generated by 25\% of the network end-nodes, while remaining end-nodes (75\%) generate traffic using to a random (uniform) distribution of destinations.
These traffic patterns model a congestion tree growing from the hot-spot end-node backwards to the source.
Note that these two options differ in the intensity of the generated congestion tree.
In the second traffic scenario (see \figurename~\ref{fig_traffic_HS4}) we generate the packets addressed to four hot-spot end-nodes (600, 3400, 5200, and 9500).
Again, we have modeled two options, where end-nodes generate 10\% and 25\% of traffic addressed to hot-spots (called ``HS10-4'' and ``HS25-4'').
In the third traffic scenario, called ``IHS'' (Intermediate Hot-Spot), we have modeled hot-spots appearing in switches in the second stage of the Fat-Tree network, and not directly connected to the end-nodes (see \figurename~\ref{fig_traffic_HSW}).
Specifically, 20\% of the end-nodes generate packets addressed to different switch output ports in the second stage. 

Finally, regarding the performance metrics, in the simulation experiments we measure the normalized throughput versus traffic load (by varying the generation rate from 0\% up to 100\% of the maximum link bandwidth).
For the sake of simplicity, in the following section we only analyze results for 100\% of traffic load.
The complete plots can be seen in \ref{results_appendix}.
Moreover, we have also measured the normalized throughput versus time during $3$ milliseconds of simulation time.
In all the experiments, we have considered a warm-up period for the network to reach the steady state, measuring the results afterwards.

\subsection{Parameters tuning for the restricted adaptive routing }
\label{sec:evaluation:parameters}

This section provides an analysis on the parameter values used in the simulator to configure the proposed restrictions to the adaptive routing algorithm.
These parameters are the \emph{low triggering threshold} (LTTh) and the \emph{high triggering threshold} (HTTh), described in Section \ref{sec:improving:restrictions:triggering}, and the $\Delta$ parameter, described in Section \ref{sec:improving:restrictions:ports-count}.
Specifically, in the assumed network configuration ($100$Gpbs links and $4$KB packets), a packet needs $320$ ns to be transmitted entirely through a link in the network.
When two packets stored at different input ports request the same output port, one of them must wait while the other is transmitted.
Note that the waiting packet needs $640$ ns to be transmitted.
An this latency increases when more packets are stored in the queue waiting to be transmitted.

After a careful tuning in the simulator, we have fixed the LTTh value to 25\% of the available credits at the next switch input port queue (or VL), i.e., the 75\% of that queue occupancy\footnote{The InfiniBand specification \cite{IBAspec2015} (Section A10.2.1.1) defines this value at 60\% of the queue occupancy.}.
This means that we will trigger adaptivity when occupancy the next queue at the neighboring switch, selected by $D$-mod-$K$ routing is becoming full.
Note that we do not want to select alternative routes very soon spreading congesting packets.
By contrast, we do not want to detect congestion very late, since congesting packets stored at the head of some queue will generate HoL blocking to those stored behind.
Note also that the LTTh value guarantees that the selection of alternative routes happens when congestion is strong enough and the queuing scheme can help to alleviate the HoL blocking, otherwise, it keeps the system using $D$-mod-$K$ in order to reduce congestion spreading.

Regarding the HTTh value used in the 2TH option, we have configured this value to 50\% of the queue occupancy level.
If the occupancy of the queue in the next switch for the output port selected by $D$-mod-$K$ reaches the HTTh value, after reaching the LTTh value, the path selection function comes back to use $D$-mod-$K$, instead of the alternative routes selection.

Regarding the $\Delta$ value, in our experiments we set this value to $3$, as this is the number VLs per buffer.
We want to use as fewer VLs as possible.
In this way the number of ports $K$ selected in the upward of downward phases to route packets is limited by $K/3$.
In the following sections, we analyze the simulation results obtained for the network configurations described before.
  
\subsection{Analysis of simulation results}
 \label{sec:evaluation:result_synthetic}
 
 Table \ref{tab:NOVOQ} summarizes the simulation results obtained for a $3$-stage $11664$-node Fat-Tree network, with switches configured without using VOQs, when generating a 100\% of traffic load in the network, for each of the routing algorithm configurations and queuing schemes, under the traffic scenarios described in Figure \ref{fig_traffic_scenarios}).
 For each traffic scenario (i.e., column in the table),
the throughput values have been obtained after the network is operating during $1$ millisecond.
 Moreover, we have colored in different degrees of gray the cells containing the three highest throughput values per traffic scenario (i.e., per column), so that it is easier to identify what is the best routing configuration for each queuing scheme.
 Note that the best routing configuration will be that one having a higher number of colored cells in the same table row.

\definecolor{mygrayp}{gray}{0.6}
\definecolor{mygrays}{gray}{0.75}
\definecolor{mygrayt}{gray}{0.9}
\newcommand{\pintap}[1]{\cellcolor{mygrayp}\textbf{#1}}
\newcommand{\pintas}[1]{\cellcolor{mygrays}\textbf{#1}}
\newcommand{\pintat}[1]{\cellcolor{mygrayt}\textbf{#1}}
 
\begin{table}[ht!]
\caption{Normalized Throughput at 100\% traffic load (after the network warm-up period) for a $11664$-node Fat-Tree network without using VOQ-based switches. Cells in grey represent the highest throughtput values obtained per column (i.e., traffic scenario). We capture each value after a $1$ms warm-up period of time. DBBM, vFtree and Flow2SL are configured with 3 VLs.}
\label{tab:NOVOQ}
\begin{subfigure}[!th]{1\textwidth}
 \centering 
\includegraphics[width=1.0\textwidth]
{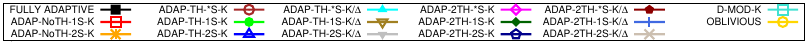}
\end{subfigure}
\centering
\footnotesize
\resizebox{.99\textwidth}{!}{%
\begin{tabular}{c | c c c c c | c c c c c | c c c c c | c c c c c}
  & \multicolumn{5}{c}{\bf 1Q}
 & \multicolumn{5}{c}{\bf DBBM}
 & \multicolumn{5}{c}{\bf vFtree}
 & \multicolumn{5}{c}{\bf Flow2SL}
\\
\parbox[t]{0.1mm}{\rotatebox[origin=c]{90}{Routing}} & \parbox[t]{0.1mm}{\rotatebox[origin=c]{90}{HS10-1}}
 & \parbox[t]{0.1mm}{\rotatebox[origin=c]{90}{HS25-1}}
 & \parbox[t]{0.1mm}{\rotatebox[origin=c]{90}{HS10-4}}
 & \parbox[t]{0.1mm}{\rotatebox[origin=c]{90}{HS25-4}}
 & \parbox[t]{0.1mm}{\rotatebox[origin=c]{90}{IHS}}
 & \parbox[t]{0.1mm}{\rotatebox[origin=c]{90}{HS10-1}}
 & \parbox[t]{0.1mm}{\rotatebox[origin=c]{90}{HS25-1}}
 & \parbox[t]{0.1mm}{\rotatebox[origin=c]{90}{HS10-4}}
 & \parbox[t]{0.1mm}{\rotatebox[origin=c]{90}{HS25-4}}
 & \parbox[t]{0.1mm}{\rotatebox[origin=c]{90}{IHS}}
 & \parbox[t]{0.1mm}{\rotatebox[origin=c]{90}{HS10-1}}
 & \parbox[t]{0.1mm}{\rotatebox[origin=c]{90}{HS25-1}}
 & \parbox[t]{0.1mm}{\rotatebox[origin=c]{90}{HS10-4}}
 & \parbox[t]{0.1mm}{\rotatebox[origin=c]{90}{HS25-4}}
 & \parbox[t]{0.1mm}{\rotatebox[origin=c]{90}{IHS}}
 & \parbox[t]{0.1mm}{\rotatebox[origin=c]{90}{HS10-1}}
 & \parbox[t]{0.1mm}{\rotatebox[origin=c]{90}{HS25-1}}
 & \parbox[t]{0.1mm}{\rotatebox[origin=c]{90}{HS10-4}}
 & \parbox[t]{0.1mm}{\rotatebox[origin=c]{90}{HS25-4}}
 & \parbox[t]{0.1mm}{\rotatebox[origin=c]{90}{IHS}}
\\
\midrule
\smallimg{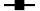} & 0 & 0 & 0 & 0 & 21 & \pintas{54} & \pintas{50} & 2 & 1 & 50 & 2 & 2 & 0 & 0 & 41 & 50 & 50 & 1 & 1 & 28\\
\smallimg{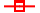} & 1 & 0 & 0 & 0 & 22 &\pintap{55} & \pintas{50} & 2 & 1 & \pintas{57} & 51 & 50 & 26 & 25 & 58 & 51 & 50 & 2 & 1 & 26\\
\smallimg{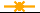} & 0 & 0 & 1 & 0 & 33 & 30 & 30 & 5 & 1 & 42 & 2 & 2 & 2 & 1 & 65 & 52 & 50 & 4 & 1 & \pintas{52}\\
\smallimg{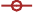} & \pintap{17} &\pintap{32} &\pintap{15} & 2 &\pintap{45} & 52 &\pintap{56} & \pintas{32} & \pintas{16} &\pintap{59} & 34 & 48 & 12 & 5 &\pintap{69} &\pintap{72} & \pintas{64} & 45 & \pintat{42} & \pintat{47}\\
\smallimg{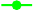} & 2 & \pintas{28} & 4 & 2 & \pintat{43} & 48 & \pintat{49} & \pintat{28} & 6 & \pintat{52} & 56 &\pintap{63} & \pintat{40} & 29 & \pintas{68} & \pintas{70} & \pintat{63} & 44 & 32 & 40\\
\smallimg{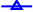} & 7 & 1 & 8 & 0 & 21 & 41 & 39 & 20 & \pintas{16} & 48 & \pintat{61} & 50 & 21 & 28 & 64 & 60 & 50 & 24 & 14 & 20\\
\smallimg{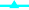} & 10 & \pintat{25} & \pintas{13} & 2 & 42 & 41 & 43 & 25 & \pintas{16} & 51 & \pintas{63} & \pintas{62} & 39 & \pintas{32} & \pintas{68} & \pintas{70} & \pintas{64} & 37 & 29 &\pintap{53}\\
\smallimg{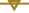} & 4 & 20 & 2 & 1 & 35 & 39 & 41 & 18 & 10 & 44 & 57 & \pintas{62} & \pintat{40} & 30 & \pintat{66} & 67 & \pintat{63} & 39 & 31 & 35\\
\smallimg{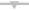} & 7 & 1 & 8 & 0 & 18 & 41 & 39 & 11 & \pintat{14} & 47 & 56 & 50 & 33 & 29 & 64 & 60 & 50 & 8 & 11 & 20\\
\smallimg{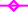} & \pintat{11} & \pintas{28} & \pintat{12} &\pintap{13} & \pintat{43} & 51 &\pintap{56} &\pintap{44} &\pintap{33} &\pintap{59} & 42 & 51 & 23 & 12 & \pintas{68} &\pintap{72} &\pintap{65} &\pintap{59} & \pintas{43} & 43\\
\smallimg{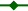} & 3 & \pintas{28} & 9 & 2 & \pintas{44} & 48 & \pintat{49} & 26 & 7 & \pintat{52} & 55 &\pintap{63} & 38 & 30 & \pintas{68} & \pintas{70} &\pintap{65} & 50 &\pintap{46} & 39\\
\smallimg{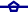} & 7 & 0 & 7 & 0 & 19 & 41 & 39 & 20 & \pintas{16} & 47 & 60 & 50 & 27 & 26 & 64 & 61 & 50 & 29 & 16 & 20\\
\smallimg{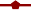} & \pintas{13} & 22 & 11 & \pintas{10} & 41 & 40 & 43 & \pintas{32} & \pintas{16} & 50 &\pintap{65} &\pintap{63} & \pintas{41} &\pintap{37} & \pintas{68} & \pintat{69} & \pintat{63} & 41 & 41 & \pintas{52}\\
\smallimg{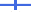} & 9 & 23 & 5 & \pintat{4} & 31 & 38 & 41 & 13 & 7 & 44 & 58 & \pintat{61} &\pintap{43} & \pintat{31} & \pintat{66} & 66 & \pintat{63} & 45 & 40 & 34\\
\smallimg{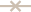} & 7 & 0 & 7 & 0 & 16 & 40 & 38 & 12 & \pintat{14} & 47 & 56 & 50 & 36 & 30 & 64 & 60 & 50 & 11 & 13 & 20\\
\smallimg{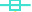} & 1 & 0 & 0 & 0 & 16 & 38 & 37 & 4 & 1 & 44 & 52 & 50 & 27 & 25 & 61 & 56 & 50 & 8 & 3 & 19\\
\smallimg{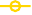} & 0 & 0 & 0 & 0 & 20 & \pintat{53} & \pintas{50} & 1 & 0 & 47 & 2 & 1 & 0 & 0 & 48 & 49 & 49 & 0 & 0 & 25\\
\bottomrule
\end{tabular}}
\end{table}

 As we can see, the  \smallimg{RLFT_adaptive_75_3_18.pdf} routing configuration achieves the best performance for the 1Q buffer scheme (i.e. without any queuing scheme), although it suffers a performance degradation under HS25-4 traffic scenario, compared to the \smallimg{RLFT_adaptive_50_3_18_sg.pdf} routing.
 DBBM achieves its best results with routing configuration denoted by symbol \smallimg{RLFT_adaptive_50_3_18_sg.pdf}, i.e., restricted routing using 2TH, *S and K options, and slightly worse results (but still good) with the \smallimg{RLFT_adaptive_75_3_18.pdf} routing.
 Moreover, vFtree achieves its best results with the routing configuration denoted by  \smallimg{RLFT_adaptive_s_3_18_50_sg.pdf}, i.e., restricted adaptive routing using 2TH, *S and K/$\Delta$ options, and similar results with the routing configuration denoted by
 \smallimg{RLFT_adaptive_fs_3_18_50_s_sg.pdf}.
 vFtree achieves better results in terms of throughput than DBBM, since the $K/\Delta$ option clearly reduces the adaptivity and makes vFtree to react better to congestion spreading.
 Flow2SL using the \smallimg{RLFT_adaptive_75_3_18.pdf} routing configuration (i.e., using 2TH, *S and $K$ options) achieves the best results under almost all the traffic patterns, outperforming vFtree and DBBM. 
 In this case, Flow2SL only uses the 2TH option to restrict adaptivity and, for this reason Flow2SL using the \smallimg{RLFT_adaptive_75_3_18.pdf} achieves worse results under HS25-4 traffic pattern than the \smallimg{RLFT_adaptive_fs_3_18_50_sg.pdf} routing configuration.
 Note that the latter configuration (applying 1S restriction option) also achieves excellent results, even outperforming vFtree.
 
 Indeed, it is worth mentioning that, for all the queuing schemes, $D$-mod-$K$ (\smallimg{RLFT_3_18.pdf}), oblivious (\smallimg{RLFT_obliv_3_18.pdf}) and fully adaptive routing (\smallimg{RLFT_adaptive_3_18.pdf}), are far from achieving good results when congestion appears.
 Even for the 1Q scheme, the use of a restricted routing algorithm (e.g., \smallimg{RLFT_adaptive_75_3_18.pdf}) is helpful, in terms of achieved throughput.
 Note that the switch architecture (without VOQs) also contributes avoid congestion spreading too much within buffers, as VLs isolates congestion from growing in other VLs of the switch.
 
 Unfortunately, this effect is not observed in other switch architectures, commonly used in commercial products, such as VOQ-based switches (see Section \ref{sec:background:congestion_management}), which natively prevent low-order HoL blocking.
 VOQ-based switches allow at every switch input port that the same VL can forward packets at the same time to different output ports, since they implement demultiplexed entries from the input-port buffers to the internal switch crossbar.
 For these reason, we have performed the same experiment as we have described before, but using switches configured with VOQs.
Table \ref{tab:VOQ} shows these results for a $3$-stage $11664$-node Fat-Tree network, when generating a 100\% of traffic load in the network, for each of the routing algorithm configurations and queuing schemes, under the traffic scenarios described in Figure \ref{fig_traffic_scenarios}.
As for the previous table, throughput values have been obtained after the network is operating during $1$ millisecond under every traffic scenario.
Again, the best routing configuration per queuing scheme is the one having the highest number of colored cells per row in the table.

\begin{table}[ht!]
\caption{Normalized Throughput at 100\% traffic load for a $11664$-node Fat-Tree network using VOQ-based switches. Cells in grey represent the highest throughtput values obtained per column (i.e., traffic scenario). We capture each value after a $1$ms warm-up period of time. DBBM, vFtree and Flow2SL are configured with 3 VLs.}
\label{tab:VOQ}
\begin{subfigure}[!th]{1\textwidth}
 \centering 
\includegraphics[width=1.0\textwidth]
{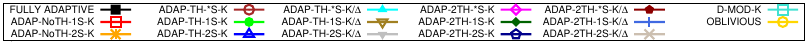}
\end{subfigure}
\centering
\footnotesize
\resizebox{.99\textwidth}{!}{%
\begin{tabular}{c | c c c c c | c c c c c | c c c c c | c c c c c}

 & \multicolumn{5}{c}{\bf 1Q}
 & \multicolumn{5}{c}{\bf DBBM}
 & \multicolumn{5}{c}{\bf vFtree}
 & \multicolumn{5}{c}{\bf Flow2SL}
\\
\parbox[t]{0.1mm}{\rotatebox[origin=c]{90}{Routing}} & \parbox[t]{0.1mm}{\rotatebox[origin=c]{90}{HS10-1}}
 & \parbox[t]{0.1mm}{\rotatebox[origin=c]{90}{HS25-1}}
 & \parbox[t]{0.1mm}{\rotatebox[origin=c]{90}{HS10-4}}
 & \parbox[t]{0.1mm}{\rotatebox[origin=c]{90}{HS25-4}}
 & \parbox[t]{0.1mm}{\rotatebox[origin=c]{90}{IHS}}
 & \parbox[t]{0.1mm}{\rotatebox[origin=c]{90}{HS10-1}}
 & \parbox[t]{0.1mm}{\rotatebox[origin=c]{90}{HS25-1}}
 & \parbox[t]{0.1mm}{\rotatebox[origin=c]{90}{HS10-4}}
 & \parbox[t]{0.1mm}{\rotatebox[origin=c]{90}{HS25-4}}
 & \parbox[t]{0.1mm}{\rotatebox[origin=c]{90}{IHS}}
 & \parbox[t]{0.1mm}{\rotatebox[origin=c]{90}{HS10-1}}
 & \parbox[t]{0.1mm}{\rotatebox[origin=c]{90}{HS25-1}}
 & \parbox[t]{0.1mm}{\rotatebox[origin=c]{90}{HS10-4}}
 & \parbox[t]{0.1mm}{\rotatebox[origin=c]{90}{HS25-4}}
 & \parbox[t]{0.1mm}{\rotatebox[origin=c]{90}{IHS}}
 & \parbox[t]{0.1mm}{\rotatebox[origin=c]{90}{HS10-1}}
 & \parbox[t]{0.1mm}{\rotatebox[origin=c]{90}{HS25-1}}
 & \parbox[t]{0.1mm}{\rotatebox[origin=c]{90}{HS10-4}}
 & \parbox[t]{0.1mm}{\rotatebox[origin=c]{90}{HS25-4}}
 & \parbox[t]{0.1mm}{\rotatebox[origin=c]{90}{IHS}}
\\
\midrule
\smallimg{RLFT_adaptive_3_18.pdf} & 0 & 0 & 0 & 0 & 36 & 60 & 50 & \pintat{2} & 1 & 60 & 2 & 2 & 0 & 0 & 59 & 60 & 50 & 2 & 1 & 36\\
\smallimg{RLFT_adaptive_fs_3_18.pdf} & 0 & 0 & 0 & 0 & 26 & 60 & 50 & 1 & 1 & 65 & \pintas{60} & \pintap{50} & 31 & \pintas{25} & 69 & 60 & 50 & 2 & 1 & 26\\
\smallimg{RLFT_adaptive_ss_3_18.pdf} & \pintap{1} & 0 &  \pintap{2} &  \pintap{1} &\pintap{81} & 60 & 50 & \pintap{4} & 1 &\pintap{80} & 3 & 2 & 5 & 2 &\pintap{81} & 60 & 50 & 6 & \pintat{2} &\pintap{81}\\
\smallimg{RLFT_adaptive_75_3_18.pdf} & 0 & 0 & 0 & 0 & 28 & 60 & 50 & \pintat{2} & 1 & \pintat{66} & 2 & 2 & 1 & 0 & 71 & 60 & 50 & 2 & 1 & 30\\
\smallimg{RLFT_adaptive_fs_3_18_75.pdf} & 0 & 0 & \pintas{1} & 0 & 30 & 60 & 50 & \pintat{2} & 1 & \pintat{66} & \pintas{60} & \pintap{50} & 31 & \pintas{25} & \pintas{72} & 60 & 50 & 2 & 1 & 30\\
\smallimg{RLFT_adaptive_ss_3_18_75.pdf} & \pintap{1} & 0 & \pintap{2} & 0 & 22 & 60 & 50 & \pintas{3} & 1 & 61 & 3 & 2 & 3 & 1 & 69 & 60 & 50 & 4 & \pintat{2} & 22\\
\smallimg{RLFT_adaptive_s_3_18_75.pdf} & \pintap{1} & 0 & 0 & 0 & \pintas{46} & 60 & 50 & \pintat{2} & 1 & \pintas{71} & \pintas{60} & \pintap{50} & 31 & \pintas{25} & \pintas{79} & 60 & 50 & 3 & \pintat{2} & \pintas{48}\\
\smallimg{RLFT_adaptive_fs_3_18_75_s.pdf} & \pintap{1} & 0 & 0 & 0 & \pintas{46} & 60 & 50 & \pintat{2} & 1 & \pintas{71} & \pintap{61} & \pintap{50} & 30 & \pintas{25} & \pintas{79} & 60 & 50 & 4 & \pintat{2} & \pintas{48}\\
\smallimg{RLFT_adaptive_ss_3_18_75_s.pdf} & \pintap{1} & 0 & \pintap{2} & 0 & 22 & 60 & 50 & \pintap{4} & 1 & 61 & \pintap{61} & \pintap{50} & \pintas{33} &\pintap{26} & 69 & \pintap{61} & 50  & \pintat{7} & \pintas{4} & 22\\
\smallimg{RLFT_adaptive_50_3_18_sg.pdf} & 0 & 0 & 0 & 0 & 28 & 60 & 50 & \pintas{3} & 1 & 65 & 2 & 2 & 1 & 0 & 71 & 60 & 50 & 2 & 1 & 30\\
\smallimg{RLFT_adaptive_fs_3_18_50_sg.pdf} & 0 & 0 & \pintas{1} & 0 & 29 & 60 & 50 & \pintat{2} & 1 & \pintat{66} & \pintas{60} & \pintap{50} & 31 & \pintas{25} & \pintas{72} & 60 & 50 & 2 & 1 & 31\\
\smallimg{RLFT_adaptive_ss_3_18_50_sg.pdf} & \pintap{1}& 0 & \pintap{2} & 0 & 22 & 60 & 50 & \pintas{3} & 1 & 61 & 3 & 2 & 4 & 1 & 69 & 60 & 50 & 4 & \pintat{2} & 22\\
\smallimg{RLFT_adaptive_s_3_18_50_sg.pdf} & \pintap{1}& 0 & \pintas{1} & 0 & \pintat{45} & 60 & 50 & \pintat{2} & 1 & \pintas{71} & \pintas{60} & \pintap{50} & 31 & \pintas{25} & \pintas{79} & 60 & 50 & 3 & 1 & \pintat{47}\\
\smallimg{RLFT_adaptive_fs_3_18_50_s_sg.pdf} & \pintap{1}& 0 & 0 & 0 & \pintat{45} & 60 & 50 & \pintat{2} & 1 & \pintas{71} & \pintap{61} & \pintap{50} & 30 & \pintas{25} & \pintas{79} & 60 & 50 & 4 & \pintat{2} & \pintat{47}\\
\smallimg{RLFT_adaptive_ss_3_18_50_s_sg.pdf} & \pintap{1}& 0 & \pintap{2} & 0 & 22 & 60 & 50 & \pintap{4} & 1 & 61 & \pintap{61} & \pintap{50} &\pintap{34} &\pintap{26} & 69 & \pintap{61} & 50  & \pintas{8} & \pintas{4} & 22\\
\smallimg{RLFT_3_18.pdf} & \pintap{1}& 0 & 1 & 0 & 22 & 60 & 50 & \pintap{4} & \pintap{2} & 61 & \pintap{61} & \pintap{50} & \pintat{32} & \pintas{25} & 69 & \pintap{61} & 50  &\pintap{12} &\pintap{6} & 22\\
\smallimg{RLFT_obliv_3_18.pdf} & 0 & 0 & 0 & 0 & 36 & 60 & 50 & 1 & 1 & 60 & 2 & 1 & 0 & 0 & 59 & 60 & 50 & 1 & 1 & 36\\
\bottomrule
\end{tabular}}
\end{table}

First of all, note that the 1Q buffer organization achieves throughput results near 0\% under traffic scenarios HS10 and HS25, both with $1$ and $4$ congestion trees (incast), since congestion is spread quickly and widely throughout the network, due to the VOQ-based switch architecture.
Only when congestion appears in-network (with the IHS traffic scenario), there are some routing configurations, such as \smallimg{RLFT_adaptive_ss_3_18.pdf}, which are able to react.
DBBM is able to react under HS10-1 and HS25-1, since it uses 3 VLs and only one congestion tree grows within one of these queues, but we do not appreciate any difference among using one or another routing configuration.
Moreover, DBBM performance is degraded under HS10-4 and HS-25, achieving a poor performance.
Again, the reason for such a bad performance is the VOQ-based switch architecture, which is not suited to DBBM when strong incast scenarios (i.e., four congestion trees) appear.
Similarly, Flow2SL performance is degraded by the HoL blocking being spread throughout all the VOQs and VLs, under HS10-4 and HS25-4.
Moreover, Flow2SL is not able to separate conveniently flows in the downward phase (see Section \ref{sec:problem}), and the use of restricted routing is not enough to prevent congestion from spreading.

Although HS10-4 and HS25-4 traffic patterns generate a corner case scenario (we have configured queuing schemes with 3 VLs, while we generate 4 congestion trees),
the vFtree scheme combined with some of our proposals for restricted adaptive routing is able to react under this very demanding traffic scenario.
More precisely, the \smallimg{RLFT_adaptive_s_3_18_50_sg.pdf} routing configuration for vFtree achieves the highest performance in most of the traffic scenarios, as it happens in Table \ref{tab:NOVOQ}, using the 2TH and K/$\Delta$ restrictions.
Note that the $D$-mod-$K$ routing achieves excellent results combined with vFtree in this scenario, since when congestion is so strong, adaptivity may be counterproductive if it is not restricted.
Moreover, note also that, in general, the proposed restrictions applied to routing configurations outperform the throughput obtained by fully adaptive and oblivious routing algorithms, such (i.e., \smallimg{RLFT_adaptive_3_18.pdf} and \smallimg{RLFT_obliv_3_18.pdf}).
 
\begin{figure}[!htb]

\begin{subfigure}[!th]{\textwidth}
 \centering 
\includegraphics[width=1.0\textwidth]{leyenda3.pdf}
\end{subfigure}

 \begin{subfigure}[!th]{0.475\textwidth}
 \centering 
\includegraphics[width=0.82\textwidth]{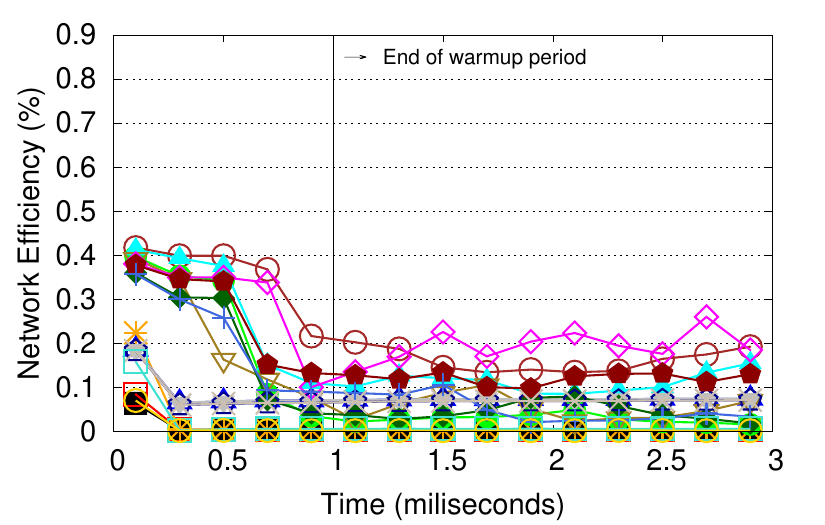}
\caption{1VC.}
\label{fig_time_RLFT_HS10_1q}
\end{subfigure}
 \begin{subfigure}[!th]{0.475\textwidth}
 \centering 
\includegraphics[width=0.82\textwidth]
{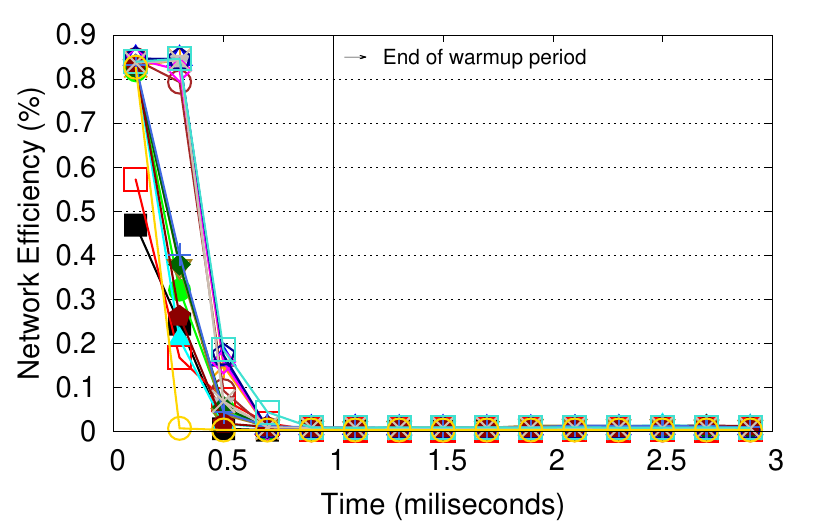}
\caption{1VC and VOQs.}
\label{fig_time_RLFT_HS10_1q-voq}
\end{subfigure}

 \begin{subfigure}[!th]{0.475\textwidth}
 \centering 
\includegraphics[width=0.82\textwidth]
{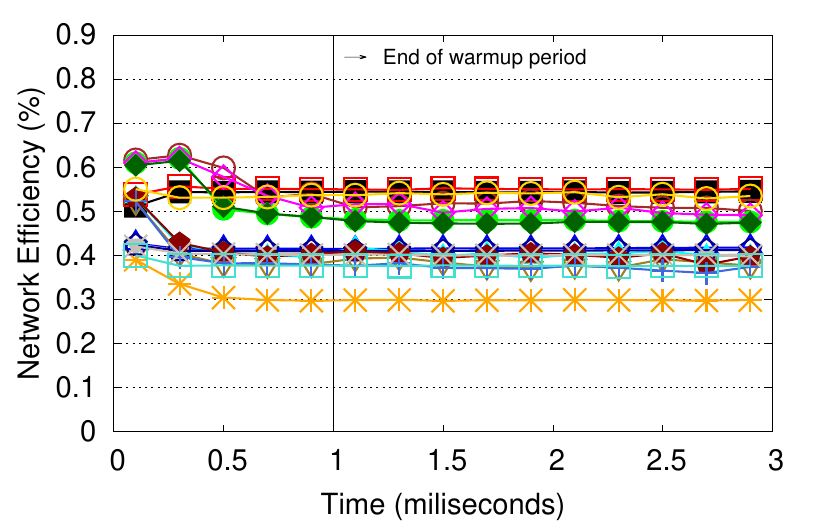}
\caption{DBBM.}
\label{fig_time_RLFT_HS10_dbbm3}
\end{subfigure}
 \begin{subfigure}[!th]{0.475\textwidth}
 \centering 
\includegraphics[width=0.82\textwidth]
{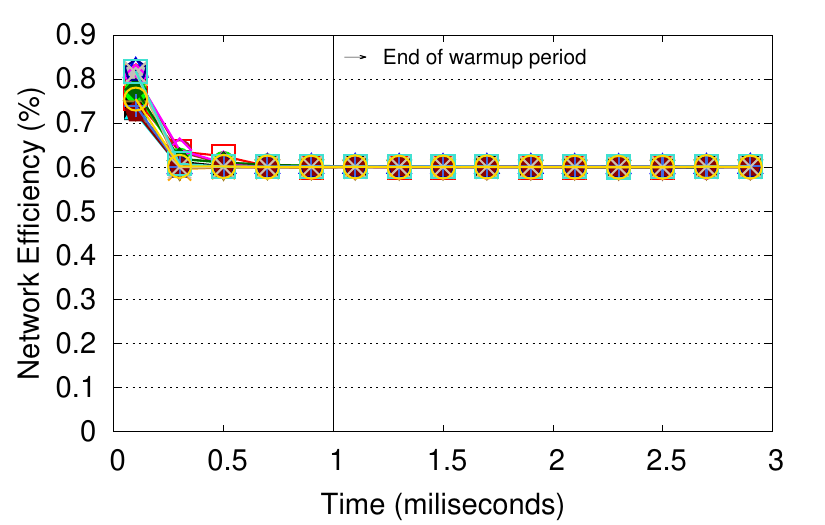}
\caption{DBBM and VOQs.}
\label{fig_time_RLFT_HS10_dbbm3-voq}
\end{subfigure}

 \begin{subfigure}[!th]{0.475\textwidth}
 \centering 
\includegraphics[width=0.82\textwidth]
{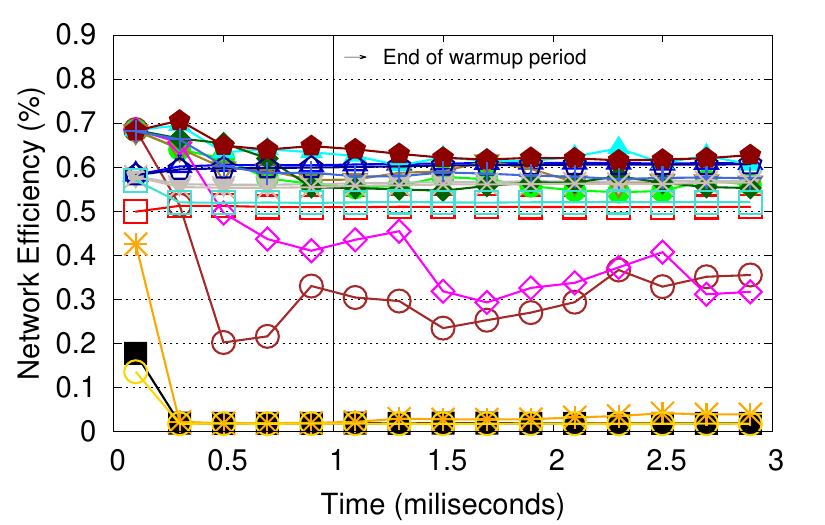}
\caption{vFtree.}
\label{fig_time_RLFT_HS10_vftree3}
\end{subfigure}
 \begin{subfigure}[!th]{0.475\textwidth}
 \centering 
\includegraphics[width=0.82\textwidth]
{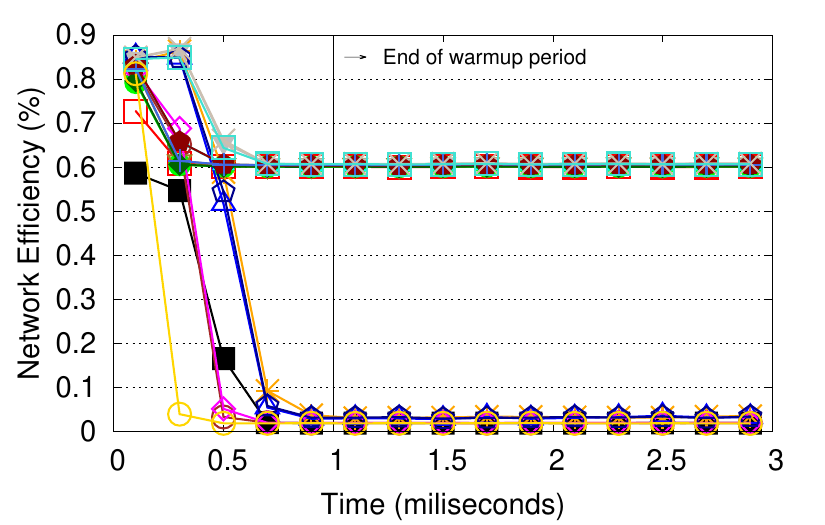}
\caption{vFtree and VOQs.}
\label{fig_time_RLFT_HS10_vftree3-voq}
\end{subfigure}

 \begin{subfigure}[!th]{0.475\textwidth}
 \centering 
\includegraphics[width=0.82\textwidth]
{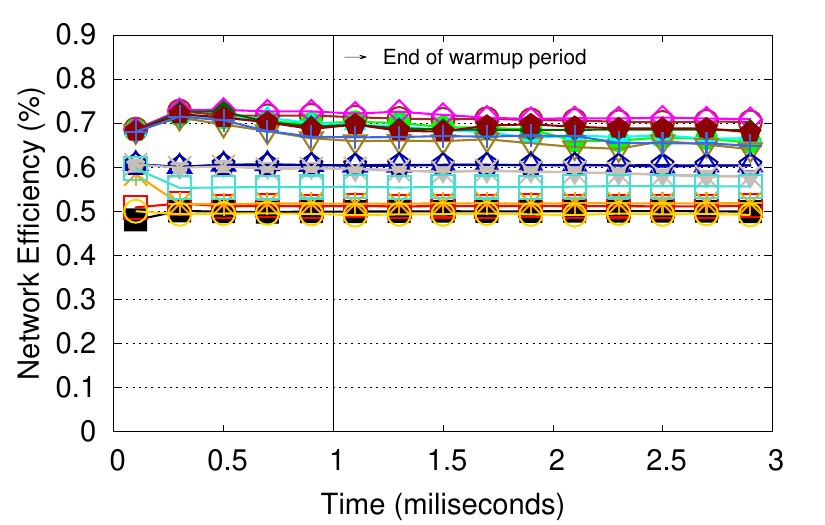}
\caption{Flow2SL.}
\label{fig_time_RLFT_HS10_flow2sl3}
\end{subfigure}
 \begin{subfigure}[!th]{0.475\textwidth}
 \centering 
\includegraphics[width=0.82\textwidth]
{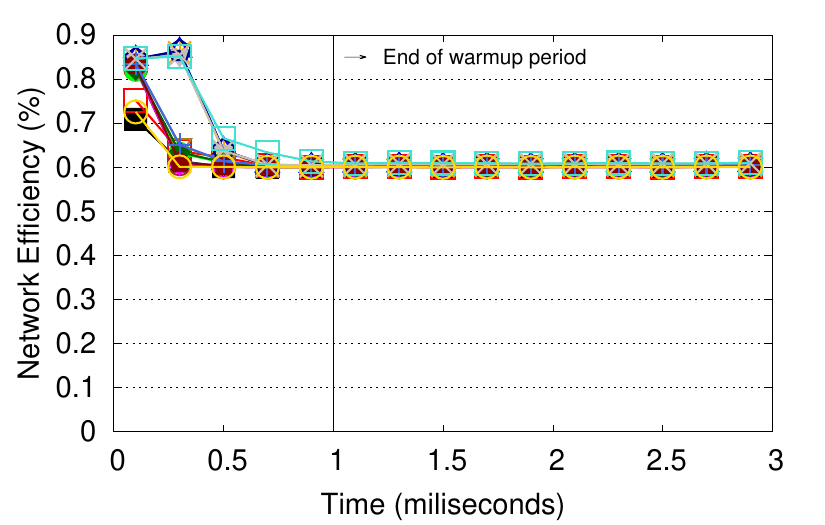}
\caption{Flow2SL and VOQs.}
\label{fig_time_RLFT_HS10_flow2sl3-voq}
\end{subfigure}

\caption{Normalized Throughput versus Time in a $11664$-node Fat-Tree under HS10-1 synthetic traffic pattern.}
\label{fig:RLFT_HS10_vTime}
\end{figure}

\begin{figure*}[!ht]
\vspace{-.5cm}
\begin{subfigure}[!th]{1\textwidth}
 \centering 
\includegraphics[width=1.0\textwidth]
{leyenda3.pdf}
\end{subfigure}

 \begin{subfigure}[!th]{0.475\textwidth}
 \centering 
\includegraphics[width=0.82\textwidth]
{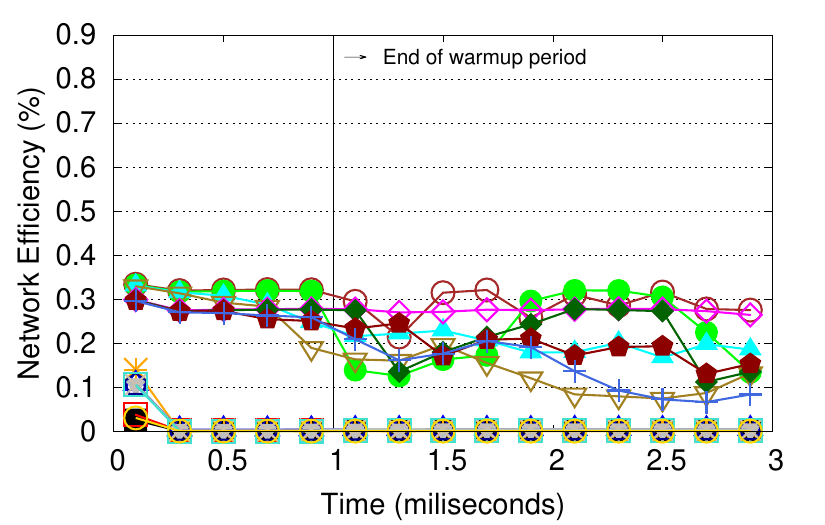}
\caption{1VC.}
\label{fig_time_RLFT_HS_1q}
\end{subfigure}
 \begin{subfigure}[!th]{0.475\textwidth}
 \centering 
\includegraphics[width=0.82\textwidth]
{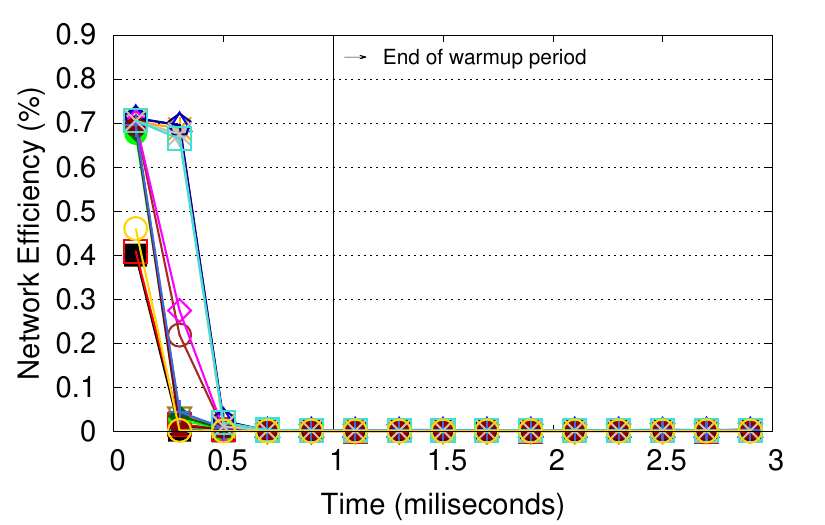}
\caption{1VC and VOQs.}
\label{fig_time_RLFT_HS_1q-voq}
\end{subfigure}

 \begin{subfigure}[!th]{0.475\textwidth}
 \centering 
\includegraphics[width=0.82\textwidth]
{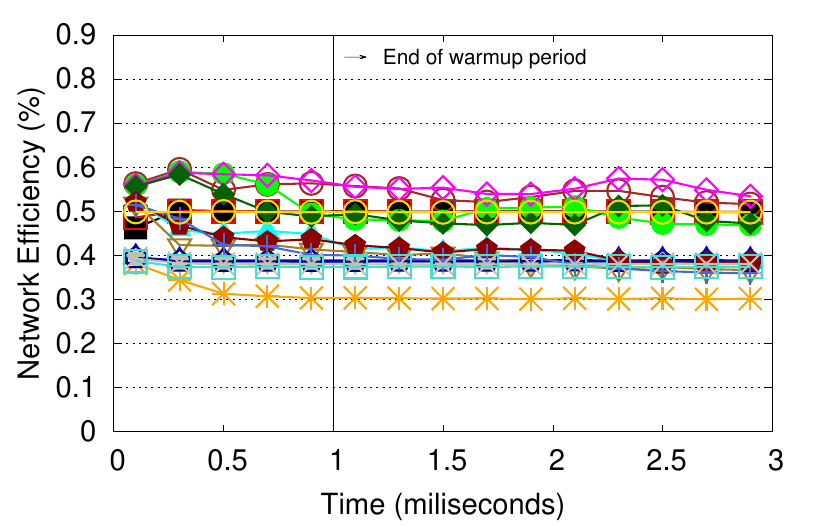}
\caption{DBBM.}
\label{fig_time_RLFT_HS_dbbm3}
\end{subfigure}
 \begin{subfigure}[!th]{0.475\textwidth}
 \centering 
\includegraphics[width=0.82\textwidth]
{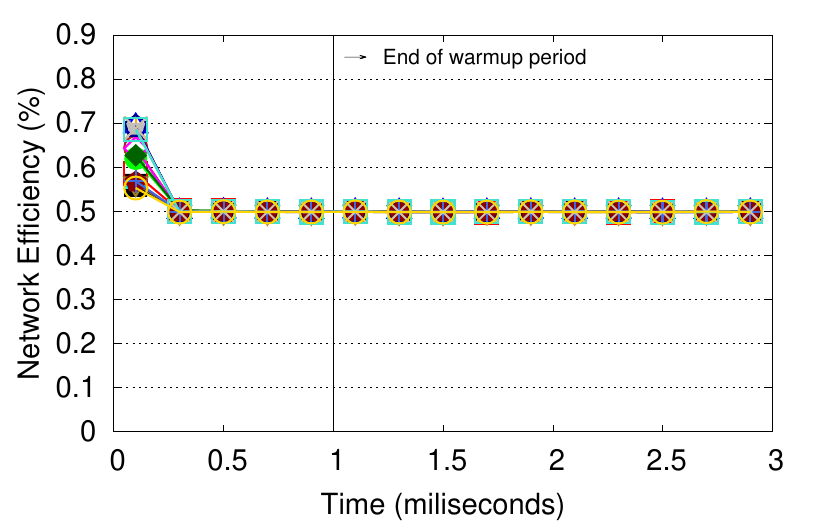}
\caption{DBBM and VOQs.}
\label{fig_time_RLFT_HS_dbbm3-voq}
\end{subfigure}

 \begin{subfigure}[!th]{0.475\textwidth}
 \centering 
\includegraphics[width=0.82\textwidth]
{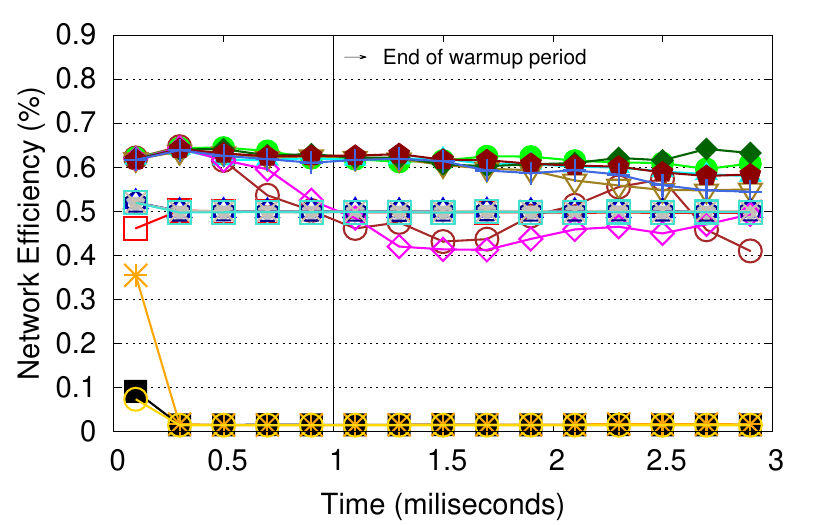}
\caption{vFtree.}
\label{fig_time_RLFT_HS_vftree3}
\end{subfigure}
 \begin{subfigure}[!th]{0.475\textwidth}
 \centering 
\includegraphics[width=0.82\textwidth]
{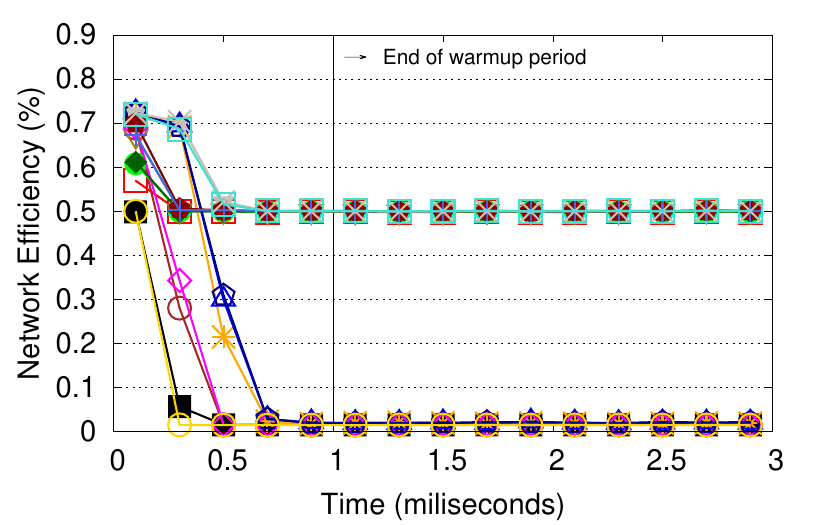}
\caption{vFtree and VOQs.}
\label{fig_time_RLFT_HS_vftree3-voq}
\end{subfigure}

 \begin{subfigure}[!th]{0.475\textwidth}
 \centering 
\includegraphics[width=0.82\textwidth]
{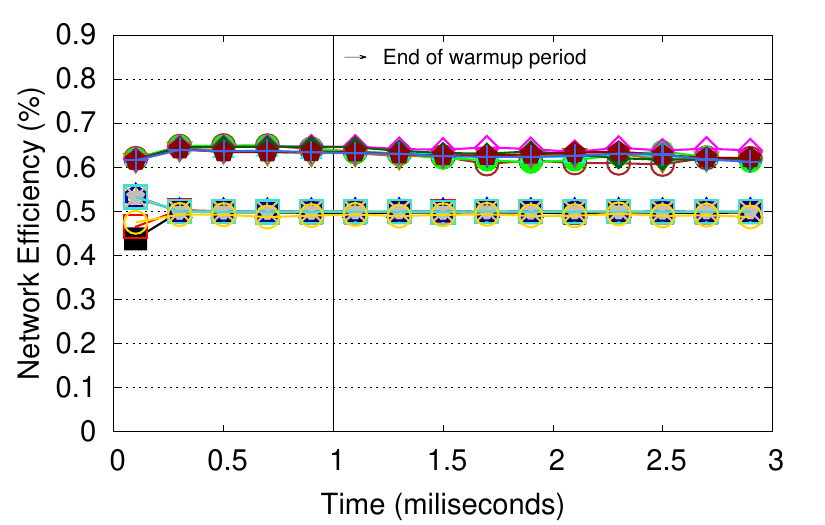}
\caption{Flow2SL.}
\label{fig_time_RLFT_HS_flow2sl3}
\end{subfigure}
 \begin{subfigure}[!th]{0.475\textwidth}
 \centering 
\includegraphics[width=0.82\textwidth]
{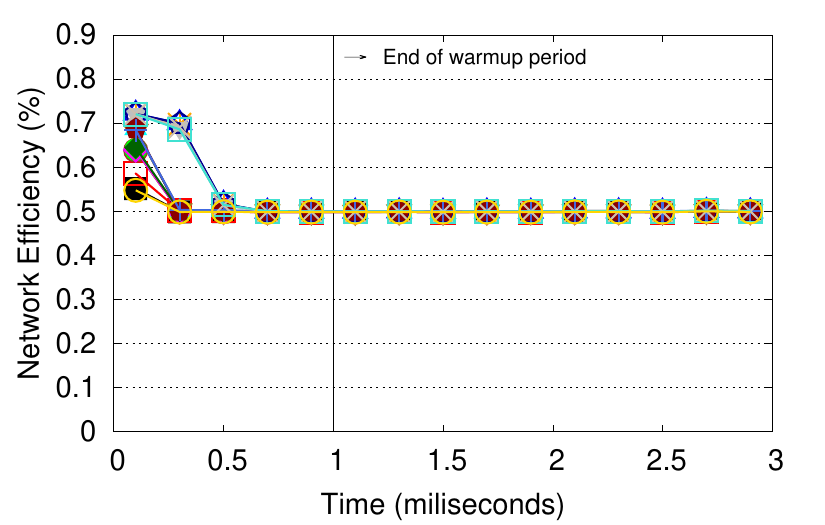}
\caption{Flow2SL and VOQs.}
\label{fig_time_RLFT_HS_flow2sl3-voq}
\end{subfigure}

\caption{Normalized Throughput versus Time in a $11664$-node Fat-Tree under HS25-1 synthetic traffic pattern.}
\label{fig:RLFT_HS25_vTime}
\end{figure*}

\begin{figure*}[!ht]
\vspace{-.5cm}
\begin{subfigure}[!th]{1\textwidth}
 \centering 
\includegraphics[width=1.0\textwidth]
{leyenda3.pdf}
\end{subfigure}

 \begin{subfigure}[!th]{0.475\textwidth}
 \centering 
\includegraphics[width=0.82\textwidth]
{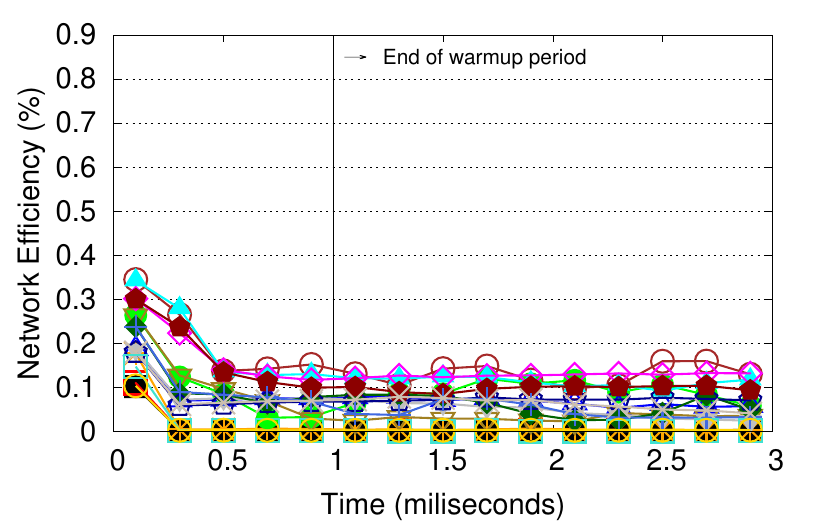}
\caption{1VC.}
\label{fig_time_RLFT_HS104_1q}
\end{subfigure}
 \begin{subfigure}[!th]{0.475\textwidth}
 \centering 
\includegraphics[width=0.82\textwidth]
{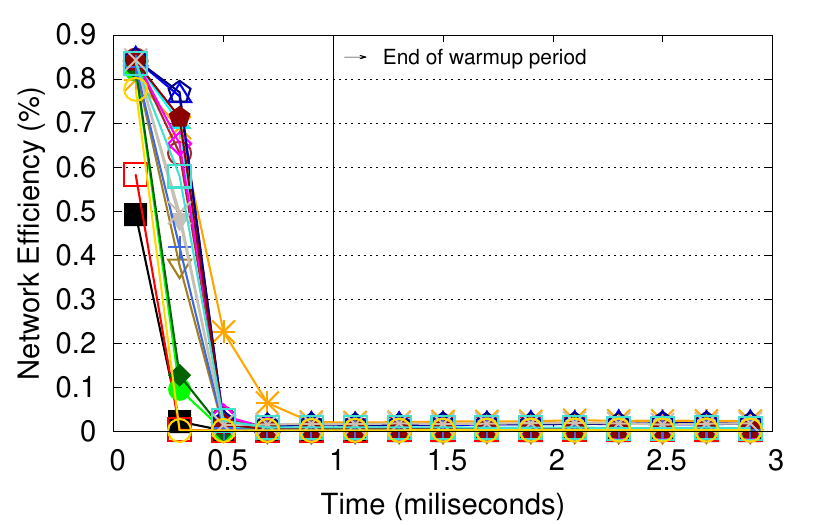}
\caption{1VC and VOQs.}
\label{fig_time_RLFT_HS104_1q-voq}
\end{subfigure}

 \begin{subfigure}[!th]{0.475\textwidth}
 \centering 
\includegraphics[width=0.82\textwidth]
{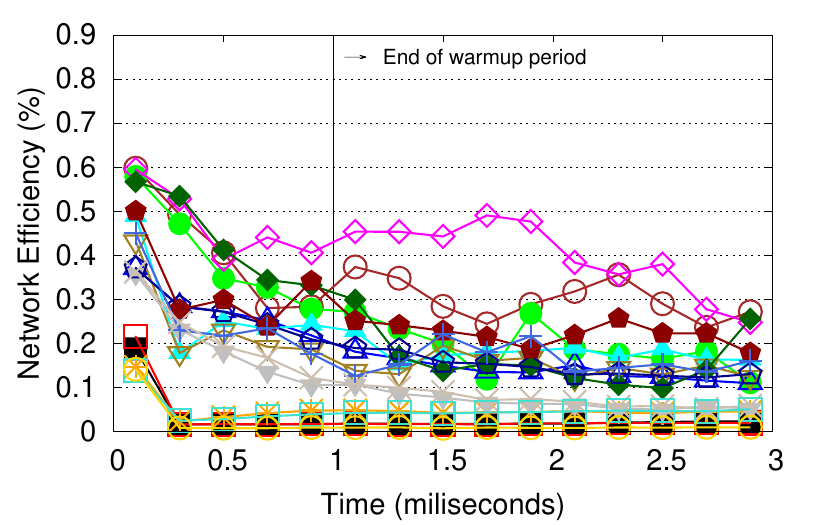}
\caption{DBBM.}
\label{fig_time_RLFT_HS104_dbbm3}
\end{subfigure}
 \begin{subfigure}[!th]{0.475\textwidth}
 \centering 
\includegraphics[width=0.82\textwidth]
{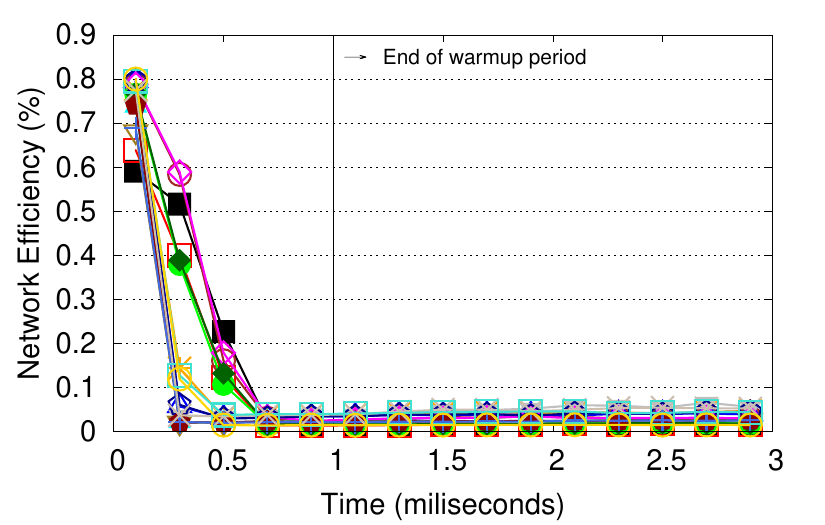}
\caption{DBBM and VOQs.}
\label{fig_time_RLFT_HS104_dbbm3-voq}
\end{subfigure}

 \begin{subfigure}[!th]{0.475\textwidth}
 \centering 
\includegraphics[width=0.82\textwidth]
{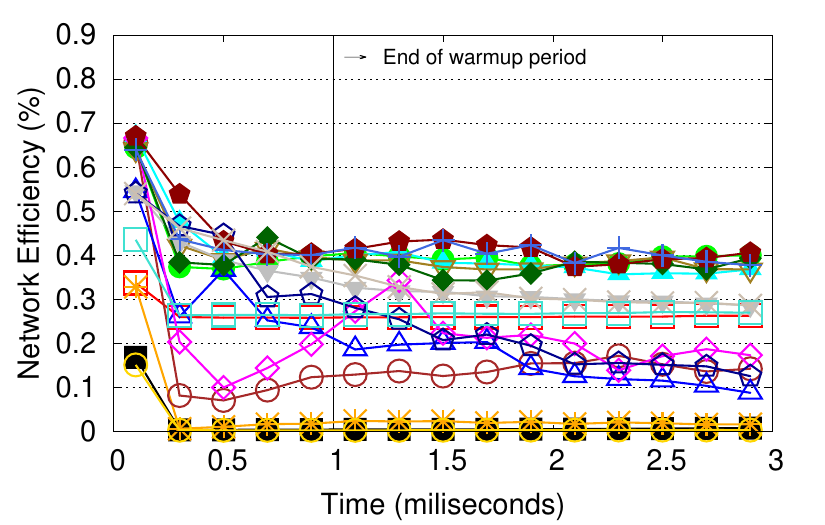}
\caption{vFtree.}
\label{fig_time_RLFT_HS104_vftree3}
\end{subfigure}
 \begin{subfigure}[!th]{0.475\textwidth}
 \centering 
\includegraphics[width=0.82\textwidth]
{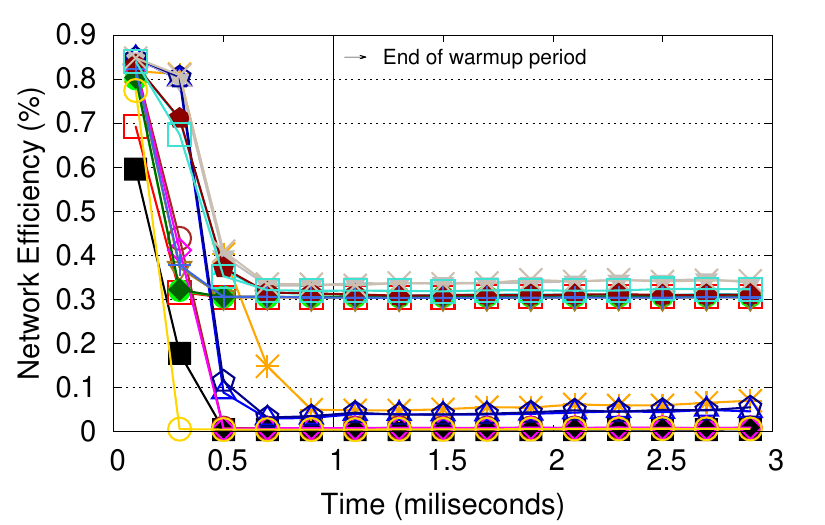}
\caption{vFtree and VOQs.}
\label{fig_time_RLFT_HS104_vftree3-voq}
\end{subfigure}

 \begin{subfigure}[!th]{0.475\textwidth}
 \centering 
\includegraphics[width=0.82\textwidth]
{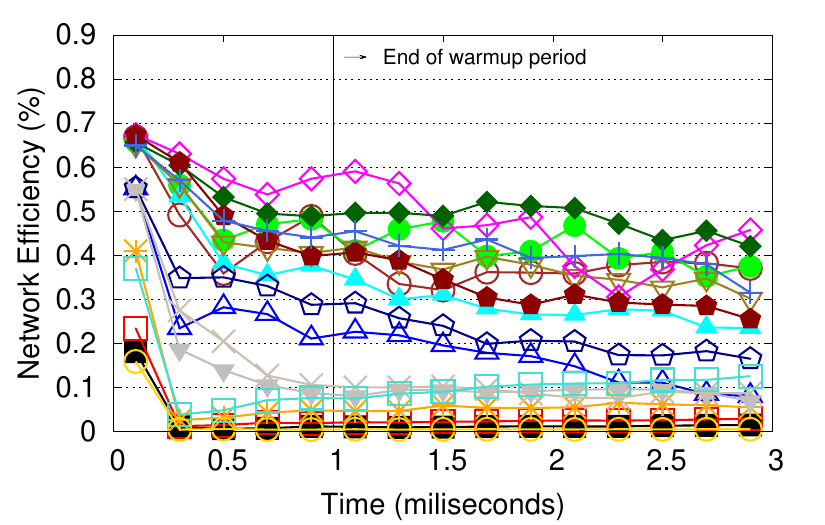}
\caption{Flow2SL.}
\label{fig_time_RLFT_HS104_flow2sl3}
\end{subfigure}
 \begin{subfigure}[!th]{0.475\textwidth}
 \centering 
\includegraphics[width=0.82\textwidth]
{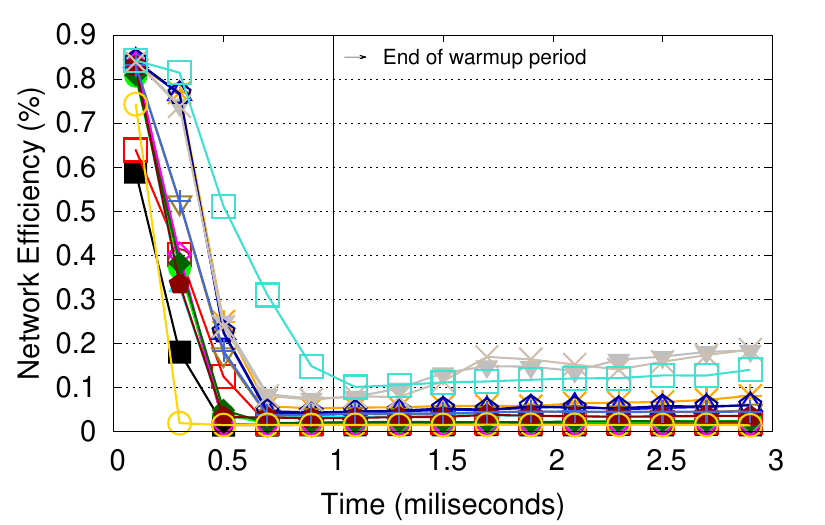}
\caption{Flow2SL and VOQs.}
\label{fig_time_RLFT_HS104_flow2sl3-voq}
\end{subfigure}

\caption{Normalized Throughput versus Time in a $11664$-node Fat-Tree network under HS10-4 synthetic traffic pattern.}
\label{fig:RLFT_HS104_vTime}
\end{figure*}

\begin{figure*}[!ht]
\vspace{-.5cm}
\begin{subfigure}[!th]{1\textwidth}
 \centering 
\includegraphics[width=1.0\textwidth]
{leyenda3.pdf}
\end{subfigure}

 \begin{subfigure}[!th]{0.475\textwidth}
 \centering 
\includegraphics[width=0.82\textwidth]
{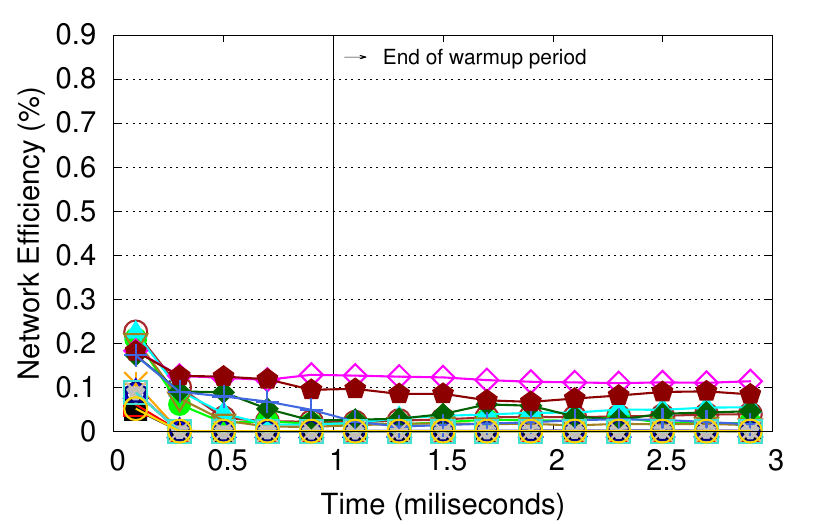}
\caption{1VC.}
\label{fig_time_RLFT_HS254_1q}
\end{subfigure}
 \begin{subfigure}[!th]{0.475\textwidth}
 \centering 
\includegraphics[width=0.82\textwidth]
{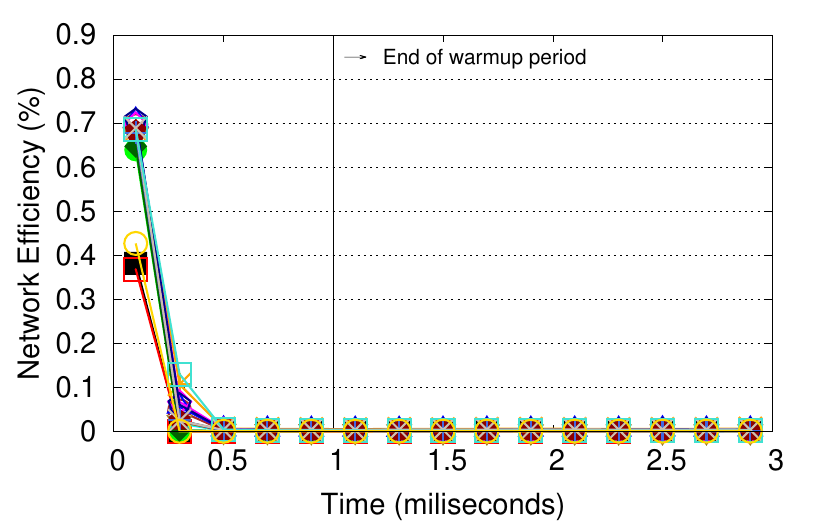}
\caption{1VC and VOQs.}
\label{fig_time_RLFT_HS254_1q-voq}
\end{subfigure}

 \begin{subfigure}[!th]{0.475\textwidth}
 \centering 
\includegraphics[width=0.82\textwidth]
{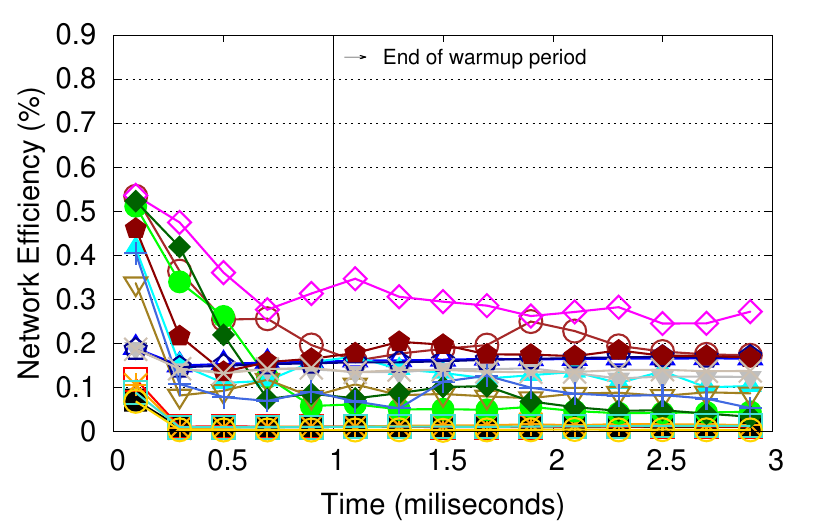}
\caption{DBBM.}
\label{fig_time_RLFT_HS254_dbbm3}
\end{subfigure}
 \begin{subfigure}[!th]{0.475\textwidth}
 \centering 
\includegraphics[width=0.82\textwidth]
{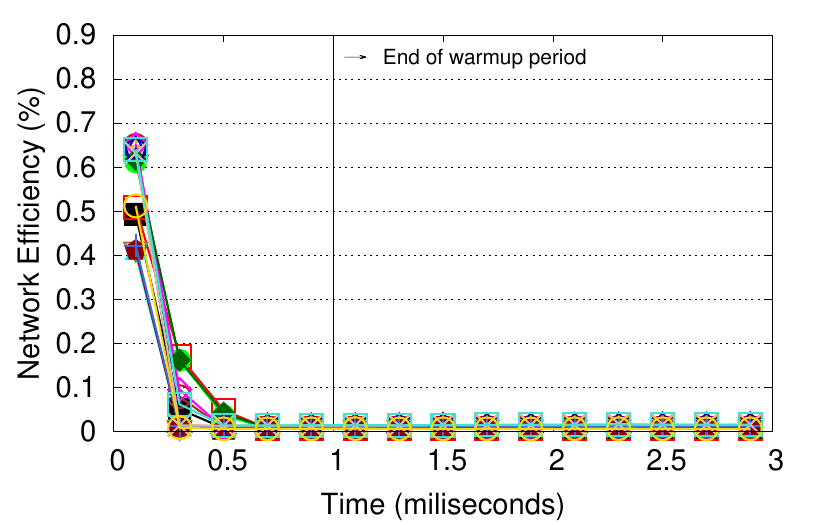}
\caption{DBBM and VOQs.}
\label{fig_time_RLFT_HS254_dbbm3-voq}
\end{subfigure}

 \begin{subfigure}[!th]{0.475\textwidth}
 \centering 
\includegraphics[width=0.82\textwidth]
{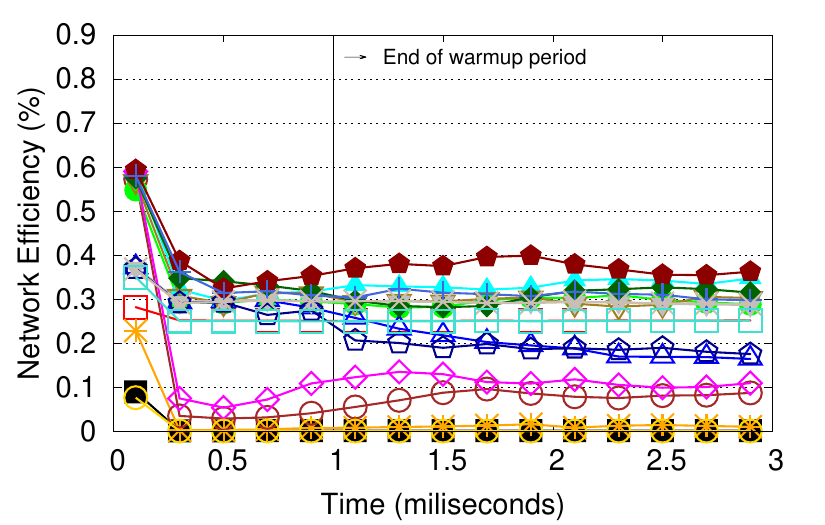}
\caption{vFtree.}
\label{fig_time_RLFT_HS254_vftree3}
\end{subfigure}
 \begin{subfigure}[!th]{0.475\textwidth}
 \centering 
\includegraphics[width=0.82\textwidth]
{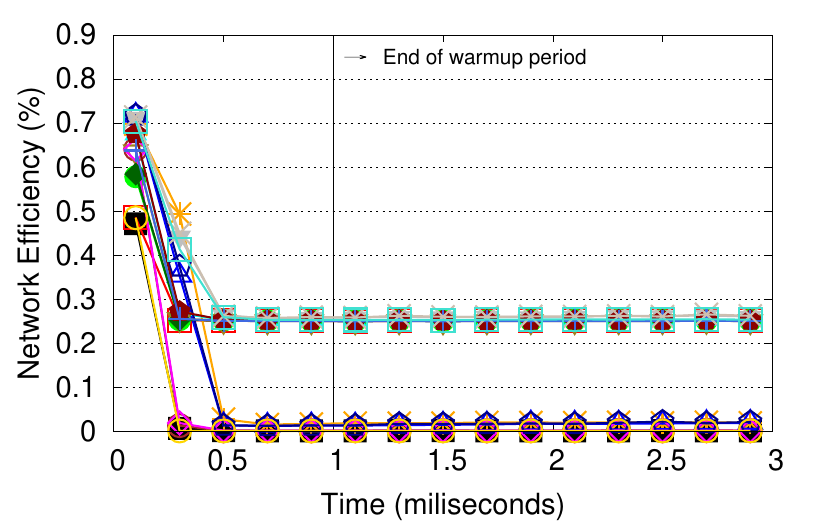}
\caption{vFtree and VOQs.}
\label{fig_time_RLFT_HS254_vftree3-voq}
\end{subfigure}

 \begin{subfigure}[!th]{0.475\textwidth}
 \centering 
\includegraphics[width=0.82\textwidth]
{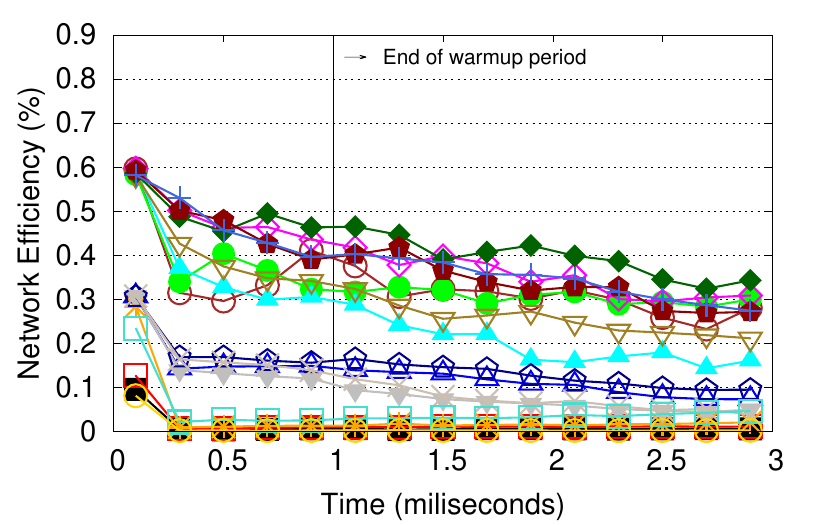}
\caption{Flow2SL.}
\label{fig_time_RLFT_HS254_flow2sl3}
\end{subfigure}
 \begin{subfigure}[!th]{0.475\textwidth}
 \centering 
\includegraphics[width=0.82\textwidth]
{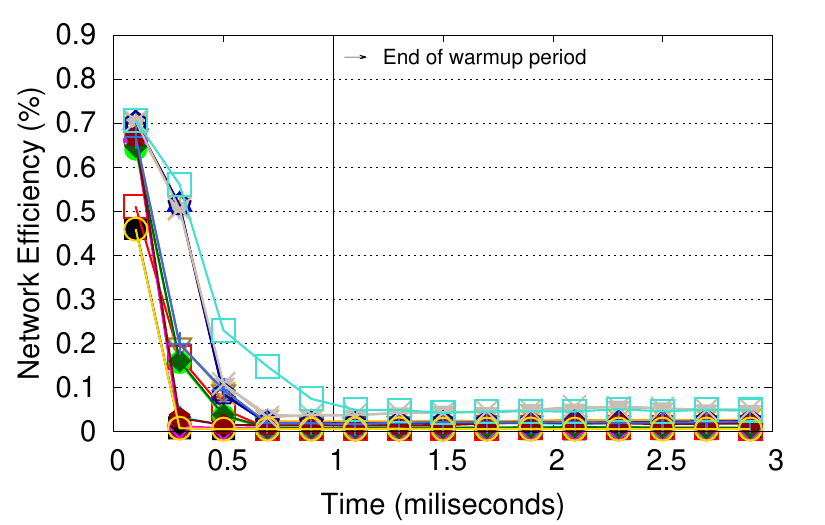}
\caption{Flow2SL and VOQs.}
\label{fig_time_RLFT_HS254_flow2sl3-voq}
\end{subfigure}

\caption{Normalized Throughput versus Time in a $11664$-node Fat-Tree network under HS25-4 synthetic traffic pattern.}
\label{fig:RLFT_HS254_vTime}
\end{figure*}

\begin{figure*}[!ht]
\vspace{-.5cm}
\begin{subfigure}[!th]{1\textwidth}
 \centering 
\includegraphics[width=1.0\textwidth]
{leyenda3.pdf}
\end{subfigure}

 \begin{subfigure}[!th]{0.475\textwidth}
 \centering 
\includegraphics[width=0.82\textwidth]
{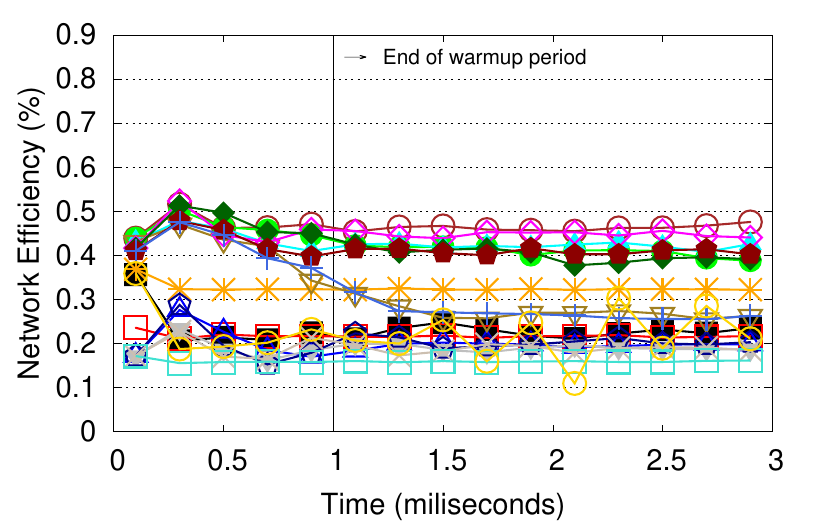}
\caption{1VC.}
\label{fig_time_RLFT_IHS_1q}
\end{subfigure}
 \begin{subfigure}[!th]{0.475\textwidth}
 \centering 
\includegraphics[width=0.82\textwidth]
{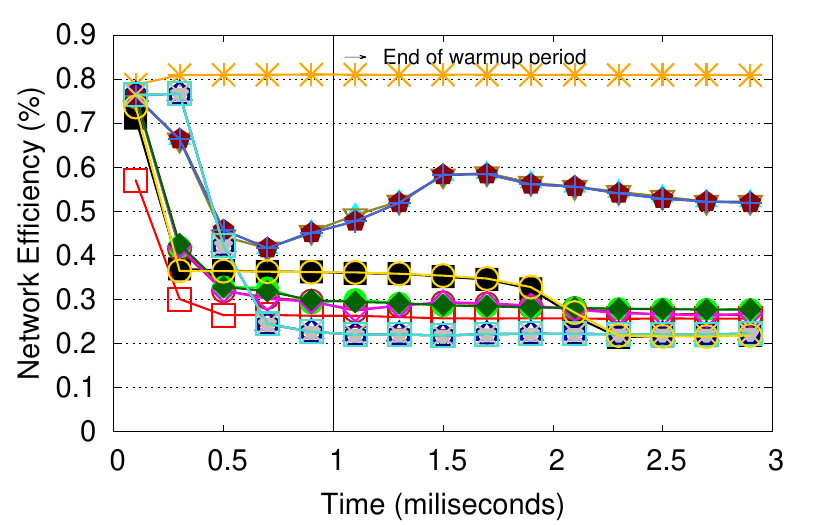}
\caption{1VC and VOQs.}
\label{fig_time_RLFT_IHS_1q-voq}
\end{subfigure}

 \begin{subfigure}[!th]{0.475\textwidth}
 \centering 
\includegraphics[width=0.82\textwidth]
{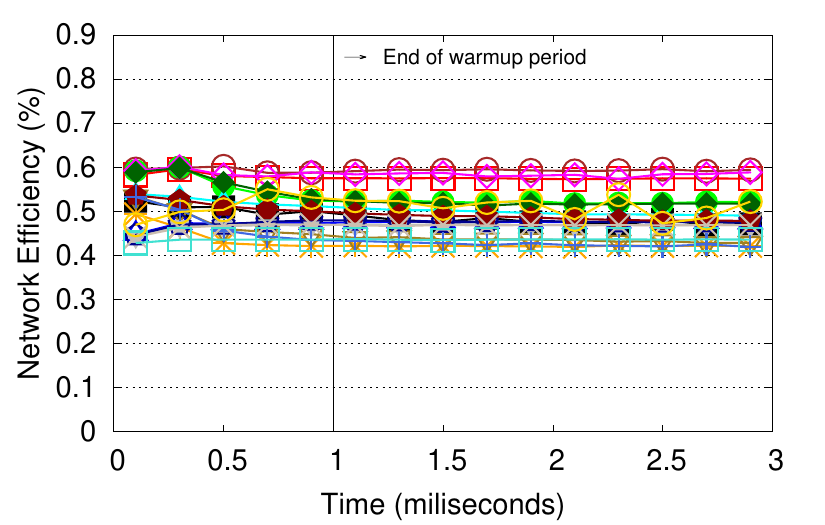}
\caption{DBBM.}
\label{fig_time_RLFT_IHS_dbbm3}
\end{subfigure}
 \begin{subfigure}[!th]{0.475\textwidth}
 \centering 
\includegraphics[width=0.82\textwidth]
{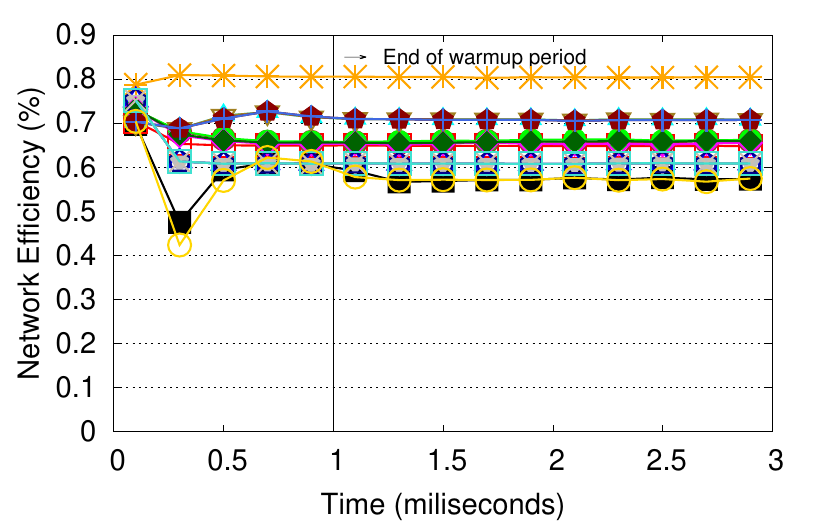}
\caption{DBBM and VOQs.}
\label{fig_time_RLFT_IHS_dbbm3-voq}
\end{subfigure}

 \begin{subfigure}[!th]{0.475\textwidth}
 \centering 
\includegraphics[width=0.82\textwidth]
{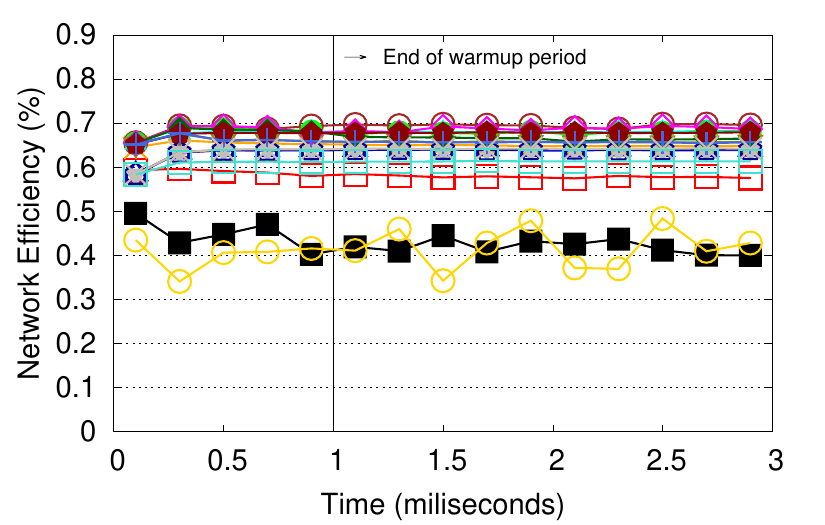}
\caption{vFtree.}
\label{fig_time_RLFT_IHS_vftree3}
\end{subfigure}
 \begin{subfigure}[!th]{0.475\textwidth}
 \centering 
\includegraphics[width=0.82\textwidth]
{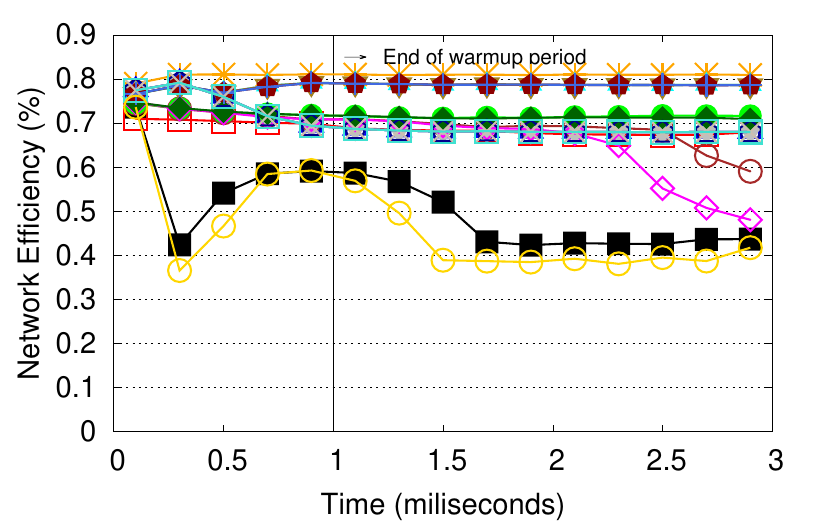}
\caption{vFtree and VOQs.}
\label{fig_time_RLFT_IHS_vftree3-voq}
\end{subfigure}

 \begin{subfigure}[!th]{0.475\textwidth}
 \centering 
\includegraphics[width=0.82\textwidth]
{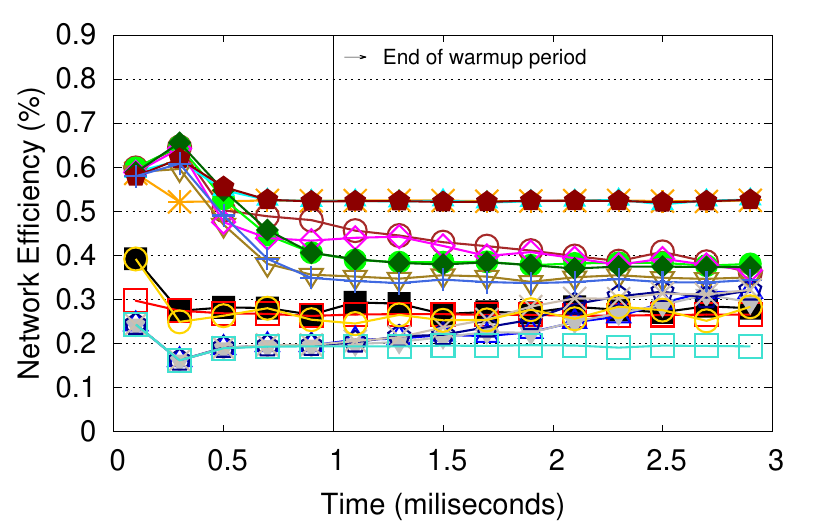}
\caption{Flow2SL.}
\label{fig_time_RLFT_IHS_flow2sl3}
\end{subfigure}
 \begin{subfigure}[!th]{0.475\textwidth}
 \centering 
\includegraphics[width=0.82\textwidth]
{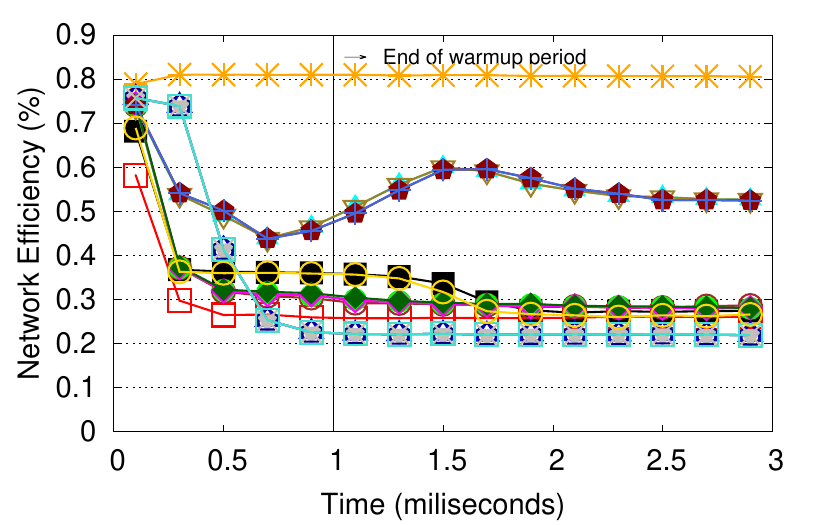}
\caption{Flow2SL and VOQs.}
\label{fig_time_RLFT_IHS_flow2sl3-voq}
\end{subfigure}

\caption{Normalized Throughput versus Time in a $11664$-node Fat-Tree network under IHS synthetic traffic pattern.}
\label{fig:RLFT_IHS_vTime}
\end{figure*}

On the other hand,
Figures \ref{fig:RLFT_HS10_vTime}, \ref{fig:RLFT_HS25_vTime}, \ref{fig:RLFT_HS104_vTime}, \ref{fig:RLFT_HS254_vTime} and \ref{fig:RLFT_IHS_vTime}
show the obtained throughput results as a function of time for the same network configurations as before.
In these figures, we also show simulation results for switches with and without VOQs.
Note that we have included in the plots the instant of time where the network warm-up period ends, and the network operation is in steady state.
Again, as it is shown in Tables \ref{tab:NOVOQ} and \ref{tab:VOQ}, switches without VOQs deal better with the congestion scenarios, and queuing schemes are able to reduce the HoL blocking better if they are combined with routing configurations using the proposed restrictions.
By contrast, switches with VOQs are more affected by congestion since VOQs, although they prevent the low-order HoL blocking, they are more affected by the high-order HoL blocking, which is quickly spread throughout the network.
Hence, vFtree using the \smallimg{RLFT_adaptive_s_3_18_50_sg.pdf} routing configuration again achieves the best results, regardless the traffic scenario and switch architecture (i.e., with and without VOQs).

\clearpage
In summary, the modeled state-of-the-art routing policies (i.e., $D$-mod-$K$, oblivious and adaptive routing) suffer a severe performance degradation under congestion, even when they are combined with queuing schemes.
Another important observation is that VOQ-based switches are more prone to spread network congestion, compared to non-VOQ-based switches.
Moreover, several routing configurations combined with queuing schemes using the proposed restrictions show a significant performance gain, compared to the state-of-the-art routing solutions, regardless the traffic pattern.
More precisely, vFtree using the \smallimg{RLFT_adaptive_s_3_18_50_sg.pdf} routing configuration obtains the best results for the simulated scenarios.
Flow2SL using the \smallimg{RLFT_adaptive_50_3_18_sg.pdf} routing configuration achieves the best results for switches without VOQs, but it suffers a dramatic performance degradation for VOQ-based switches when congestion is strong (i.e., HS10-4 and HS25-4 traffic patterns).
As a future work, we will optimize Flow2SL to perform a better mapping when restricted adaptive routing is used in VOQ-based switches.

 \section{Conclusions}
 \label{sec:conclusion}

Congestion is an important threat for the performance of interconnection networks in current HPC and Datacenter systems.
For this reason, we need to devise strategies to reduce the dramatic impact of congestion and its effects (i.e., HoL blocking and buffer hogging).
One common approach is the use of queuing schemes that divide the buffering space at network switches into several queues or virtual channels (VCs).
Moreover, other strategies for reducing congestion effects propose the use of adaptive or oblivious routing algorithms in order to dissolve congestion trees.
In previous studies we have analyzed that, when congestion appears, adaptive routing policies may spread congestion throughout the network, making queuing schemes useless.
In this paper, we have proposed a new approach to reduce congestion spreading when combining adaptive routing with queuing schemes.
Specifically, we have proposed three criteria used to restrict routing adaptivity and improve the path selection policy, which can be used together to reduce congestion spreading and improve the network performance.
The experiment results, obtained by means of simulations of large Fat-Tree networks under different congested traffic scenarios show that our approach efficiently improves the network performance under congestion situations, when adaptive routing is used in combination with queuing schemes.

 \section*{Acknowledgment}
This work has been jointly supported by the Spanish MINECO and European Commission (FEDER funds) under the project TIN2015-66972-C5-2-R (MINECO/FEDER),
by the Spanish Ministry of Science, Innovation \& Universities under the project RTI2018-098156-B-C52(MCI/AEI/FEDER, UE), and by Junta de Comunidades de Castilla-La Mancha under the projects PEII-2014-028-P and SBPLY/17/180501/000498.
Jose Rocher-Gonzalez is funded by Intel Corporation under the project CONV180068 signed by the University of Castilla-La Mancha (UCLM).
Jesus Escudero-Sahuquillo is funded by UCLM  and the European Commission (FSE funds), with a contract for accessing the Spanish System of Science, Technology and Innovation, for the implementation of the UCLM research program (UCLM resolution date: 31/07/2014). 


 \section*{References}

%
\bibliographystyle{elsarticle-harv} 
\bibliography{bibliography} 




\end{document}


\appendix
\section{Extended Simulation Results}
\label{results_appendix}

\begin{figure*}[!htb]

\begin{subfigure}[!th]{\textwidth}
 \centering 
\includegraphics[width=1.0\textwidth]{leyenda3.pdf}
\end{subfigure}

 \begin{subfigure}[!th]{0.47\textwidth}
 \centering 
\includegraphics[width=0.82\textwidth]{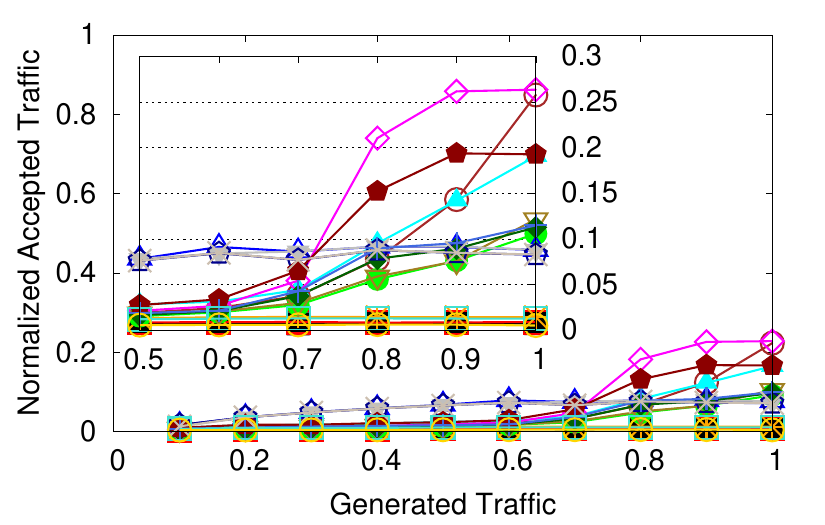}
\caption{1VC.}
\label{fig_RLFT_HS10_1q}
\end{subfigure}
 \begin{subfigure}[!th]{0.47\textwidth}
 \centering 
\includegraphics[width=0.82\textwidth]
{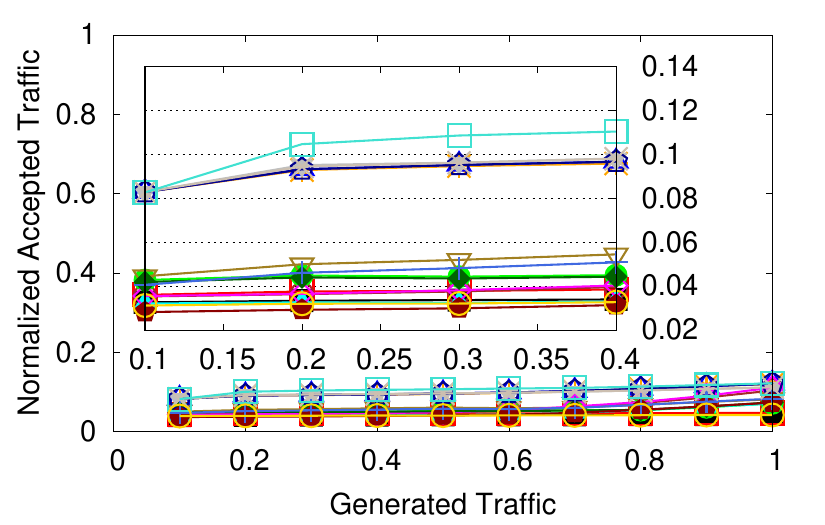}
\caption{1VC and VOQs.}
\label{fig_RLFT_HS10_1q-voq}
\end{subfigure}

 \begin{subfigure}[!th]{0.47\textwidth}
 \centering 
\includegraphics[width=0.82\textwidth]
{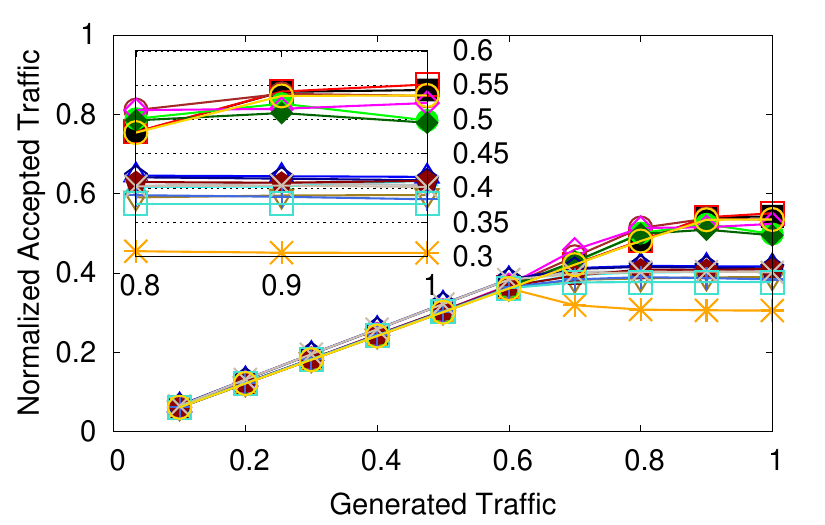}
\caption{DBBM.}
\label{fig_RLFT_HS10_dbbm3}
\end{subfigure}
 \begin{subfigure}[!th]{0.47\textwidth}
 \centering 
\includegraphics[width=0.82\textwidth]
{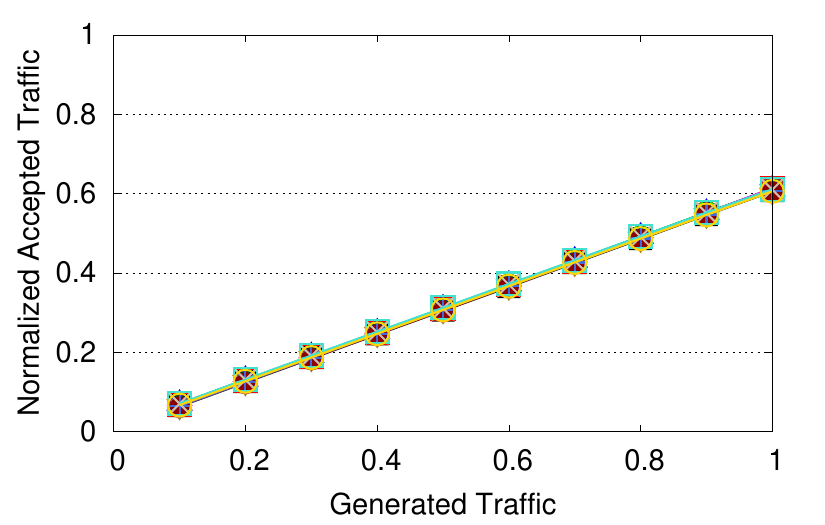}
\caption{DBBM and VOQs.}
\label{fig_RLFT_HS10_dbbm3-voq}
\end{subfigure}

 \begin{subfigure}[!th]{0.47\textwidth}
 \centering 
\includegraphics[width=0.82\textwidth]
{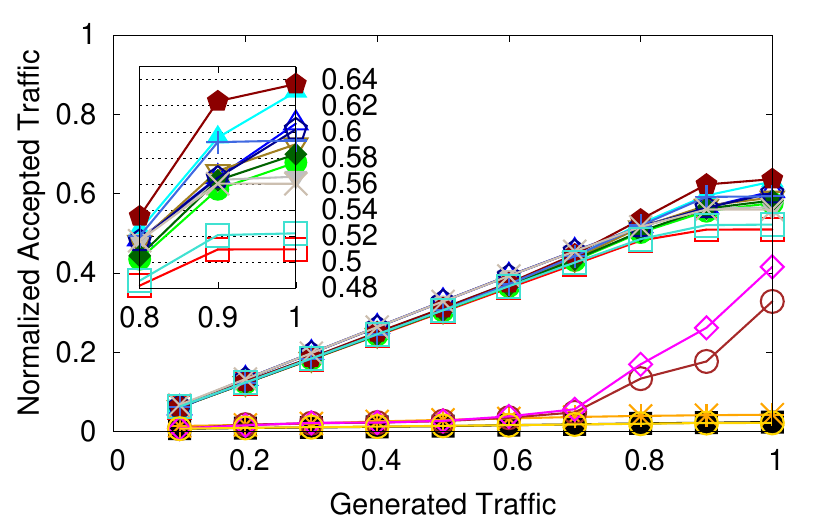}
\caption{vFtree.}
\label{fig_RLFT_HS10_vftree3}
\end{subfigure}
 \begin{subfigure}[!th]{0.47\textwidth}
 \centering 
\includegraphics[width=0.82\textwidth]
{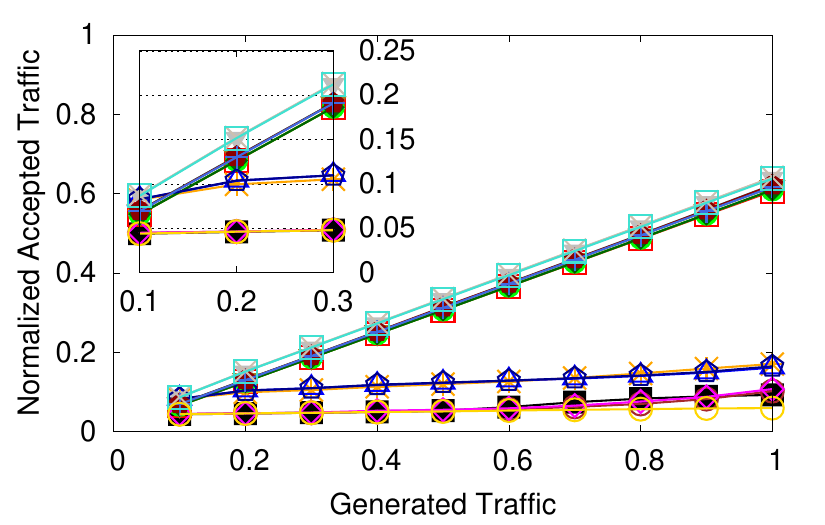}
\caption{vFtree and VOQs.}
\label{fig_RLFT_HS10_vftree3-voq}
\end{subfigure}

 \begin{subfigure}[!th]{0.47\textwidth}
 \centering 
\includegraphics[width=0.82\textwidth]
{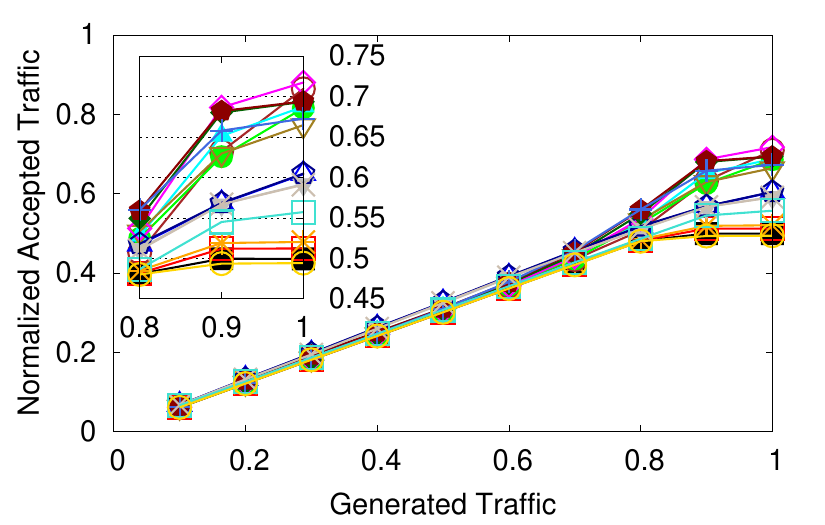}
\caption{Flow2SL.}
\label{fig_RLFT_HS10_flow2sl3}
\end{subfigure}
 \begin{subfigure}[!th]{0.47\textwidth}
 \centering 
\includegraphics[width=0.82\textwidth]
{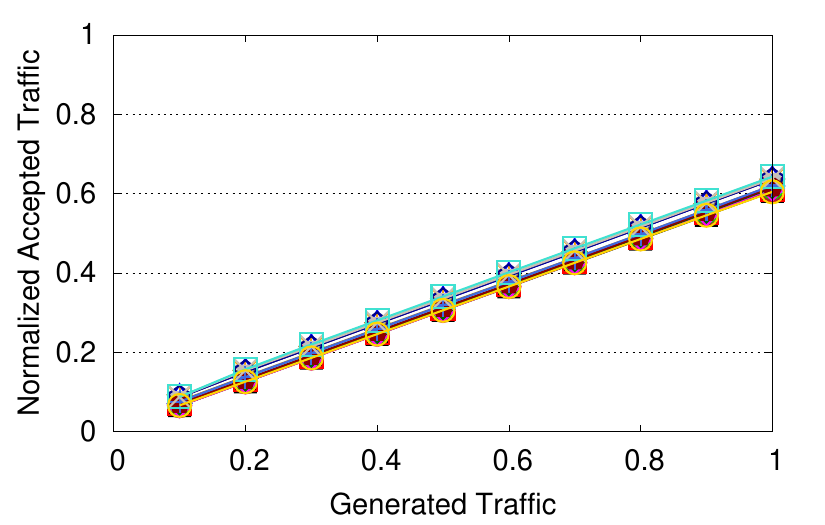}
\caption{Flow2SL and VOQs.}
\label{fig_RLFT_HS10_flow2sl3-voq}
\end{subfigure}

\caption{Normalized Throughput versus Generated Traffic Load in a $11664$-node RLFT under HS10-1 synthetic traffic pattern.}
\label{fig:RLFT_HS10}
\end{figure*}

\begin{figure*}[!htb]
\vspace{-.5cm}
\begin{subfigure}[!th]{1\textwidth}
 \centering 
\includegraphics[width=1.0\textwidth]
{leyenda3.pdf}
\end{subfigure}

 \begin{subfigure}[!th]{0.47\textwidth}
 \centering 
\includegraphics[width=0.82\textwidth]
{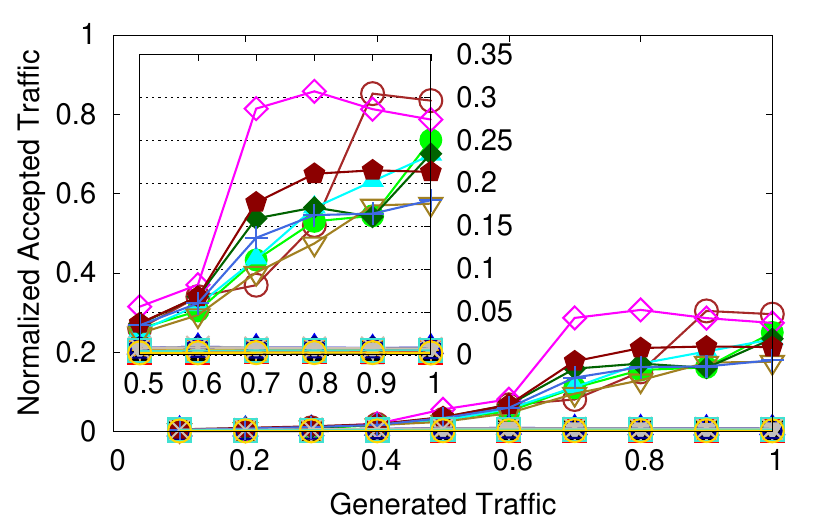}
\caption{1VC.}
\label{fig_RLFT_HS_1q}
\end{subfigure}
 \begin{subfigure}[!th]{0.47\textwidth}
 \centering 
\includegraphics[width=0.82\textwidth]
{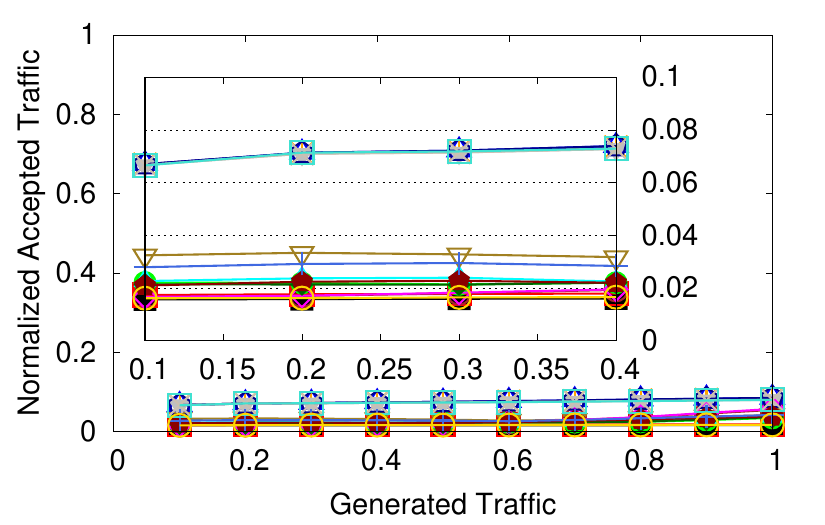}
\caption{1VC and VOQs.}
\label{fig_RLFT_HS_1q-voq}
\end{subfigure}

 \begin{subfigure}[!th]{0.47\textwidth}
 \centering 
\includegraphics[width=0.82\textwidth]
{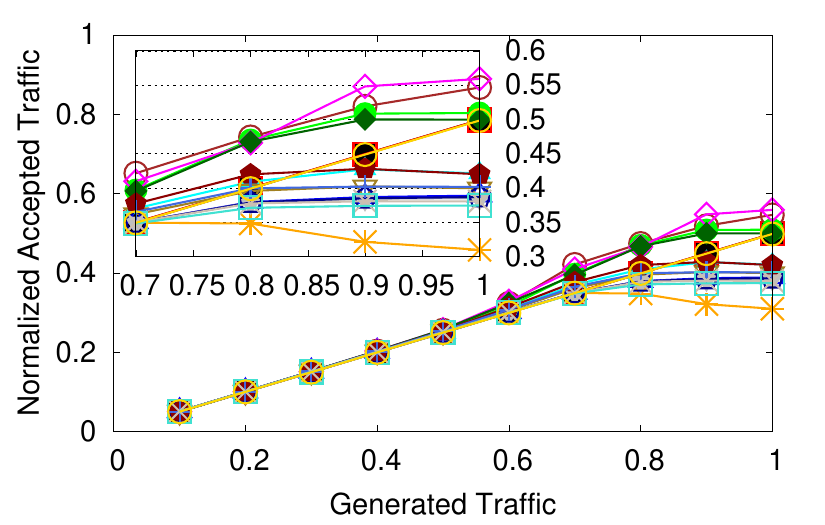}
\caption{DBBM.}
\label{fig_RLFT_HS_dbbm3}
\end{subfigure}
 \begin{subfigure}[!th]{0.47\textwidth}
 \centering 
\includegraphics[width=0.82\textwidth]
{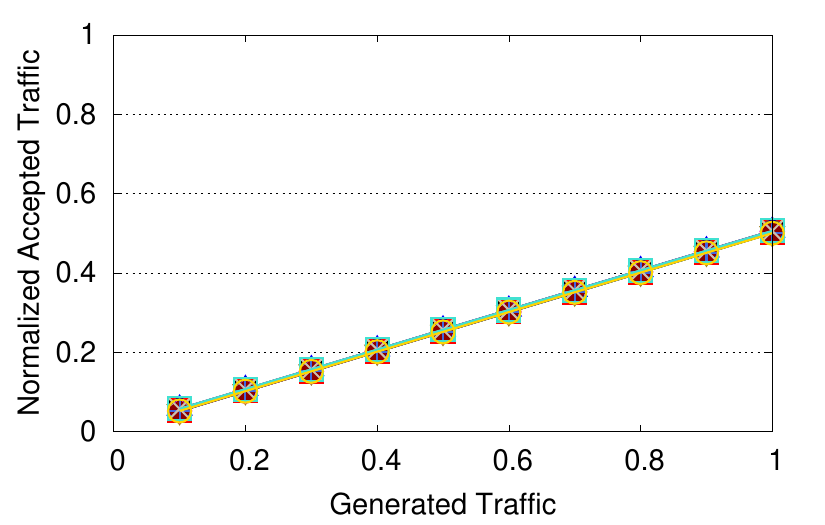}
\caption{DBBM and VOQs.}
\label{fig_RLFT_HS_dbbm3-voq}
\end{subfigure}

 \begin{subfigure}[!th]{0.47\textwidth}
 \centering 
\includegraphics[width=0.82\textwidth]
{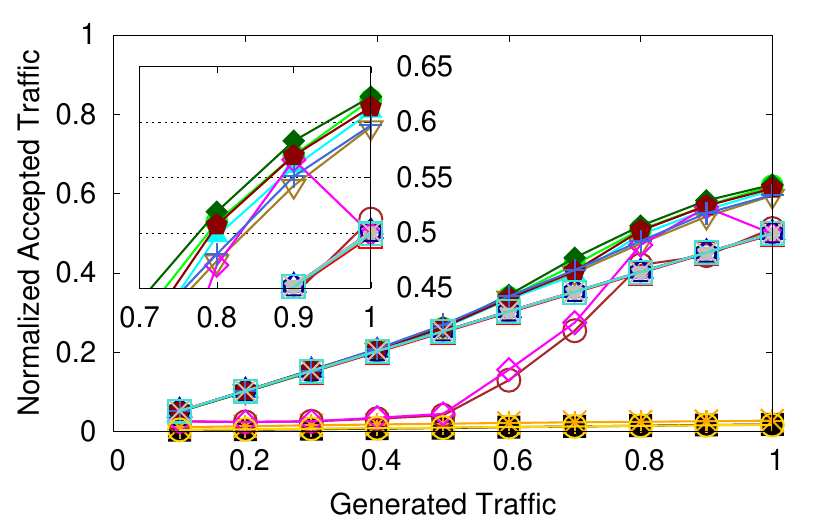}
\caption{vFtree.}
\label{fig_RLFT_HS_vftree3}
\end{subfigure}
 \begin{subfigure}[!th]{0.47\textwidth}
 \centering 
\includegraphics[width=0.82\textwidth]
{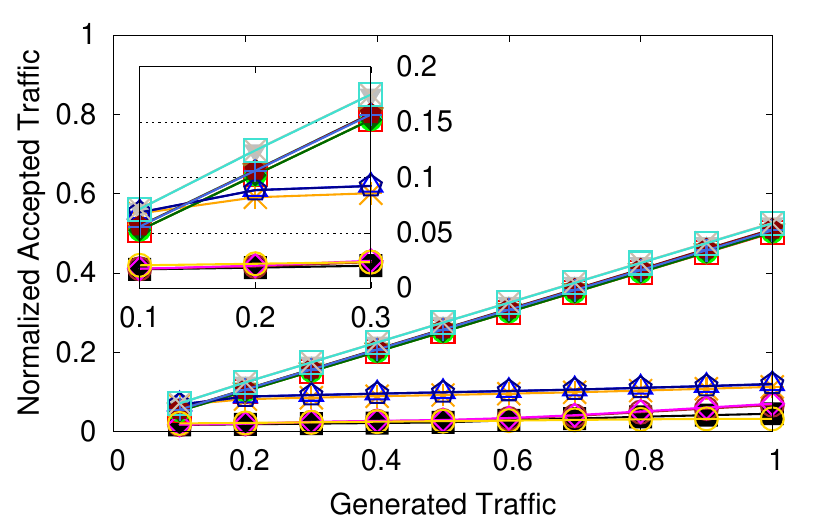}
\caption{vFtree and VOQs.}
\label{fig_RLFT_HS_vftree3-voq}
\end{subfigure}

 \begin{subfigure}[!th]{0.47\textwidth}
 \centering 
\includegraphics[width=0.82\textwidth]
{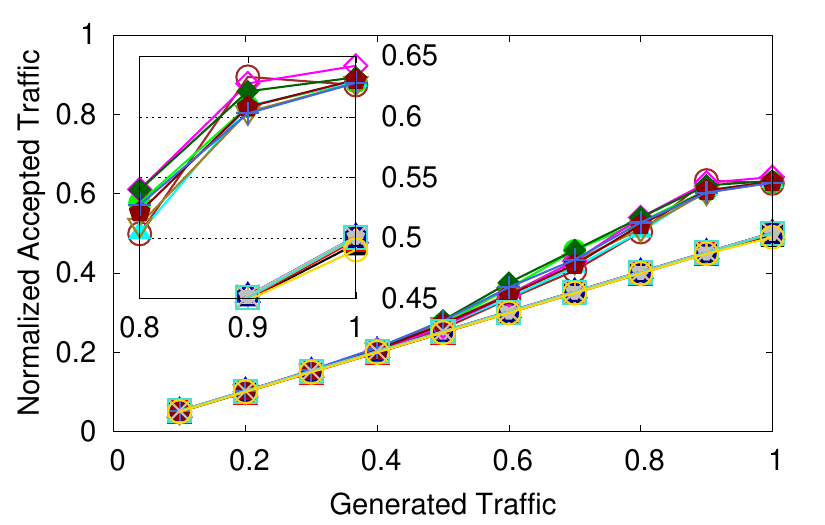}
\caption{Flow2SL.}
\label{fig_RLFT_HS_flow2sl3}
\end{subfigure}
 \begin{subfigure}[!th]{0.47\textwidth}
 \centering 
\includegraphics[width=0.82\textwidth]
{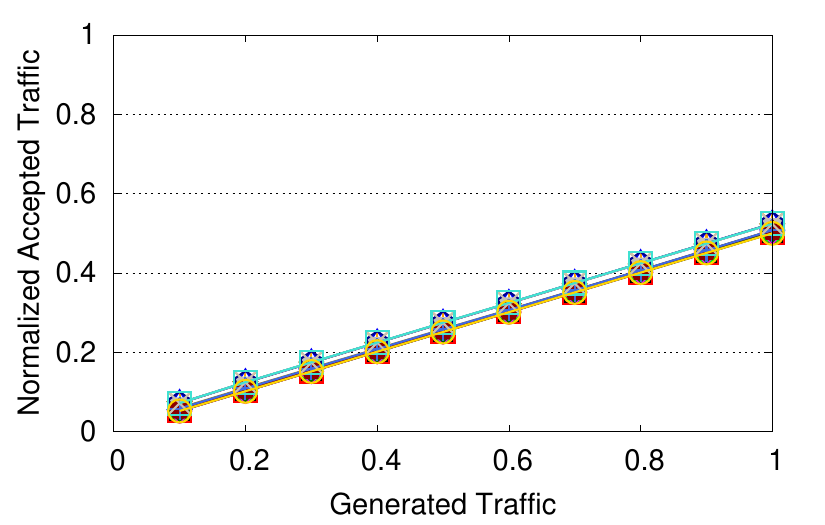}
\caption{Flow2SL and VOQs.}
\label{fig_RLFT_HS_flow2sl3-voq}
\end{subfigure}

\caption{Normalized Throughput versus Generated Traffic Load in a $11664$-node RLFT under HS25-1 synthetic traffic pattern.}
\label{fig:RLFT_HS25}
\end{figure*}

Figure \ref{fig:RLFT_HS10} and \ref{fig:RLFT_HS25} show the simulation results when 10\% and 25\% of the end-nodes generate congested traffic addressed to a single end-node (i.e., traffic pattern HS10-1 and HS25-1). 

As we can see in these figures, the 1-VC configuration cannot deal with congestion just applying restricted adaptive routing, due to the HoL blocking (see Figures \ref{fig_RLFT_HS10_1q},  \ref{fig_RLFT_HS10_1q-voq}, \ref{fig_RLFT_HS_1q} and \ref{fig_RLFT_HS_1q-voq}).

Note that when no VOQs are used, the adaptive routing using \emph{triggering thresholds} is able to raise a bit the performance, since this restriction delays the adaptivity decisions. The results show that better performance is achieved in a more congested scenario. 
This strange effect is due to the congestion tree. In fact, a higher incast congestion scenario creates a larger congestion tree that reaches the level 1 switches in the routing upward phase. Note that the performance raise in Figure \ref{fig_RLFT_HS10_1q} and \ref{fig_RLFT_HS_1q} when they reach 70\% and 50\% of the generated traffic. 
As a consequence, the performance of adaptive algorithms, restricted only to the second stage,  drops to 0.
By contrast, when VOQs are used, 1VC performance is close to 0\% as congestion trees grow into the VOQs, regardless the used routing algorithm.

DBBM (see Figures \ref{fig_RLFT_HS10_dbbm3}, \ref{fig_RLFT_HS10_dbbm3-voq}, \ref{fig_RLFT_HS_dbbm3} and \ref{fig_RLFT_HS_dbbm3-voq}) outperforms 1VC regardless the routing configurations.
When no VOQs are used,the results are homogeneous until the percentage of injected traffic exceeds 60\% and they show that K/$\Delta$ and 2S restrictions do not work as good as the other routing configurations.
The 2S restriction does not work well due to routing configuration uses $D$-mod-$K$ in the first stage of the Fat Tree, so that DBBM maps all the destinations to the same VC (see \figurename~\ref{fig_RLFT_3_2_destro}). Something similar happens with K/$\Delta$ restriction. This restriction uses a module to decide which upward port is chosen in the routing function \ref{algorithmAdaptive}. Since the module for choosing the port and mapping in the queue matches, then all adapted packets are mapped in the same queue. That's why the worst results are obtained when packets are adapted only in the second stage without any restrictions.
In other words, DBBM improves the deterministic routing when it adapts the routes in the first stage and manages to balance the use of queues.
However, when VOQs are used, low order HoL-blocking disappears and the results converge regardless the used routing algorithm.

When we use vFtree (see Figures \ref{fig_RLFT_HS10_vftree3}, \ref{fig_RLFT_HS10_vftree3-voq}, \ref{fig_RLFT_HS_vftree3-voq} and \ref{fig_RLFT_HS_vftree3}) the network efficiency drops significantly when adaptive routing is applied in the second stage (see the top-right switch \figurename~\ref{fig_RLFT_3_2_adapt} and downward stages in red), since all the destinations are mapped to all the VCs so that the HoL blocking probability increases.
In this case, the K/$\Delta$ restrictions avoid this defect and achieve productivity even when packets are adapted just in the second stage.
Note that *S-K configurations are able to raise the performance when the injection traffic is over 70\% and 60\%.  
This effect is more noticeable in the VOQ architecture because of its higher productivity.

When we use Flow2SL without VOQs (see Figures \ref{fig_RLFT_HS10_flow2sl3}, \ref{fig_RLFT_HS_flow2sl3}), we can see that better results are obtained when adaptive routing is applied in the first stage (FS and S*)  with triggering restrictions.
On the other hand, when switches implement VOQs (see Figures \ref{fig_RLFT_HS10_flow2sl3-voq}, \ref{fig_RLFT_HS_flow2sl3-voq}), the obtained results are similar.
The reason is that VOQs spread the congestion throughout all the VCs in the same buffer, since flow-control is performed at VC level, but not at VOQ level.

\begin{figure*}[!htb]
\vspace{-.5cm}
\begin{subfigure}[!th]{1\textwidth}
 \centering 
\includegraphics[width=1.0\textwidth]
{leyenda3.pdf}
\end{subfigure}

 \begin{subfigure}[!th]{0.47\textwidth}
 \centering 
\includegraphics[width=0.82\textwidth]
{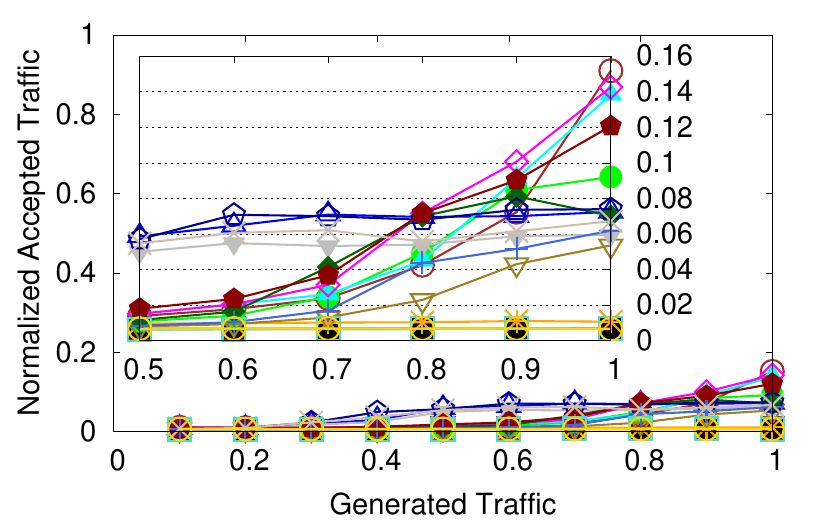}
\caption{1VC.}
\label{fig_RLFT_HS104_1q}
\end{subfigure}
 \begin{subfigure}[!th]{0.47\textwidth}
 \centering 
\includegraphics[width=0.82\textwidth]
{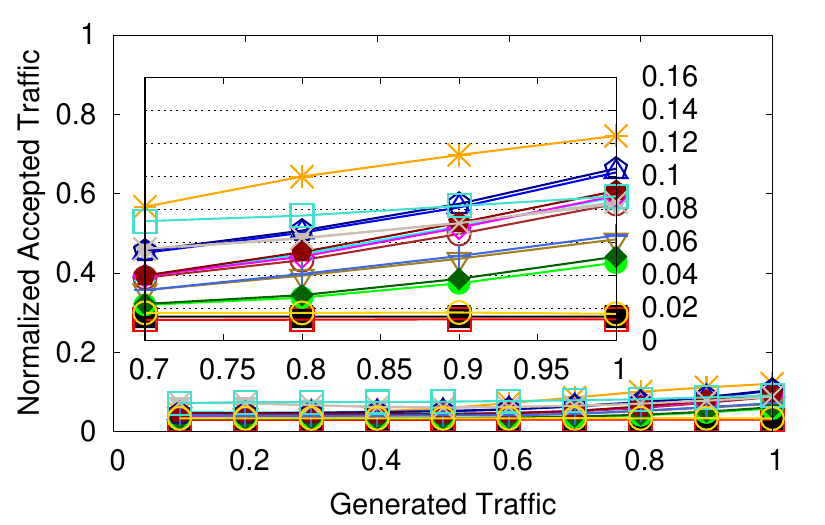}
\caption{1VC and VOQs.}
\label{fig_RLFT_HS104_1q-voq}
\end{subfigure}

 \begin{subfigure}[!th]{0.47\textwidth}
 \centering 
\includegraphics[width=0.82\textwidth]
{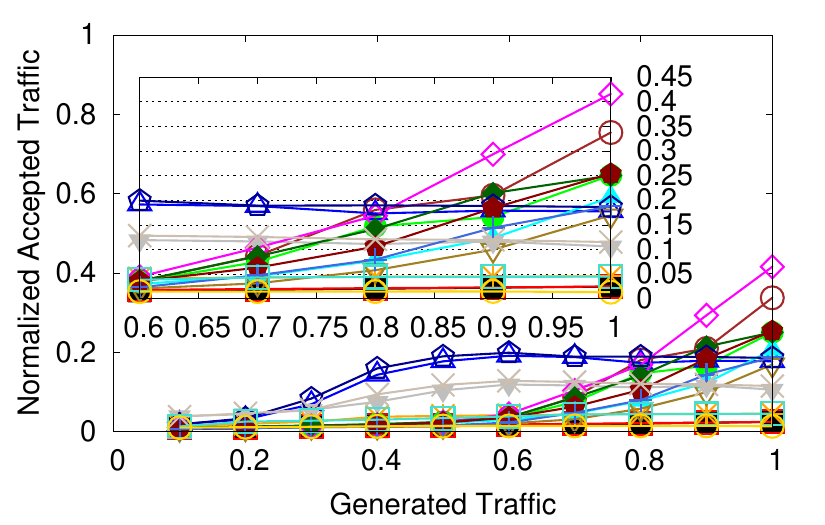}
\caption{DBBM.}
\label{fig_RLFT_HS104_dbbm3}
\end{subfigure}
 \begin{subfigure}[!th]{0.47\textwidth}
 \centering 
\includegraphics[width=0.82\textwidth]
{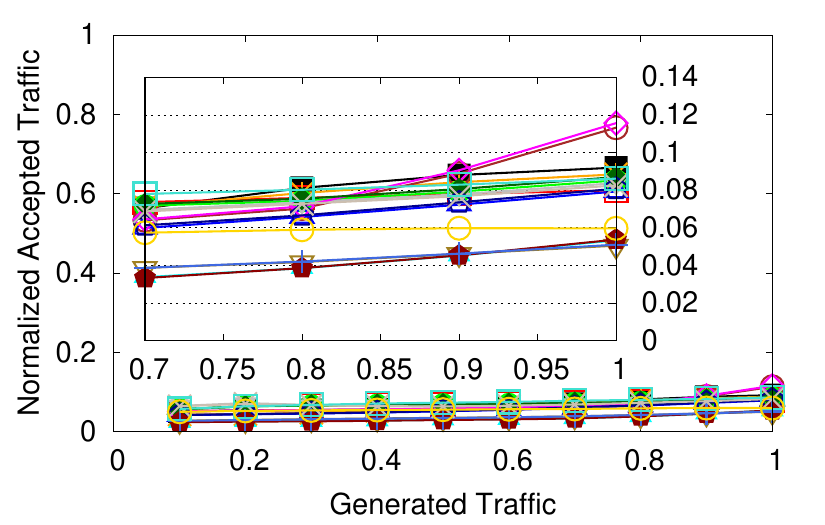}
\caption{DBBM and VOQs.}
\label{fig_RLFT_HS104_dbbm3-voq}
\end{subfigure}

 \begin{subfigure}[!th]{0.47\textwidth}
 \centering 
\includegraphics[width=0.82\textwidth]
{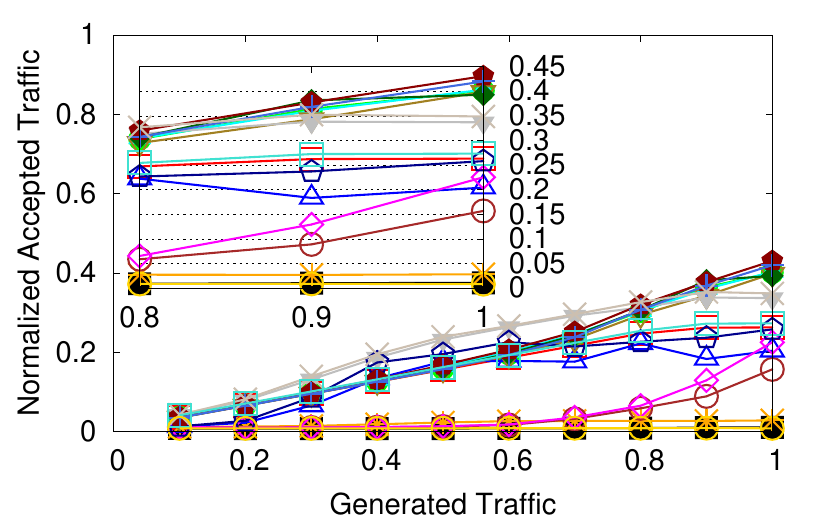}
\caption{vFtree.}
\label{fig_RLFT_HS104_vftree3}
\end{subfigure}
 \begin{subfigure}[!th]{0.47\textwidth}
 \centering 
\includegraphics[width=0.82\textwidth]
{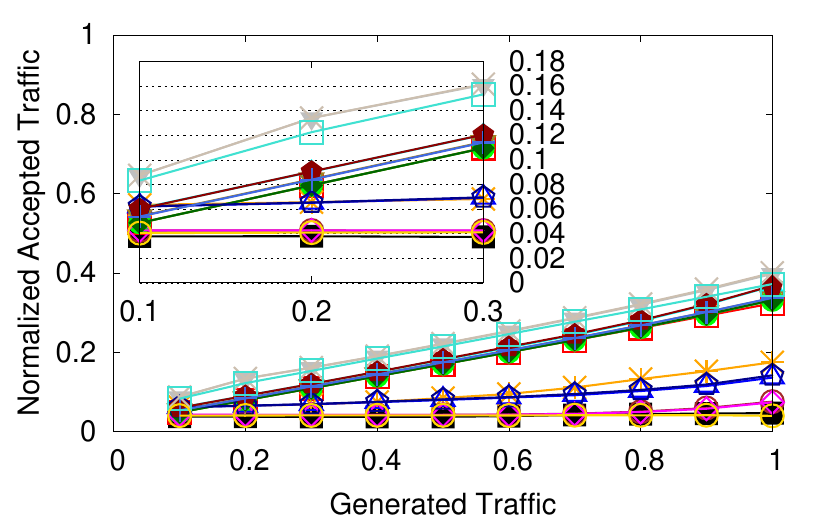}
\caption{vFtree and VOQs.}
\label{fig_RLFT_HS104_vftree3-voq}
\end{subfigure}

 \begin{subfigure}[!th]{0.47\textwidth}
 \centering 
\includegraphics[width=0.82\textwidth]
{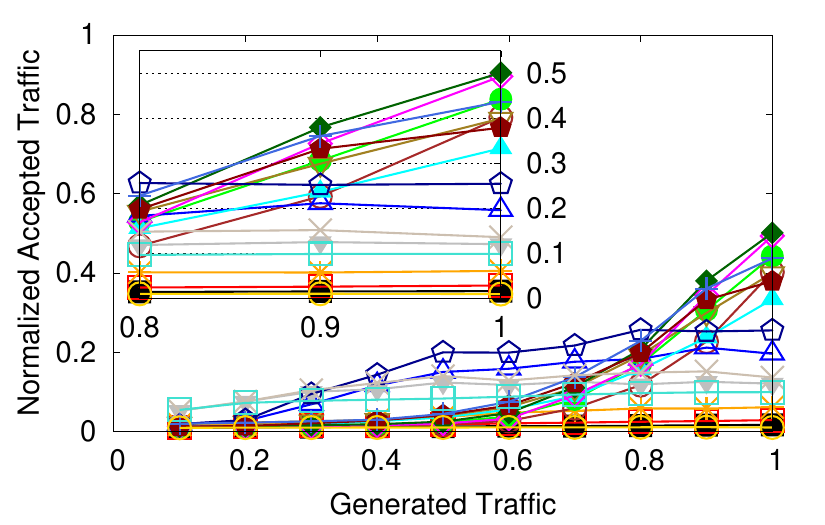}
\caption{Flow2SL.}
\label{fig_RLFT_HS104_flow2sl3}
\end{subfigure}
 \begin{subfigure}[!th]{0.47\textwidth}
 \centering 
\includegraphics[width=0.82\textwidth]
{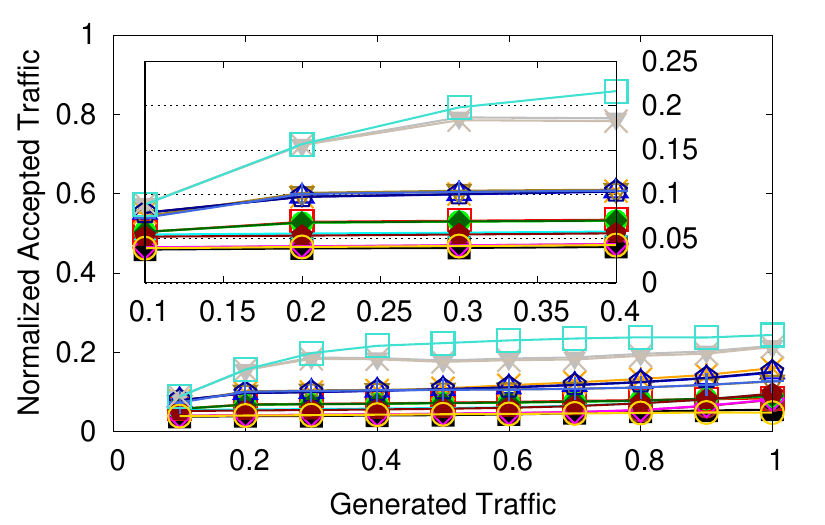}
\caption{Flow2SL and VOQs.}
\label{fig_RLFT_HS104_flow2sl3-voq}
\end{subfigure}

\caption{Normalized Throughput versus Generated Traffic Load in a $11664$-node RLFT under HS10-4 synthetic traffic pattern.}
\label{fig:RLFT_HS104}
\end{figure*}

\begin{figure*}[!htb]
\vspace{-.5cm}
\begin{subfigure}[!th]{1\textwidth}
 \centering 
\includegraphics[width=1.0\textwidth]
{leyenda3.pdf}
\end{subfigure}

 \begin{subfigure}[!th]{0.47\textwidth}
 \centering 
\includegraphics[width=0.82\textwidth]
{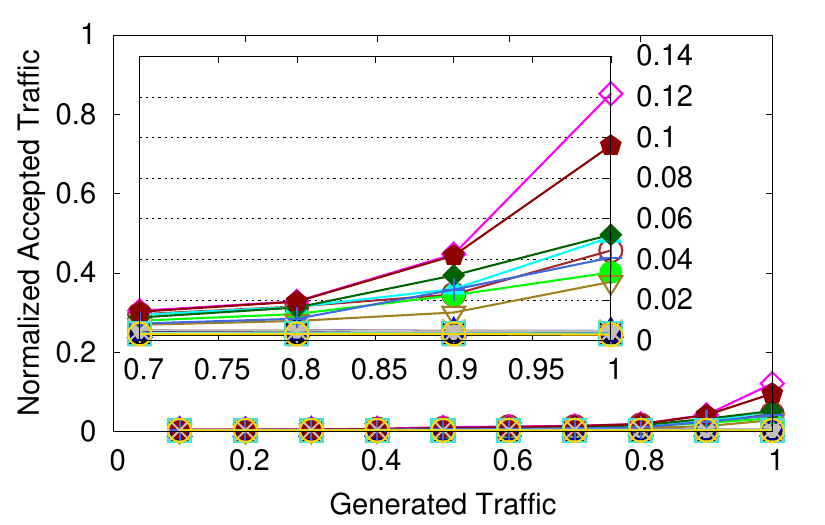}
\caption{1VC.}
\label{fig_RLFT_HS254_1q}
\end{subfigure}
 \begin{subfigure}[!th]{0.47\textwidth}
 \centering 
\includegraphics[width=0.82\textwidth]
{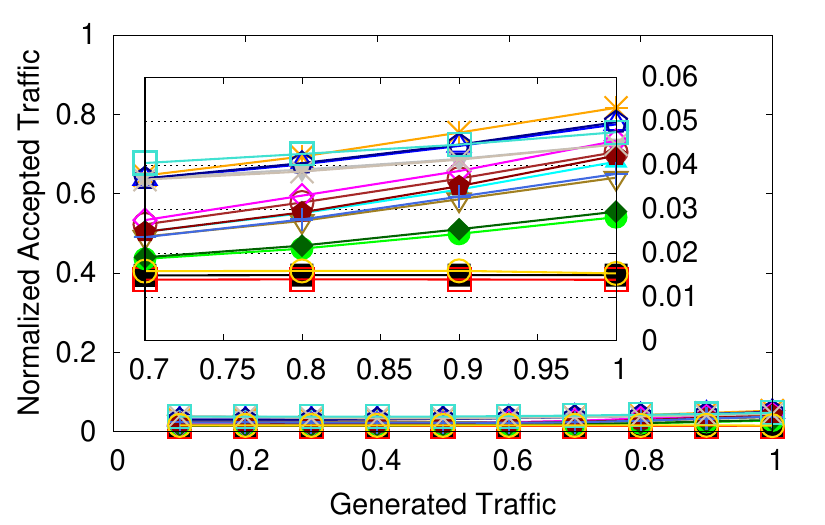}
\caption{1VC and VOQs.}
\label{fig_RLFT_HS254_1q-voq}
\end{subfigure}

 \begin{subfigure}[!th]{0.47\textwidth}
 \centering 
\includegraphics[width=0.82\textwidth]
{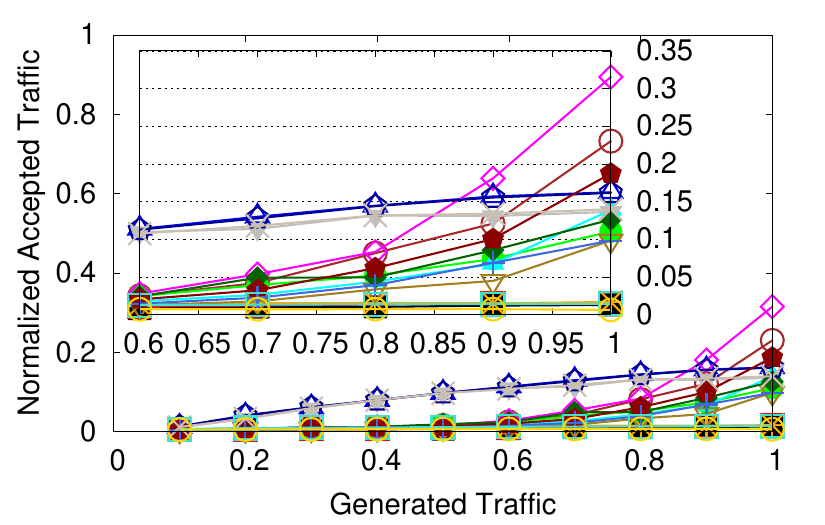}
\caption{DBBM.}
\label{fig_RLFT_HS254_dbbm3}
\end{subfigure}
 \begin{subfigure}[!th]{0.47\textwidth}
 \centering 
\includegraphics[width=0.82\textwidth]
{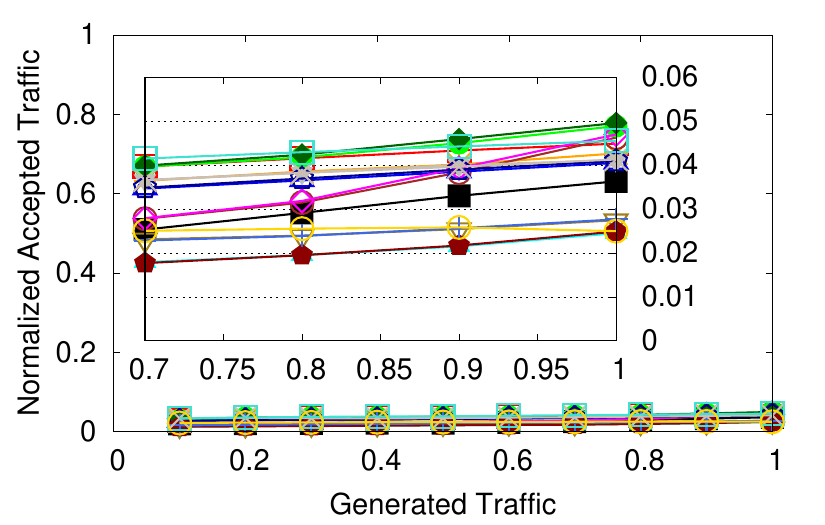}
\caption{DBBM and VOQs.}
\label{fig_RLFT_HS254_dbbm3-voq}
\end{subfigure}

 \begin{subfigure}[!th]{0.47\textwidth}
 \centering 
\includegraphics[width=0.82\textwidth]
{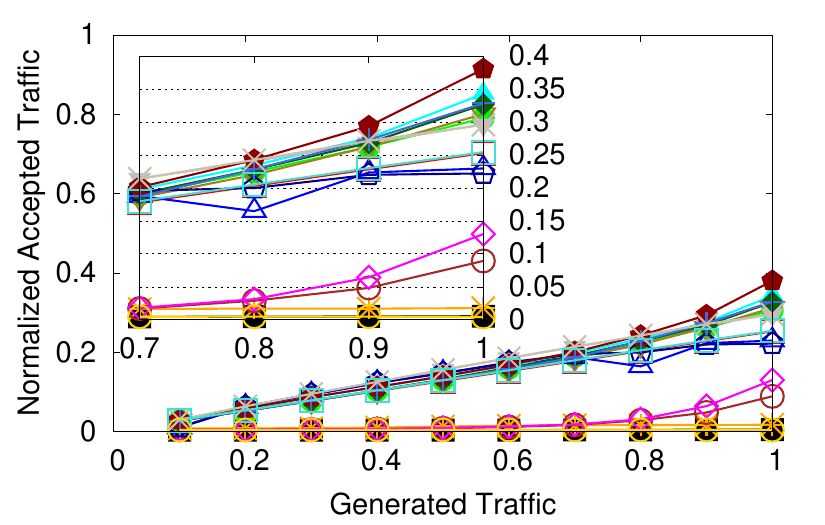}
\caption{vFtree.}
\label{fig_RLFT_HS254_vftree3}
\end{subfigure}
 \begin{subfigure}[!th]{0.47\textwidth}
 \centering 
\includegraphics[width=0.82\textwidth]
{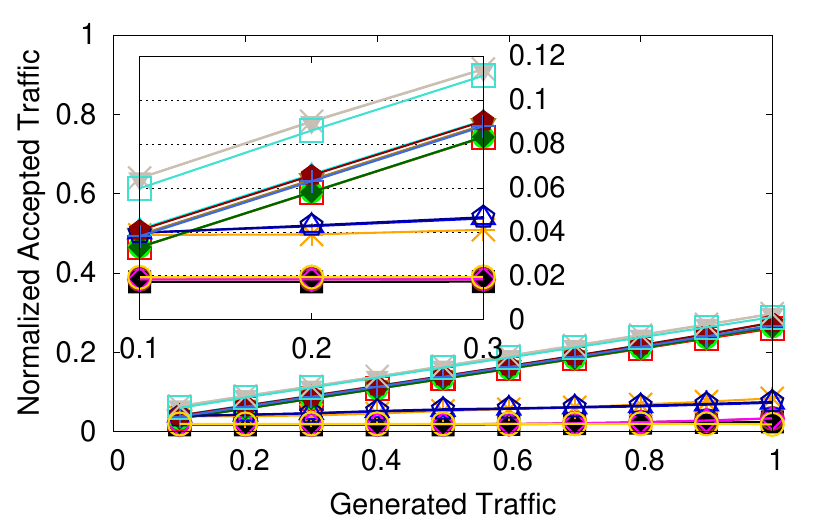}
\caption{vFtree and VOQs.}
\label{fig_RLFT_HS254_vftree3-voq}
\end{subfigure}

 \begin{subfigure}[!th]{0.47\textwidth}
 \centering 
\includegraphics[width=0.82\textwidth]
{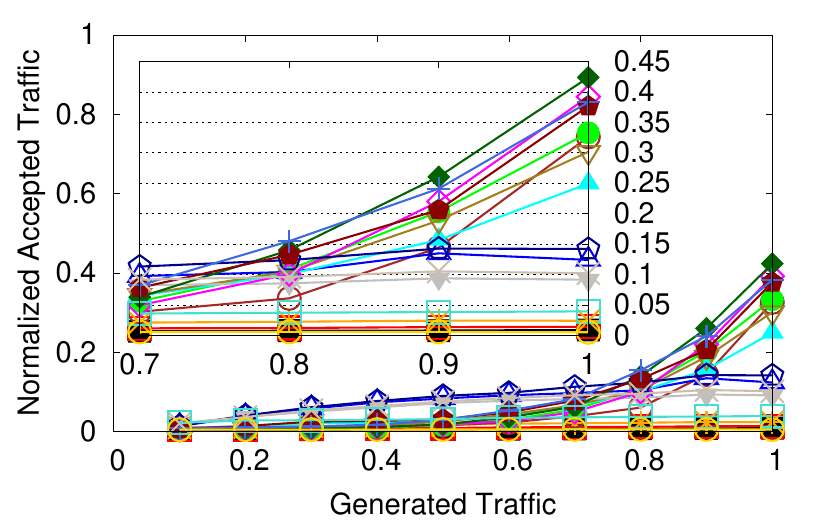}
\caption{Flow2SL.}
\label{fig_RLFT_HS254_flow2sl3}
\end{subfigure}
 \begin{subfigure}[!th]{0.47\textwidth}
 \centering 
\includegraphics[width=0.82\textwidth]
{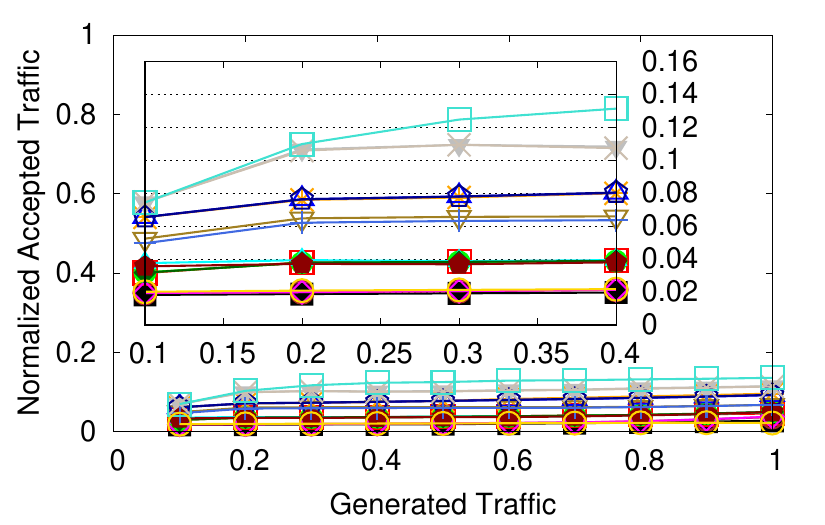}
\caption{Flow2SL and VOQs.}
\label{fig_RLFT_HS254_flow2sl3-voq}
\end{subfigure}

\caption{Normalized Throughput versus Generated Traffic Load in a $11664$-node RLFT under HS25-4 synthetic traffic pattern.}
\label{fig:RLFT_HS254}
\end{figure*}

Figures \ref{fig:RLFT_HS104} and \ref{fig:RLFT_HS254} show experiment results for the same scenarios described before when we generate traffic creating four congestion trees  (i.e., traffic HS10-4 and HS25-4 depicted in \figurename~\ref{fig_traffic_HS4}).

This is a very strong congestion situation, since we generate four congestion trees whose branches affect to different VCs at the same time and thus reduces the effectiveness of the static queuing scheme.
Therefore, this situation makes complex, even for restricted routing combined with queuing schemes, to reduce the HoL blocking.
When no VOQs are used, the SQS show similar behaviour as described above but they yield less performance.
By contrast, when VOQs are used, the only queuing scheme that works is vFtree (see Figures \ref{fig_RLFT_HS_vftree3} and \ref{fig_RLFT_HS_vftree3}) due to its mapping properties. It works well with deterministic routing but a little more productivity can be achieved when we adapt packets using K/$\Delta$ restriction even when there are four strong congestion trees.

\begin{figure*}[!htb]
\vspace{-.5cm}
\begin{subfigure}[!th]{1\textwidth}
 \centering 
\includegraphics[width=1.0\textwidth]
{leyenda3.pdf}
\end{subfigure}

 \begin{subfigure}[!th]{0.47\textwidth}
 \centering 
\includegraphics[width=0.82\textwidth]
{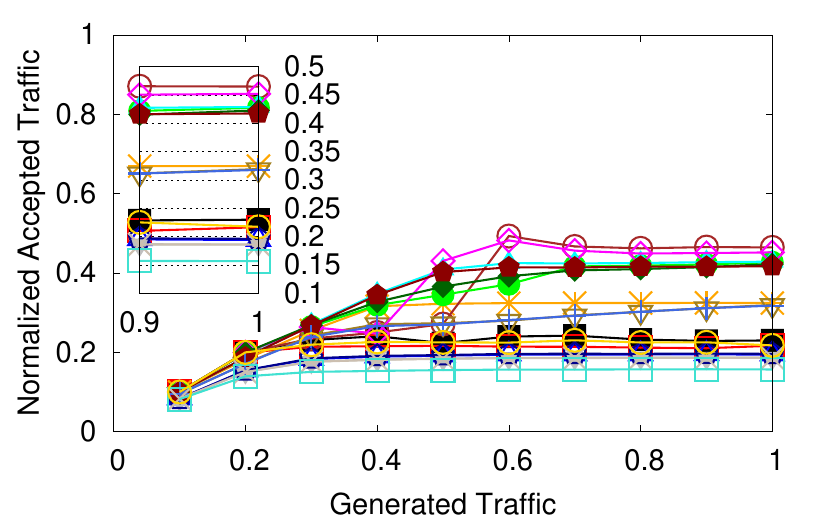}
\caption{1VC.}
\label{fig_RLFT_IHS_1q}
\end{subfigure}
 \begin{subfigure}[!th]{0.47\textwidth}
 \centering 
\includegraphics[width=0.82\textwidth]
{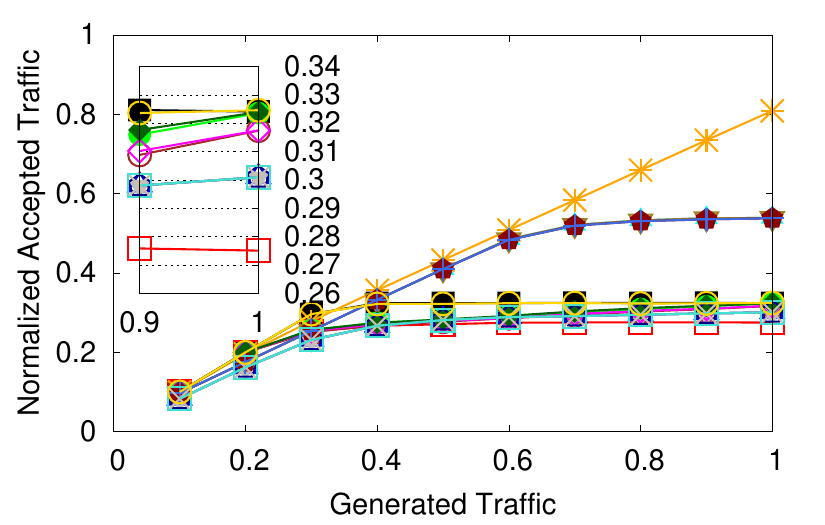}
\caption{1VC and VOQs.}
\label{fig_RLFT_IHS_1q-voq}
\end{subfigure}

 \begin{subfigure}[!th]{0.47\textwidth}
 \centering 
\includegraphics[width=0.82\textwidth]
{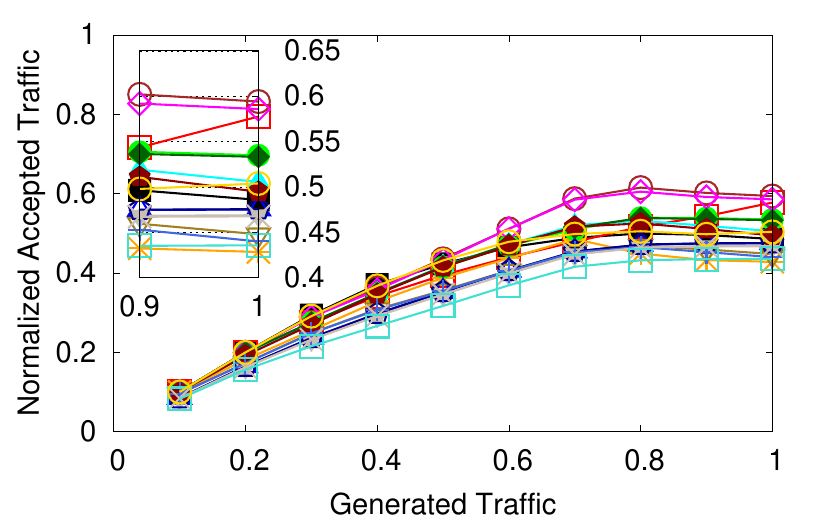}
\caption{DBBM.}
\label{fig_RLFT_IHS_dbbm3}
\end{subfigure}
 \begin{subfigure}[!th]{0.47\textwidth}
 \centering 
\includegraphics[width=0.82\textwidth]
{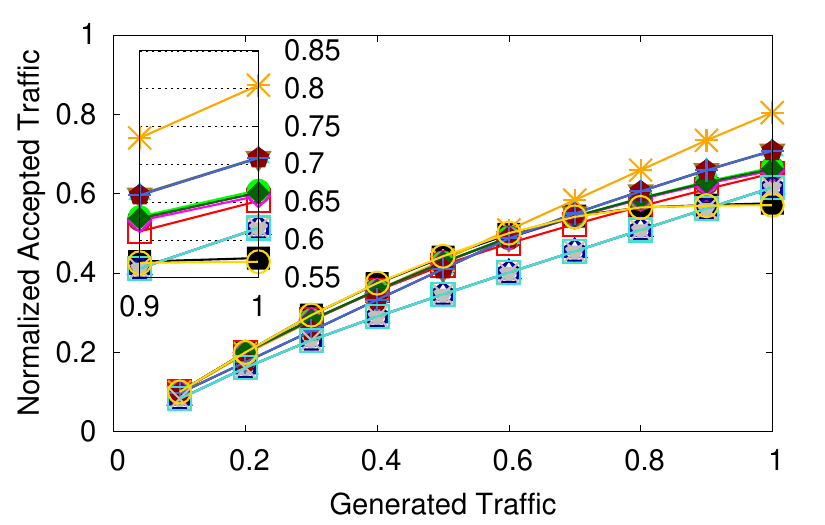}
\caption{DBBM and VOQs.}
\label{fig_RLFT_IHS_dbbm3-voq}
\end{subfigure}

 \begin{subfigure}[!th]{0.47\textwidth}
 \centering 
\includegraphics[width=0.82\textwidth]
{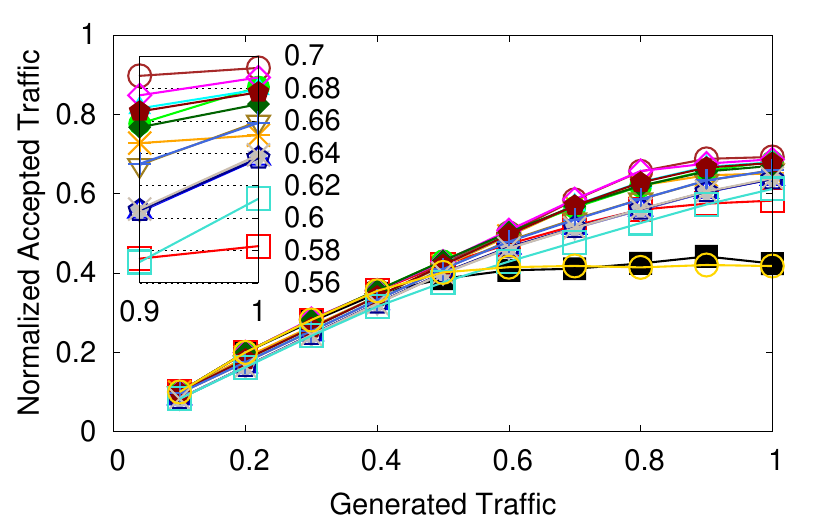}
\caption{vFtree.}
\label{fig_RLFT_IHS_vftree3}
\end{subfigure}
 \begin{subfigure}[!th]{0.47\textwidth}
 \centering 
\includegraphics[width=0.82\textwidth]
{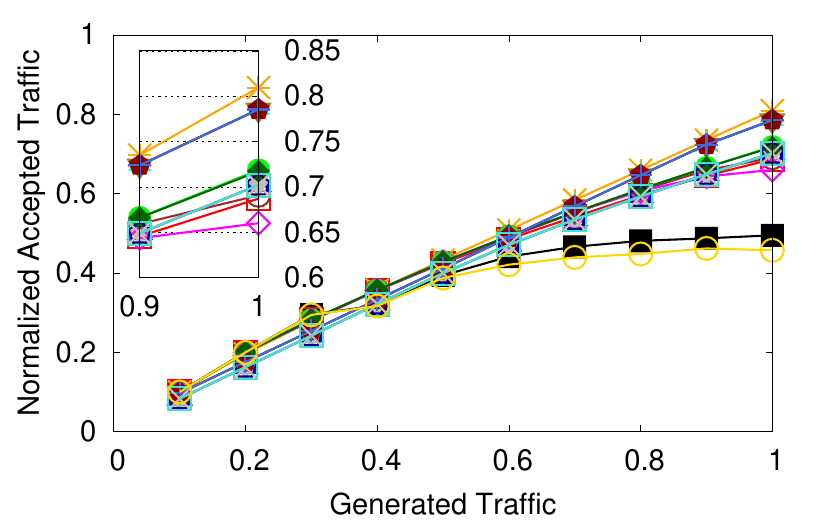}
\caption{vFtree and VOQs.}
\label{fig_RLFT_IHS_vftree3-voq}
\end{subfigure}

 \begin{subfigure}[!th]{0.47\textwidth}
 \centering 
\includegraphics[width=0.82\textwidth]
{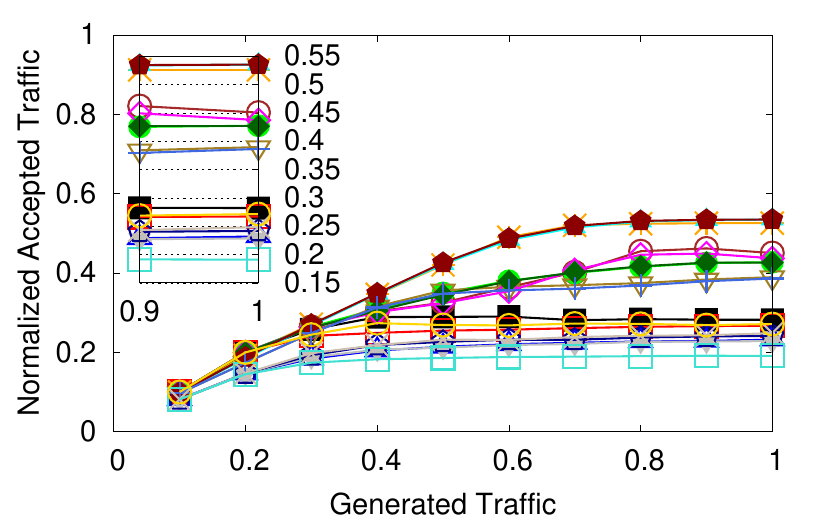}
\caption{Flow2SL.}
\label{fig_RLFT_IHS_flow2sl3}
\end{subfigure}
 \begin{subfigure}[!th]{0.47\textwidth}
 \centering 
\includegraphics[width=0.82\textwidth]
{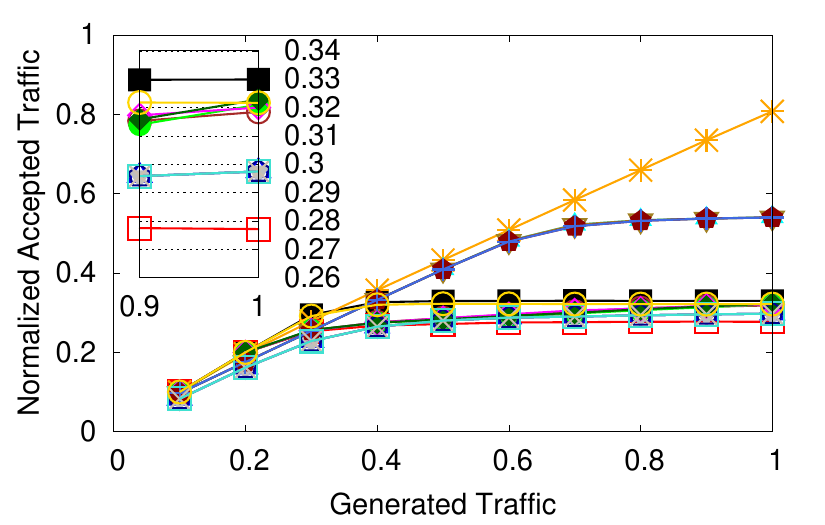}
\caption{Flow2SL and VOQs.}
\label{fig_RLFT_IHS_flow2sl3-voq}
\end{subfigure}

\caption{Normalized Throughput versus Generated Traffic Load in a $11664$-node RLFT under IHS synthetic traffic pattern.}
\label{fig:RLFT_IHS}
\end{figure*}

Figure \ref{fig:RLFT_IHS} shows simulation results for traffic generating hot-spots in intermediate stages of the topology (see \figurename~\ref{fig_traffic_HSW}).
In particular, hot-spots (i.e., congestion tree roots) are generated in the output ports of some switches placed at the second stage of the RLFT.
As we can see in the figures, this traffic pattern generates a congestion situation where $D$-mod-$K$, oblivious and fully-adaptive routing algorithms (both with and without restrictions) can deal with the HoL blocking appearing in the network.
For switches without VOQs, vFtree queuing scheme achieves the best results combined with adaptive routing using combined restrictions, such as ADAP-NoTH-SS-K, ADAP-TH-AS-K or ADAP-2TH-AS-$K/3$ outperform 1VC, DBBM and Flow2SL. 
This behavior is the same for vFtree when using switches with VOQs.
Note that switches with VOQs using the configuration ADAP-NoTH-SS-K achieve excellent results regardless the use of queuing schemes (see Figures  \ref{fig_RLFT_IHS_1q-voq}, \ref{fig_RLFT_IHS_dbbm3-voq},  \ref{fig_RLFT_IHS_vftree3-voq}, \ref{fig_RLFT_IHS_flow2sl3-voq}).
Although the use of restricted routing and queuing schemes are not necessary in this traffic scenario when using VOQs, note that their use preserves the performance gains shown in the previous traffic scenarios when using restricted adaptive routing and queuing schemes.
Therefore, the use of restricted adaptive routing combined with queuing schemes significantly increases the network performance under congested scenarios, compared to when we use deterministic and oblivious routing.

\begin{figure*}[!htb]
\vspace{-.5cm}
\begin{subfigure}[!th]{1\textwidth}
 \centering 
\includegraphics[width=1.0\textwidth]
{leyenda3.pdf}
\end{subfigure}

 \begin{subfigure}[!th]{1\textwidth}
 \centering 
\includegraphics[width=0.8\textwidth]
{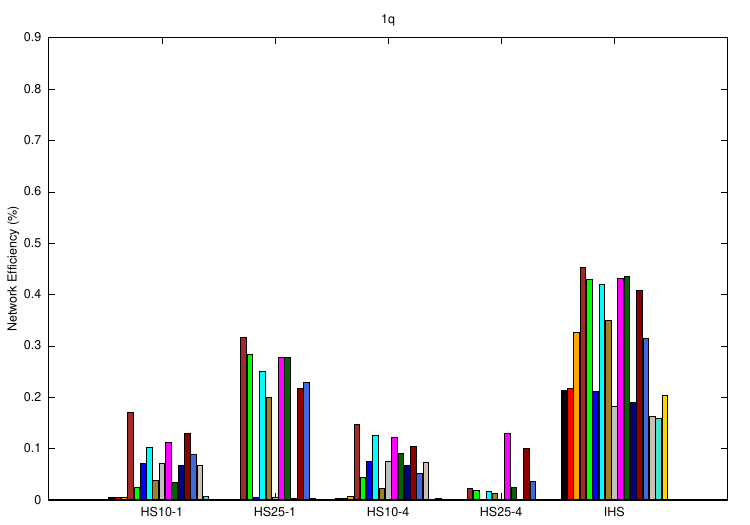}
\caption{1VC.}
\label{fig_histo_1q}
\end{subfigure}
 \begin{subfigure}[!th]{1\textwidth}
 \centering 
\includegraphics[width=0.8\textwidth]
{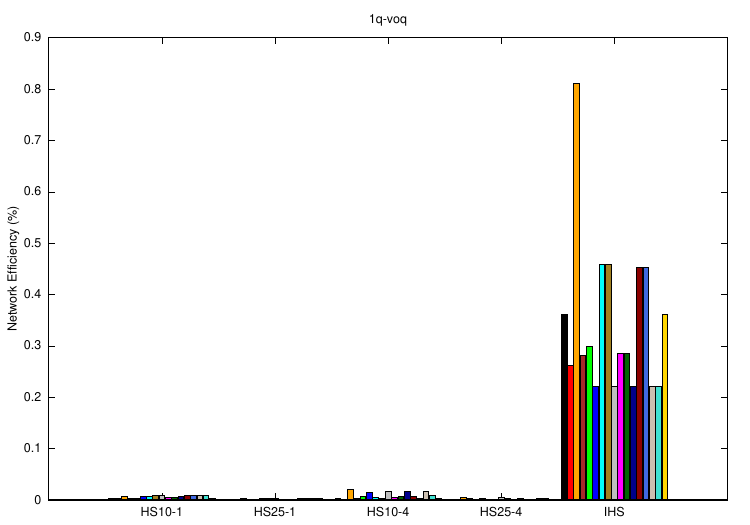}
\caption{1VC and VOQs.}
\label{fig_histo_1q-voq}
\end{subfigure}
\caption{Normalized Throughput after warmup period in a $11664$-node RLFT using 1Q and 1Q-VOQ.}
\label{fig:RLFT_histo_1q}
\end{figure*}

\begin{figure*}[!htb]
\vspace{-.5cm}
\begin{subfigure}[!th]{1\textwidth}
 \centering 
\includegraphics[width=1.0\textwidth]
{leyenda3.pdf}
\end{subfigure}

 \begin{subfigure}[!th]{1\textwidth}
 \centering 
\includegraphics[width=0.8\textwidth]
{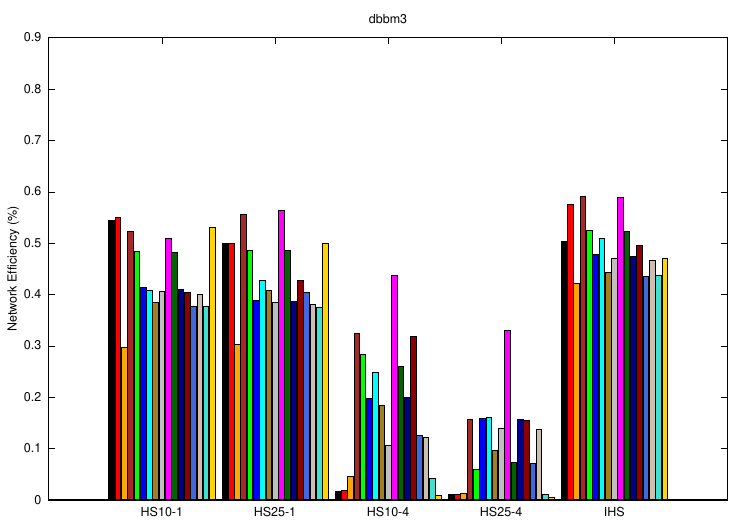}
\caption{DBBM.}
\label{fig_histo_dbbm}
\end{subfigure}
 \begin{subfigure}[!th]{1\textwidth}
 \centering 
\includegraphics[width=0.8\textwidth]
{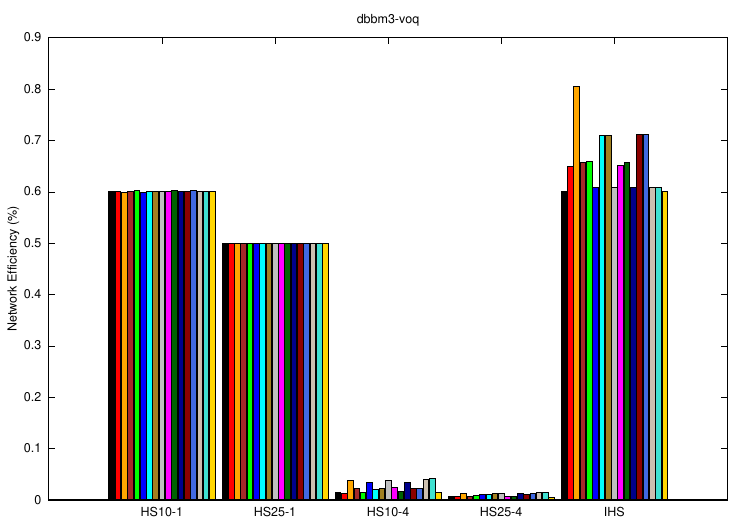}
\caption{DBBM and VOQs.}
\label{fig_histo_dbbm-voq}
\end{subfigure}
\caption{Normalized Throughput after warmup period in a $11664$-node RLFT using DBBM and DBBM-VOQ.}
\label{fig:RLFT_histo_dbbm}
\end{figure*}

\begin{figure*}[!ht]
\vspace{-.5cm}
\begin{subfigure}[!th]{1\textwidth}
 \centering 
\includegraphics[width=1.0\textwidth]
{leyenda3.pdf}
\end{subfigure}

 \begin{subfigure}[!th]{1\textwidth}
 \centering 
\includegraphics[width=0.8\textwidth]
{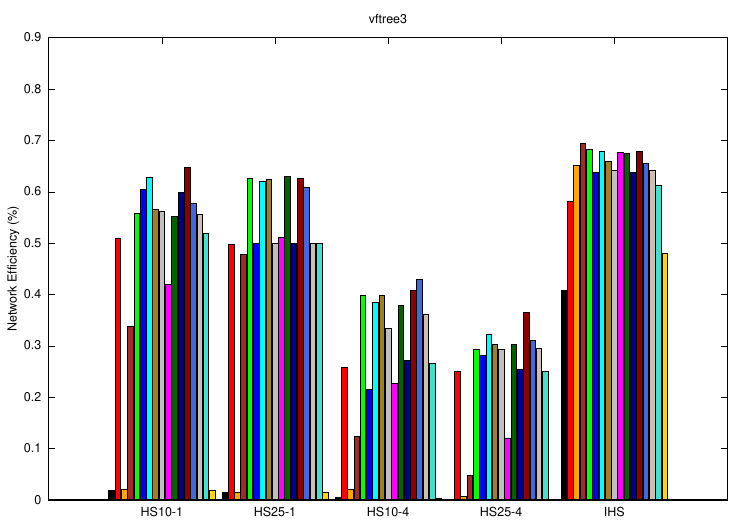}
\caption{vFtree.}
\label{fig_histo_vftree}
\end{subfigure}
 \begin{subfigure}[!th]{1\textwidth}
 \centering 
\includegraphics[width=0.8\textwidth]
{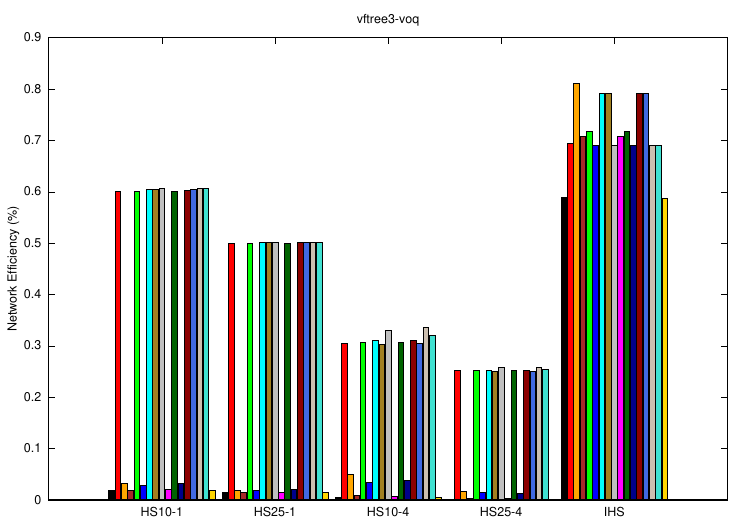}
\caption{vFtree and VOQs.}
\label{fig_histo_vftree-voq}
\end{subfigure}
\caption{Normalized Throughput after warmup period in a $11664$-node RLFT using vFtree and vFtree-VOQ.}
\label{fig:RLFT_histo_vftree}
\end{figure*}

\begin{figure*}[!ht]
\vspace{-.5cm}
\begin{subfigure}[!th]{1\textwidth}
 \centering 
\includegraphics[width=1.0\textwidth]
{leyenda3.pdf}
\end{subfigure}

 \begin{subfigure}[!th]{1\textwidth}
 \centering 
\includegraphics[width=0.8\textwidth]
{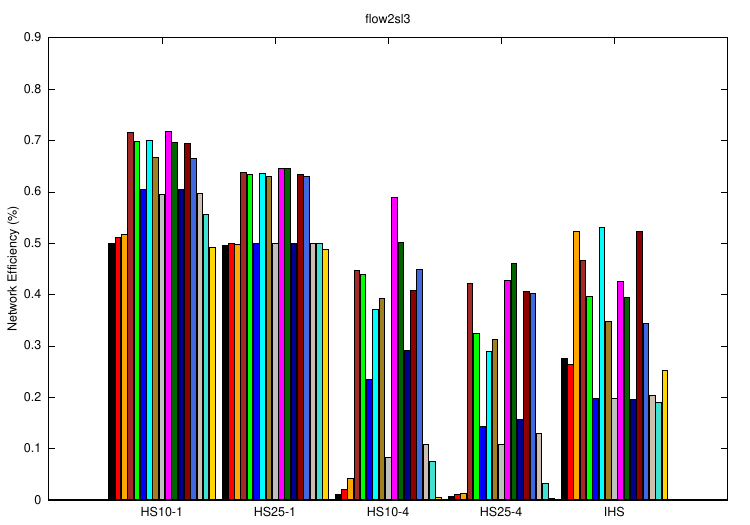}
\caption{Flow2SL.}
\label{fig_histo_flow2sl}
\end{subfigure}
 \begin{subfigure}[!th]{1\textwidth}
 \centering 
\includegraphics[width=0.8\textwidth]
{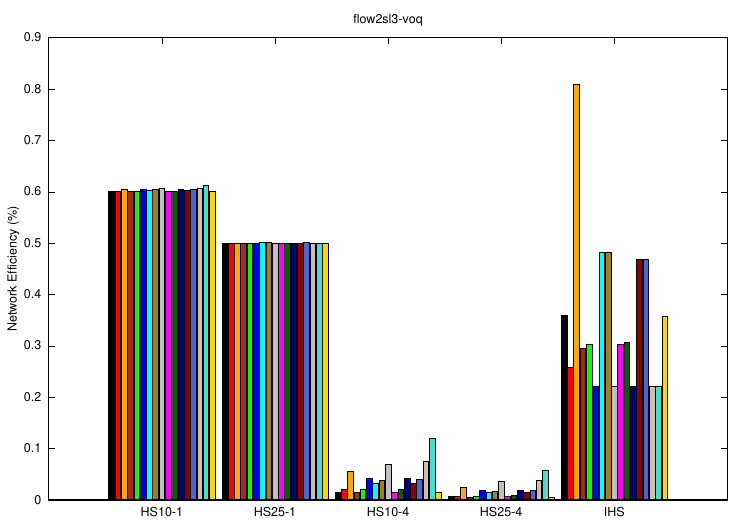}
\caption{Flow2SL and VOQs.}
\label{fig_histo_flow2sl-voq}
\end{subfigure}
\caption{Normalized Throughput after warmup period in a $11664$-node RLFT using Flow2SL and Flow2SL-VOQ.}
\label{fig:RLFT_histo_flow2sl}
\end{figure*}

Figures \ref{fig:RLFT_histo_1q}, \ref{fig:RLFT_histo_dbbm}, \ref{fig:RLFT_histo_vftree} and \ref{fig:RLFT_histo_flow2sl} show in histograms the same results as the Tables \ref{tab:NOVOQ} and \ref{tab:VOQ}. This allows the comparison of different switch architectures for a given queuing scheme

